\documentclass[fleqn,usenatbib]{mnras}

\usepackage{newtxtext,newtxmath}

\usepackage[T1]{fontenc}

\DeclareRobustCommand{\VAN}[3]{#2}
\let\VANthebibliography\thebibliography
\def\thebibliography{\DeclareRobustCommand{\VAN}[3]{##3}\VANthebibliography}

\usepackage{graphicx}	
\usepackage{amsmath}	
\usepackage{xspace}	
\usepackage{color}
\usepackage{lineno}
\usepackage{tablefootnote}
\usepackage{fleqn,natbib}
\usepackage{flushend}
\usepackage{cuted}
\usepackage{threeparttable}
\usepackage{xcolor}
\usepackage{dcolumn}
\usepackage{multicol}
\usepackage{adjustbox}
\usepackage{float}
\usepackage{multirow,longtable,supertabular}

\newcommand{\thisgrb}{GRB~190530A\xspace}
\newcommand{\AstroSat}{{\em AstroSat}\xspace}
\newcommand{\fermi}{{\em Fermi}\xspace}
\newcommand{\kw}{{\em Konus}-Wind\xspace}

\newcommand{\fermiT}{{T$_{\rm 0}$}\xspace}
\newcommand{\keV}{{\rm keV}\xspace}

\newcommand{\swift}{{\em Swift}\xspace}
\newcommand{\tninty}{{$T_{\rm 90}$}\xspace}
\newcommand{\mvts}{{$t_{\rm mvts}$}\xspace}
\newcommand{\lmin}{{$\Gamma_{\rm min}$}\xspace}
\newcommand{\Ep}{$E_{\rm p}$\xspace}
\newcommand{\sw}[1]{\texttt{#1}}

\graphicspath{{./}{figures/}}

\title[Spectro-polarimetric results of \thisgrb]{Probing into emission mechanisms of \thisgrb using time-resolved spectra and polarization studies: Synchrotron Origin?}

\author[Rahul Gupta et al.]{Rahul Gupta,$^{1, 2}$\thanks{E-mail: rahulbhu.c157@gmail.com, rahul@aries.res.in}
S. Gupta,$^{3,4}$
T. Chattopadhyay,$^{5}\thanks{tanmoyc@stanford.edu}$
V. Lipunov,$^{6}$ 
A. J. Castro-Tirado,$^{7,8}$ 
D. Bhattacharya,$^{4}$
\newauthor S. B. Pandey,$^{1}\thanks{shashi@aries.res.in}$
S. R. Oates,$^{9}$
Amit Kumar,$^{1,10}$ 
Y.-D. Hu,$^{7,11}$ 
A. F. Valeev,$^{12,13}$
P. Yu. Minaev,$^{14,15}$
H. Kumar,$^{3}$
\newauthor J. Vinko,$^{16,17,18}$
Dimple,$^{1, 2}$
V. Sharma,$^{19}$
A. Aryan,$^{1, 2}$
A. Castell\'on,$^{20}$
A. Gabovich,$^{6}$ 
A. Moskvitin,$^{13}$
A. Ordasi,$^{16}$
\newauthor A. P\'al,$^{16,17}$
A. Pozanenko,$^{14,21}$
B.-B. Zhang,$^{22,23}$
B. Kumar,$^{1}$
D. Svinkin,$^{24}$
D. Saraogi,$^{3}$
D. Vlasenko,$^{6}$
\newauthor E. Fern\'andez-Garc\'ia,$^{7}$
E. Gorbovskoy,$^{6}$
G. C. Anupama,$^{25}$
K. Misra,$^{1}$
K. S\'arneczky,$^{16}$
L. Kriskovics,$^{16}$
\newauthor M. \'A. Castro-Tirado,$^{7}$
M. D. Caballero-Garc\'ia,$^{7}$
N. Tiurina,$^{6}$
P. Balanutsa,$^{6}$
R. R. Lopez,$^{26}$
R. S\'anchez-Ram\'irez,$^{27}$
\newauthor R. Szak\'ats,$^{16}$
S. Belkin,$^{14}$
S. Guziy,$^{28,29}$
S. Iyyani,$^{4}$
S. N. Tiwari,$^{2}$
Santosh V. Vadawale,$^{30}$
T. Sun,$^{7,31}$
V. Bhalerao,$^{3}$
\newauthor V. Kornilov,$^{6}$ 
and V. V. Sokolov$^{12}$
}
\date{Accepted XXX. Received YYY; in original form ZZZ}

\pubyear{2021}

\begin{document}
\label{firstpage}
\pagerange{\pageref{firstpage}--\pageref{lastpage}}
\maketitle

\begin{abstract}
Multi-pulsed \thisgrb, detected by the GBM and LAT onboard \fermi, is the sixth most fluent GBM burst detected so far. This paper presents the timing, spectral, and polarimetric analysis of the prompt emission observed using \AstroSat and \fermi to provide insight into the prompt emission radiation mechanisms. The time-integrated spectrum shows conclusive proof of two breaks due to peak energy and a second lower energy break. 
Time-integrated (55.43 $\pm$ 21.30 \%) as well as time-resolved polarization measurements, made by the Cadmium Zinc Telluride Imager (CZTI) onboard \AstroSat, show a hint of high degree of polarization. The presence of a hint of high degree of polarization and the values of low energy spectral index ($\alpha_{\rm pt}$) do not run over the synchrotron limit for the first two pulses, supporting the synchrotron origin in an ordered magnetic field. However, during the third pulse, $\alpha_{\rm pt}$ exceeds the synchrotron line of death in few bins, and a thermal signature along with the synchrotron component in the time-resolved spectra is observed. Furthermore, we also report the earliest optical observations constraining afterglow polarization using the MASTER (P $<$ 1.3 \%) and the redshift measurement ($z$= 0.9386) obtained with the 10.4m GTC telescopes. The broadband afterglow can be described with a forward shock model for an ISM-like medium with a wide jet opening angle. We determine a circumburst density of $n_{0} \sim$ 7.41, kinetic energy $E_{\rm K} \sim$ 7.24 $\times 10^{54}$ erg, and radiated $\gamma$-ray energy $E_{\rm \gamma, iso} \sim$ 6.05 $\times 10^{54}$ erg, respectively.
\end{abstract}




\section{Introduction}
\label{intro}

Gamma-ray bursts (GRBs) can be divided into two main categories depending on their gamma-ray duration. Long GRBs (LGRBs, \tninty $>$ 2 s, \citealt{1993ApJ...405..273W}) are thought to be due to the core-collapse of massive stars and are accompanied with the broad-lined Ic supernovae \citep{2006ARA&A..44..507W}. Short GRBs (SGRBs, \tninty $\leq$ 2 s, \citealt{2017ApJ...848L..13A, 2017ApJ...848L..14G}) are thought to be the merger of compact binaries such as two neutron stars (NSs) or a NS and a black hole (BH). Gravitational waves (GW) have also recently been detected to accompany a SGRB \citep{2017ApJ...848L..13A, 2017ApJ...850L...1L}. Since the first detection of GRBs using Vela satellite \citep{1973ApJ...182L..85K} in the 1960s, the prompt emission of GRBs (initial intense, highly variable, and short-lived $\gamma$-ray/hard X-ray emission phase) has been widely studied by several space-based missions, such as Burst and Transient Source Experiment (BATSE) onboard Compton Gamma Ray Observatory (CGRO, \citealt{2013ApJS..208...21G}), Burst alert telescope (BAT, \citealt{2005SSRv..120..143B}) onboard {\em Neil Gehrels Swift observatory} \citep{2004ApJ...611.1005G}, Gamma-ray Burst Monitor (GBM, \citealt{2009ApJ...702..791M}) \& Large Area Telescope (LAT, \citealt{2009ApJ...697.1071A}) onboard \fermi-Gamma-Ray Space Telescope \footnote{\url{https://fermi.gsfc.nasa.gov/}}. However, the radiation mechanism producing the prompt emission of GRBs is still a mystery \citep{2015AdAst2015E..22P}. 
Regardless of the GRB progenitor, according to the standard fireball shock model \citep{2015PhR...561....1K}, a relativistic jet is produced by the central engine \citep{2004RvMP...76.1143P}. The prompt emission is thought to be produced by internal dissipation within the relativistic jet, either via internal shocks or the dissipation of magnetic fields in a Poynting flux-dominated outflow \citep{2015AdAst2015E..22P}. The mechanism responsible for the GRB prompt emission has been a matter of intense discussion for years and is still under debate. Some authors explain the observed prompt emission spectrum using synchrotron radiation model \citep{2019A&A...628A..59O, 2020NatAs...4..174B, 2020NatAs...4..210Z}, while on the other hand, photospheric models (low-energy blackbody emission) similarly can equally describe the observed prompt emission spectrum \citep{2005ApJ...628..847R, 2011MNRAS.415.3693R}. 

The prompt emission spectrum of a GRB is typically described by the empirical \sw{Band} function \citep{Band:1993}. However, discrepancies from this standard spectral model have been observed, such as the presence of an additional thermal component due to photospheric emission \citep{2011MNRAS.415.3693R, 2011MNRAS.416.2078P}; the presence of an additional non-thermal power-law component extending to high energies primarily due to an inverse Compton origin \citep{2010ApJ...716.1178A}; the presence of a sub-GeV spectral \sw{cut-off} (along with the traditional \sw{Band} function), due to pair production within the emitting region \citep{2018ApJ...864..163V, 2020ApJ...903....9C}; and the presence of multiple components due to the overlap from different emission sites in the same burst \citep{2015ApJ...812..156B, 2019ApJ...876...76T}. Recently, \cite{2019A&A...625A..60R} systematically investigated the ten brightest SGRBs and LGRBs detected by \fermi to search for evidence of incomplete cooling of electrons in their prompt emission spectra. They found an additional low-energy break (below the peak energy (\Ep)) in eight LGRBs in their sample. Interestingly, before and after this break, spectral indices are consistent with the photon indices of the synchrotron spectrum (respectively -2/3 and -3/2 below and above the break), supporting a synchrotron origin.

An effective technique to examine the emission mechanisms of GRBs is to study the spectral evolution of the prompt emission. The characteristics of the evolution of \Ep and $\alpha_{\rm pt}$ have been studied by many authors \citep{1986ApJ...301..213N, 1995ApJ...439..307F, 1997ApJ...479L..39C, 2012ApJ...756..112L}. Three general patterns in the evolution of \Ep have been observed: 
(i) an `intensity-tracking' evolution, where \Ep increases/decreases as the flux increases/decreases \citep{1999ApJ...512..693R}; (ii) a `hard-to-soft' evolution, where \Ep decreases continuously \citep{1986ApJ...301..213N}; (iii) a `soft-to-hard' evolution or disordered evolution, where \Ep increases continuously or does not show any correlation with intensity \citep{1994ApJ...422..260K}. The $\alpha_{\rm pt}$ also evolves with time but does not display any typical trends \citep{1997ApJ...479L..39C}. Recently, \cite{2019ApJ...884..109L} and \cite{2021MNRAS.505.4086G} found that both \Ep and $\alpha_{\rm pt}$ track flux (``double-tracking'') for GRB 131231A and GRB 140102A.

Spectral information from the prompt emission together with prompt emission polarization is a powerful tool that can provide a clear view about the long-debated mystery of the emission mechanisms of GRBs. However, we should always be aware of the challenges of polarization measurements. The first detection of prompt emission polarization was reported by {\it RHESSI} satellite for GRB 021206 \citep[highly linearly polarized;][]{2003Natur.423..415C}. This result was challenged in a subsequent study \citep{2004MNRAS.350.1288R}. Since then, prompt emission polarization measurements have only been performed a handful of bursts using: {\it INTEGRAL} \citep{2007A&A...466..895M, 2009ApJ...695L.208G, 2013MNRAS.431.3550G, 2014MNRAS.444.2776G}, GAP onboard {\it IKAROS} \citep{2011PASJ...63..625Y, 2011ApJ...743L..30Y,2012ApJ...758L...1Y}, {\it POLAR} onboard Tiangong-2 space laboratory \citep{2019NatAs...3..258Z, 2019A&A...627A.105B, 2020A&A...644A.124K}, and CZTI onboard \AstroSat \citep{2016ApJ...833...86R, 2019ApJ...884..123C, 2018ApJ...862..154C,2019ApJ...874...70C, 2019ApJ...882L..10S, 2020MNRAS.493.5218S}. 

Recent studies on prompt emission polarization suggest the presence of time-varying (rapid changes in the polarization angle) linear polarization within the burst \citep{2009ApJ...695L.208G, 2019ApJ...882L..10S, 2017Natur.547..425T, 2020A&A...644A.124K}. It indicates that observations of time-integrated polarization could be an artifact of summing over the varying polarization signal \citep{2020A&A...644A.124K}. Therefore, a detailed time-resolved polarization is crucial to understand the radiation mechanisms of GRBs \citep{2021Galax...9...82G}. In this work, we present the time-integrated as well as time-resolved spectro-polarimetric results for the sixth most fluent GBM burst, \thisgrb\footnote{GRB 130427A, GRB 160625B, GRB 160821A, GRB 171010A, and GRB 190114C have higher fluence than \thisgrb, and more importantly all of them are well-studied bursts.} Our spectro-polarimetric analysis is based on observations of \thisgrb performed by \fermi and \AstroSat-CZTI.

The relativistically moving outflow is eventually decelerated by the circumburst medium resulting in the production of external shocks. These shocks are responsible for producing the broadband and well-studied afterglow emission phase (e.g., see \citealt{2015PhR...561....1K} for a review). The external shocks comprise of two different shocks: a long-lived forward shock (FS) that spreads into the circumburst medium and creates a multiwavelength afterglow, and a short-lived reverse shock (RS) that travels backwards through the ejecta and creates a short-lived optical flash  \citep{2003ApJ...595..950Z}. For most GRBs, the FS component alone can usually explain the observed afterglow. Application of the external FS model to the afterglow emission provides detailed info about the late time multiwavelength afterglow, circumburst medium, jet geometry, and blastwave kinetic energy \citep{2007MNRAS.379..331P, Wang_2015}.

In this article, we present the multiwavelength observations and analysis of \thisgrb, including prompt spectro-polarimetric observations using \AstroSat CZTI and optical afterglow observations taken with a variety of telescopes (see \S~\ref{optical observations}). The very bright prompt emission along with LAT GeV photons inspired us to investigate this burst in detail. We find a hint (detection significance $\leq$ 3 $\sigma$) of high degree of polarization at \keV energy range and a ``double-tracking'' characteristic spectral evolution during the prompt emission phase of \thisgrb. In addition, we also constrain a limiting value on afterglow polarization using MASTER observations, making it the first GRB for which both the prompt and afterglow polarization have been investigated. This paper is organized as follows. In \S~\ref{multiwaveength observations and data reduction}, we discuss the multiwavelength observations and data reduction. The main results are presented in \S~\ref{results}. This is followed by discussion in \S~\ref{discussion}, and finally summary \& conclusion in \S~\ref{conclusion}. All the uncertainties are quoted at 1 $\sigma$ throughout this paper unless mentioned otherwise. The temporal ($\rm \alpha$) and spectral indices ($\rm \beta$) for the afterglow are given by the expression $\rm F (t,\nu)\propto t^{-\alpha}\nu^{-\beta}$. We consider the Hubble parameter $\rm H_{0}$ = 70 km $\rm s^{-1}$ $\rm Mpc^{-1}$, density parameters $\rm \Omega_{\Lambda}= 0.73$, and $\rm \Omega_m= 0.27$ \citep{2011ApJS..192...14J}.

\section{\bf multiwavelength Observations and Data Reduction}
\label{multiwaveength observations and data reduction}

\begin{figure}
\centering
\includegraphics[scale=0.36]{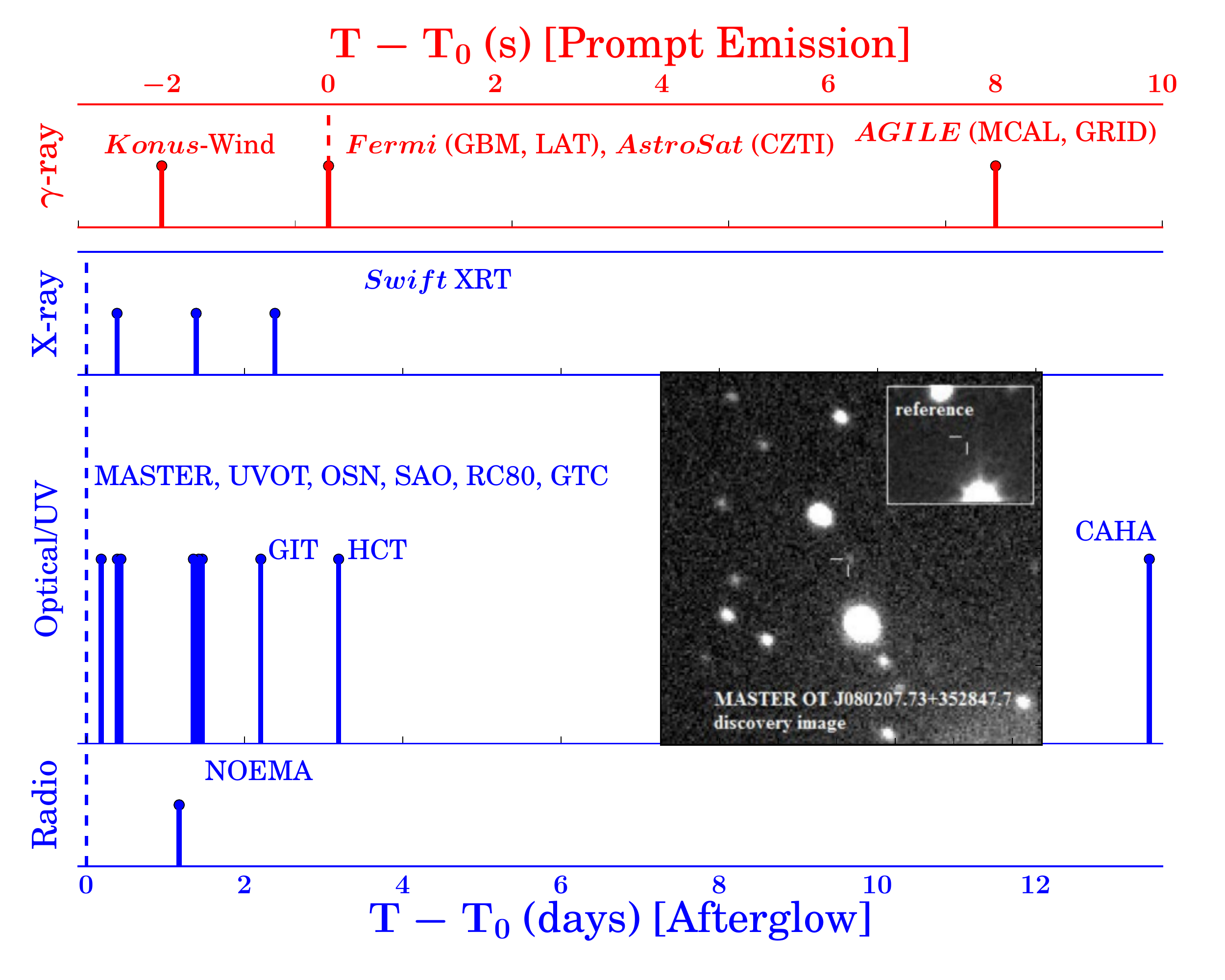}
\caption{{\bf A timeline of events for \thisgrb:} The epochs of prompt (red) and afterglow (blue) observations were taken by various space-based and ground-based facilities. The sky image is the MASTER-Kislovodsk discovery image of the optical afterglow, MASTER OT J080207.73+352847.7, and inset is the reference image (observed at 2010-12-07 01:15:41 UT with unfiltered limiting magnitude m$_{\rm lim}$=21.9 mag). The red and blue vertical dashed line indicates the \fermiT.}
\label{timeline}
\end{figure}

In this section, we present the multiwavelength observations and data reduction for \thisgrb. In Figure \ref{timeline}, we provide a timeline depicting when the various space and ground-based observatories performed observations.

\subsection{\bf $\bf \gamma$-ray/hard X-ray observations}

\begin{figure}
\centering
\includegraphics[scale=0.35]{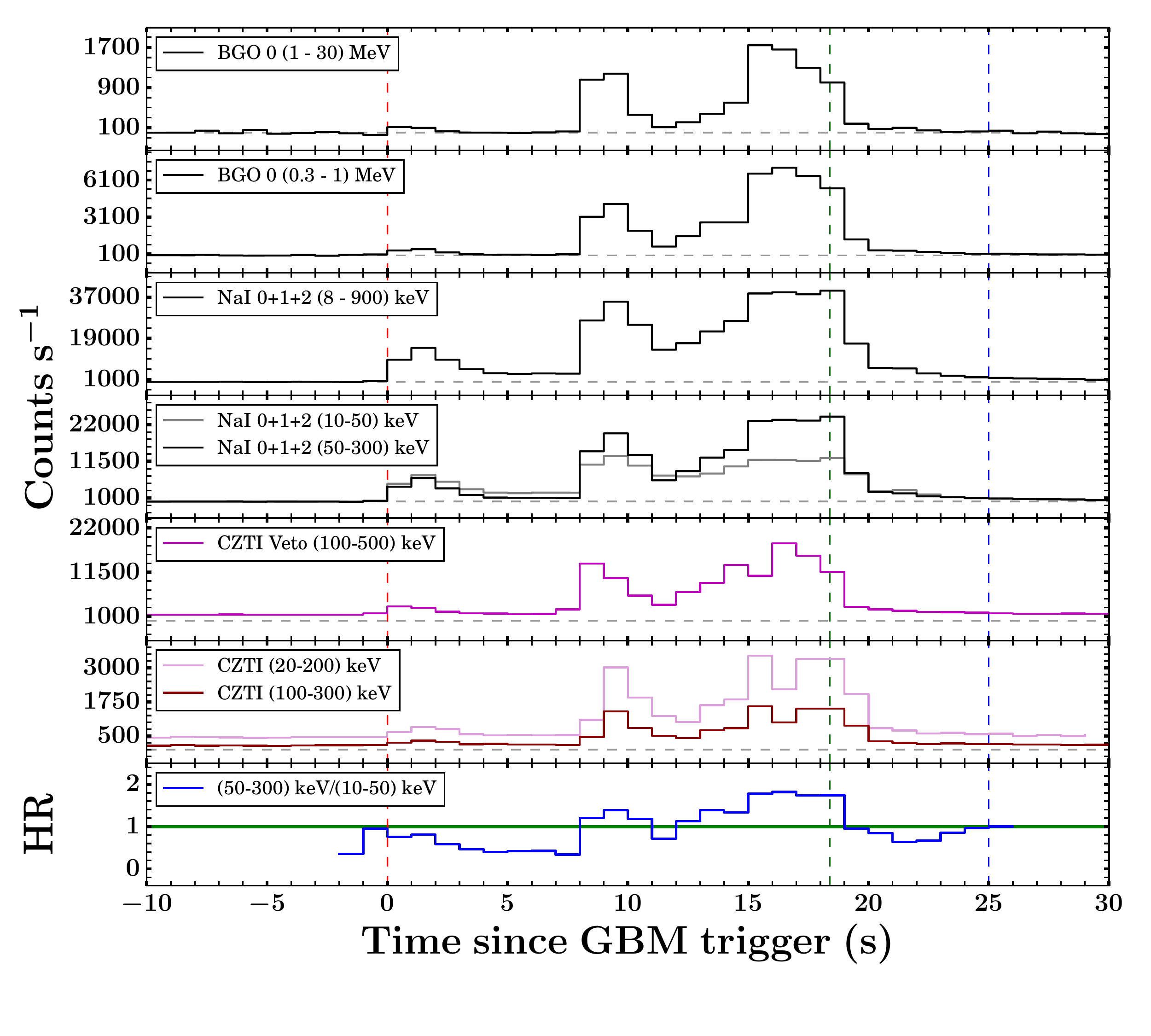}
\includegraphics[scale=0.35]{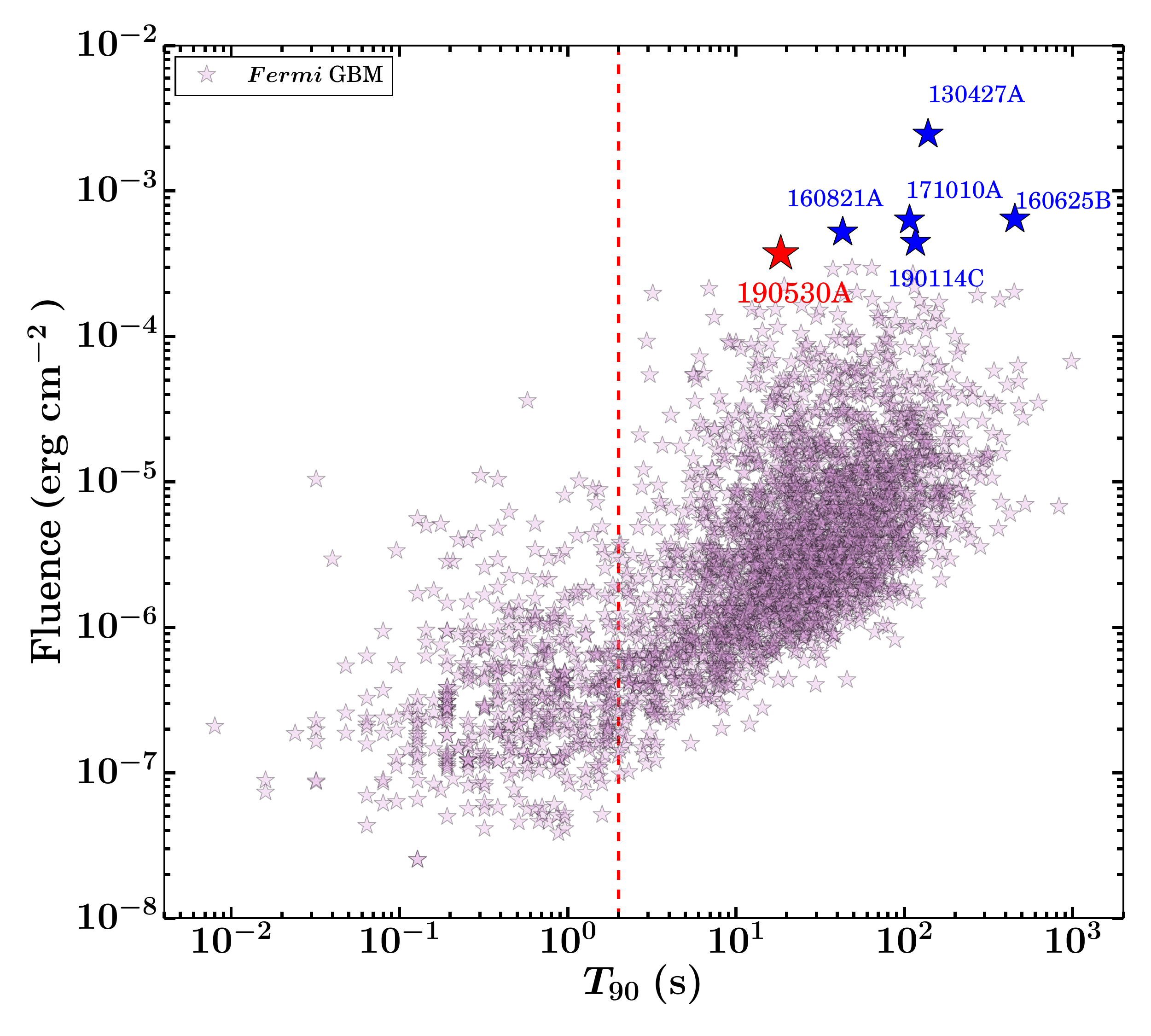}
\caption{{\bf Top: Energy-resolved prompt emission light curves of \thisgrb :} The background-subtracted 1 s binned light curves of \fermi GBM and \AstroSat CZTI detectors provide in multiple energy channels (given in the first six panels). 
The \fermi trigger time (\fermiT) and \tninty durations for the \fermi GBM detector in the 50 - 300 \keV energy range are given by the red, and green vertical dashed lines, respectively. The start and stop times used for the time-averaged spectral analysis are provided by \fermiT and the blue vertical dashed line. The horizontal grey solid lines differentiate between signal and background (at a count rate equal to zero). {\bf Evolution of hardness ratio (HR) :} The bottom panel shows the evolution of HR in hard (50 - 300 $\keV$) to soft (10 - 50 \keV) energy channels of the NaI 1 detector. The horizontal green solid line corresponds to HR equal to one. {\bf Bottom: Fluence distribution for GRBs:} \tninty duration as a function of energy fluence for \fermi detected GRBs in the observer frame. \thisgrb (shown with a red star) is the sixth most fluent GBM burst. The other five most fluent bursts are also highlighted with blue stars. The vertical black dashed line represents the classical boundary between long and short bursts.}
\label{promptlc}
\end{figure}

\thisgrb simultaneously triggered GBM \citep{2009ApJ...702..791M} and LAT \citep{2009ApJ...697.1071A} onboard \fermi at 10:19:08 UT on May 30, 2019 (\fermiT). The best on-ground \fermi GBM position is RA, DEC = 116.9, 34.0 degrees (J2000) with an uncertainty radius of 1$^{\circ}$ \citep{2019GCN.24676....1F, 2019GCN.24679....1L}. The GBM light curve comprises of multiple bright emission peaks with a \tninty duration of 18.4 s (in 50 - 300 \keV energy channel, see Figure \ref{promptlc}). For the time interval \fermiT to \fermiT + 20 s, the time-averaged \fermi GBM spectrum is best fitted with \sw{band} (GRB) function with a low energy spectral index ($\alpha_{\rm pt}$) = -1.00 $\pm$ 0.01, a high energy spectral index ($\beta_{\rm pt}$) = -3.64 $\pm$ 0.12 and a spectral peak energy (\Ep) = 900 $\pm$ 10 \keV. For this time interval, the fluence is $3.72 \pm 0.01 \times 10^{-4}$~erg~cm$^{-2}$, which is calculated in the 10 \keV ~- 10 MeV energy band \citep{2019GCN.24692....1B}. With this fluence, \thisgrb is the sixth brightest burst observed by \fermi-GBM (see Figure \ref{promptlc}, other GBM data points are obtained from GRBweb page\footnote{\url{https://user-web.icecube.wisc.edu/~grbweb_public/index.html}}). This brightness also implies this GRB is suitable for detailed analysis. The best \fermi-LAT on-ground position (RA, DEC = 120.76, 35.5 degrees (J2000) with an uncertainty radius of 0.12$^{\circ}$) was at 63$^{\circ}$ from the LAT boresight angle at the time of \fermiT. The \fermi LAT data show a significant increase in the event rate that is temporally correlated with the GBM \keV emission with high significance \citep{2019GCN.24679....1L}. \thisgrb also triggered \AstroSat-CZTI with a \tninty duration of 23.9 s in the CZTI energy channel \citep{2019GCN.24694....1G}. \thisgrb was also detected by several other $\gamma$-ray/hard X-ray space missions, including: the Mini-CALorimeter and Gamma-Ray Imaging Detector onboard AGILE \citep{2019GCN.24678....1L, 2019GCN.24683....1V}, Insight-HXMT/HE \citep{2019GCN.24714....1Y}  and \kw \citep{2019GCN.24715....1F}. \kw obtained a total energy fluence of $5.57 \pm 0.15 \times 10^{-4}$~erg~cm$^{-2}$ in the 20 - 10000 \keV energy band; it is amongst the highest fluence event detected by \kw \citep{2019GCN.24715....1F}. The prompt emission characteristics of \thisgrb are listed in Table \ref{tab:prompt_properties}.

\begin{table}
\caption{Prompt emission characteristics of \thisgrb. \tninty: Duration from GBM observations in 50 - 300 \keV; \mvts : minimum variability time scale in 8 - 900 \keV; HR: ratio of the counts in hard (50 - 300 \keV) to the counts in soft (10 - 50  \keV) energy range; \Ep:  peak energy obtained using joint \fermi GBM and LAT observations from \fermiT to \fermiT+25 s; $F_{\rm p}$: peak flux in $\rm 10^{-6} erg ~cm^{-2}$ using GBM data in the 1 \keV -10 MeV energy range in the rest frame; $E_{\rm \gamma, iso}$: Isotropic $\gamma$-ray energy in the rest frame; $L_{\rm p, iso}$: Isotropic $\gamma$-ray peak luminosity in the rest frame; $z$: redshift of the burst obtained using GTC spectrum.}
\label{tab:prompt_properties}
\begin{center}
\begin{tabular}{|c|c|c|}
\hline
\bf {Prompt Properties} & \bf {\thisgrb }& \bf {Detector} \\
\hline 
\hline
\tninty (s) & 18.43 $\pm$ 0.36 & GBM \\ \hline 
\mvts (s) &  $ \sim $ 0.50 & GBM \\ \hline 
HR  &   1.35 & GBM\\ \hline
\Ep (\keV) & $888.36_{-11.94}^{+12.71}$ & GBM+LAT \\ \hline
$F_{\rm p}$  & 135.38 & GBM \\ \hline
$E_{\rm \gamma, iso}$ ($\rm erg$) & $6.05 \times 10^{54}$ &-\\ \hline
$L_{\rm p, iso}$ ($\rm erg ~s^{-1}$) & 6.26 $ \times 10^{53}$ & -\\ \hline
Redshift $z$ & 0.9386 & GTC  \\ \hline
\end{tabular}
\end{center}
\end{table}

\subsubsection{\bf \fermi Large Area Telescope analysis}
\label{section:LAT}

\begin{figure}
\centering
\includegraphics[scale=0.39]{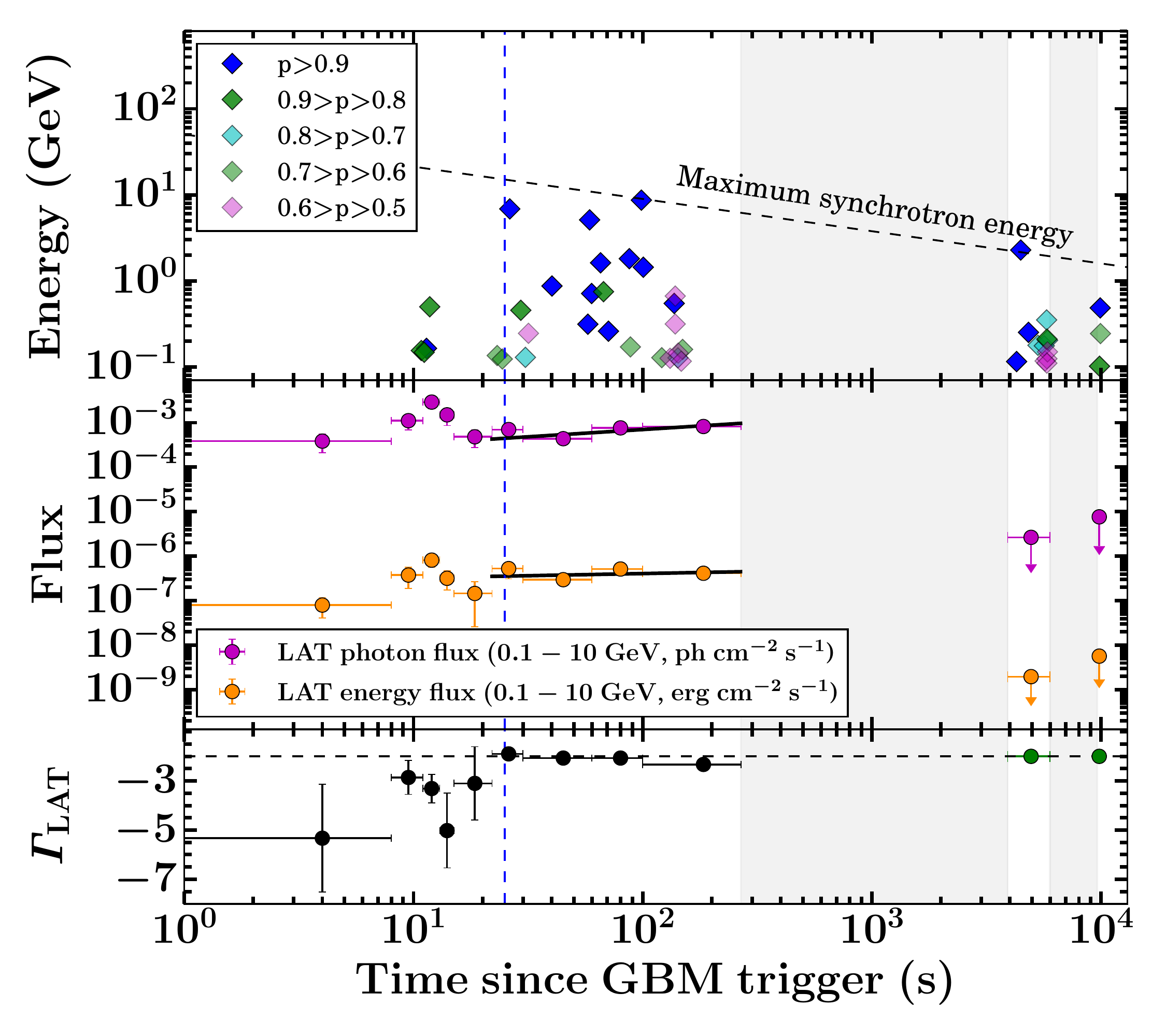}
\caption{\textit {Top panel}: Temporal distribution of \fermi-LAT photons with energies $> 100$ MeV and their association probabilities with \thisgrb. \textit{Middle panel}: Evolution of the \fermi LAT energy and photon fluxes in 0.1 - 10 GeV energy range. For the last two temporal bins, the LAT photon index was fixed to $-2$ to get an upper limit on the flux values. The black lines indicate the simple power-law fit to the extended (the photons detected after the end of the prompt emission) \fermi LAT photon and the energy flux light curves. \textit{Bottom panel}: Temporal evolution of the \fermi LAT photon indices in the 0.1 - 10 GeV range. The vertical blue dashed line represents the end epoch of the prompt emission phase (at \fermiT+25 s). Grey regions show the intervals having angle between the GRB position and the boresight of the \fermi LAT (off-axis angle) greater than 65$^{\circ}$.}
\label{fig:LAT_LCs}
\end{figure}

For \thisgrb, the \fermi LAT data from \fermiT to \fermiT+10ks was retrieved from the \fermi LAT data server\footnote{\url{https://fermi.gsfc.nasa.gov/cgi-bin/ssc/LAT/LATDataQuery.cgi}} using the \sw{gtburst} \footnote{\url{https://fermi.gsfc.nasa.gov/ssc/data/analysis/scitools/gtburst.html}} GUI software. We analyzed the \fermi LAT data using the same software. To carry out an unbinned likelihood investigation, we selected a region of interest (ROI) of $\rm 12^{\circ}$ around the enhanced \swift XRT position \citep{2019GCN.24689....1M}. We cleaned the LAT data by placing an energy cut, selecting only those photons in the energy range 100 MeV - 300 GeV. In addition, we applied an angular cut of 120$^{\circ}$ between the GRB location and zenith of the satellite to reduce the contamination of photons arriving from the Earth limb, based on the navigation plot. For the full-time intervals, we employed the \sw{P8R3\_SOURCE\_V2} response (useful for longer durations $\rm \sim 10^{3} ~s$), and for short temporal bins, we used the \sw{ P8R2\_TRANSIENT020E\_V6} response (useful for small durations $<$ 100 s). We calculated the probability of the high-energy photons being related to the source with the help of the \sw{gtsrcprob} tool. In Figure \ref{fig:LAT_LCs}, we have shown the temporal distribution of the LAT photons for a total duration of 10 ks since \fermiT. LAT observed the high energy emission simultaneously with the \fermi GBM. LAT detected few photons with energy above 1 GeV and the highest-energy photon with an energy of 8.7 GeV (the emitted photon energy is 16.87 GeV in the rest frame at $z$ = 0.9386), which is observed 96 s after \fermiT \citep{2019GCN.24679....1L}. In the time interval of our analysis (\fermiT to \fermiT+10ks), we calculated the energy and photon flux in 100 MeV - 10 GeV energy range of $(5.60 \pm 1.02) ~\times 10^{-9}$  $\rm ~erg ~cm^{-2} ~s^{-1}$ and $(9.78 \pm 1.73) ~ \times 10^{-6}$  $\rm ~ph. ~cm^{-2} ~ s^{-1}$, respectively. For this temporal window, the LAT photon index ($\it \Gamma_{\rm LAT}$) is $-2.21 \pm 0.14$ with a test-statistic (TS) of detection 149. The LAT spectral index, $\beta_{\rm LAT} = \it \Gamma_{\rm LAT} + 1$, is $-1.21 \pm 0.14$. Furthermore, to probe the origin of the LAT photons (see Table A1 in the appendix), we examine the time-resolved analysis using the \fermi LAT observations in \S~\ref{lat_TRS}. In \S~\ref{Comparison with LAT catalogue}, we compare the high energy properties of \thisgrb with a well-studied sample of \fermi LAT catalogue.

\subsubsection{\bf \fermi Gamma-ray Burst Monitor and Joint spectral analysis}

We obtained the time-tagged event (TTE) mode \fermi GBM data from the \fermi GBM trigger catalogue\footnote{\url{https://heasarc.gsfc.nasa.gov/W3Browse/fermi/fermigtrig.html}} using \sw{gtburst} software. TTE data have high time precision in all the 128 energy channels. We studied the temporal and spectral prompt emission properties of \thisgrb using the three brightest sodium iodide detectors (NaI 0, 1, and 2) with source observing angles, NaI 0: $\rm 39^\circ$ degree, NaI 1: $\rm 15^\circ$ degree, NaI 2: $\rm 34^\circ$ degree, respectively. We also selected the brightest bismuth germanate detector (BGO 0) as this BGO detector is closer to the direction of the burst (an observing angle of $\rm 49^\circ$ degree). The angle restrictions are to ignore the systematics coming due to uncertainty in the response at large angles.

We used \sw{RMFIT} version 4.3.2 software\footnote{\url{https://fermi.gsfc.nasa.gov/ssc/data/analysis/rmfit/}} to create the energy-resolved prompt emission light curve using \fermi GBM observations.
The \fermi GBM energy-resolved (background-subtracted) light curves along with the evolution of hardness ratio (HR) are presented in Figure \ref{promptlc}. The prompt emission light curve consists of three bright overlappings peaked structures, a soft and faint peak (lasting up to $\sim$ 4 s after \fermiT) followed by two merging hard peaks with a total duration of $\sim$ 18 s. The hardness ratio (HR) evolution indicates that the peaks are in increasing HR (softer to harder trend), which is also evident from the very low signal for the first peak in the BGO data.

For the spectral analysis, we used the same NaI and BG0 detectors as used for the temporal analysis. We reduced the time-averaged \fermi GBM spectra (from \fermiT to \fermiT+ 25 s) using \sw{Make spectra for XSPEC} tool of \sw{gtburst} software from \emph{Fermi Science Tools}. The background (around the burst main emission) is fitted by selecting two temporal intervals, one interval before the GRB emission and another after the GRB emission. We performed the modelling of the joint GBM and LAT time-averaged spectra using the Multi-Mission Maximum Likelihood framework \citep[\sw{3ML}\footnote{\url{https://threeml.readthedocs.io/en/latest/}}]{2015arXiv150708343V} software to investigate the possible emission mechanisms of \thisgrb. We began by modelling the time-averaged GBM spectrum with the \sw{Band} or \sw{GRB} function \citep{Band:1993}, and included various other models such as \sw{Black Body} in addition to the \sw{Band} function to search for thermal component in the burst; a power-law with two breaks (\sw{bkn2pow}\footnote{\url{https://heasarc.gsfc.nasa.gov/xanadu/xspec/manual/node140.html}}), and cutoff-power law model (\sw{cutoffpl}) or their combinations based upon model fit, residuals of the data, and their parameters (see Table A2 of the appendix). The \sw{bkn2pow} is a continual model that consist of two sharp spectral breaks (hereafter $E_{\rm break, 1}$, and $E_{\rm break, 2}$ or \Ep, respectively) and three power-laws indices (hereafter $\alpha_{1}$, $\alpha_{2}$, and $\alpha_{3}$ respectively). Where $\alpha_{1}$ is power-law index below the $E_{\rm break, 1}$, $\alpha_{2}$ is power-law index between $E_{\rm break, 1}$ and $E_{\rm break, 2}$, and $\alpha_{3}$ is power-law index above the $E_{\rm break, 2}$, respectively. The statistics Bayesian information criteria \citep[BIC;][]{Kass:1995}, and Log (likelihood) is used for optimization, testing, and to find the best fit model of the various models used. Furthermore, we also calculated the goodness of fit value using the \sw{GoodnessOfFit}\footnote{\url{https://threeml.readthedocs.io/en/v2.2.4/notebooks/gof_lrt.html}} class of the Multi-Mission Maximum Likelihood framework. We consider GBM spectrum over the energy range of 8 - 900 \keV (NaI detectors) and 250 - 30000 \keV (BGO detectors) for the spectral analysis. However, we ignore the 33–40 \keV energy range due to the presence of the iodine K-edge at 33.17 \keV while analyzing the NaI data. We consider 100 MeV - 100 GeV energy channels for the \fermi LAT observations.  

\begin{figure}
\centering
\includegraphics[scale=0.45]{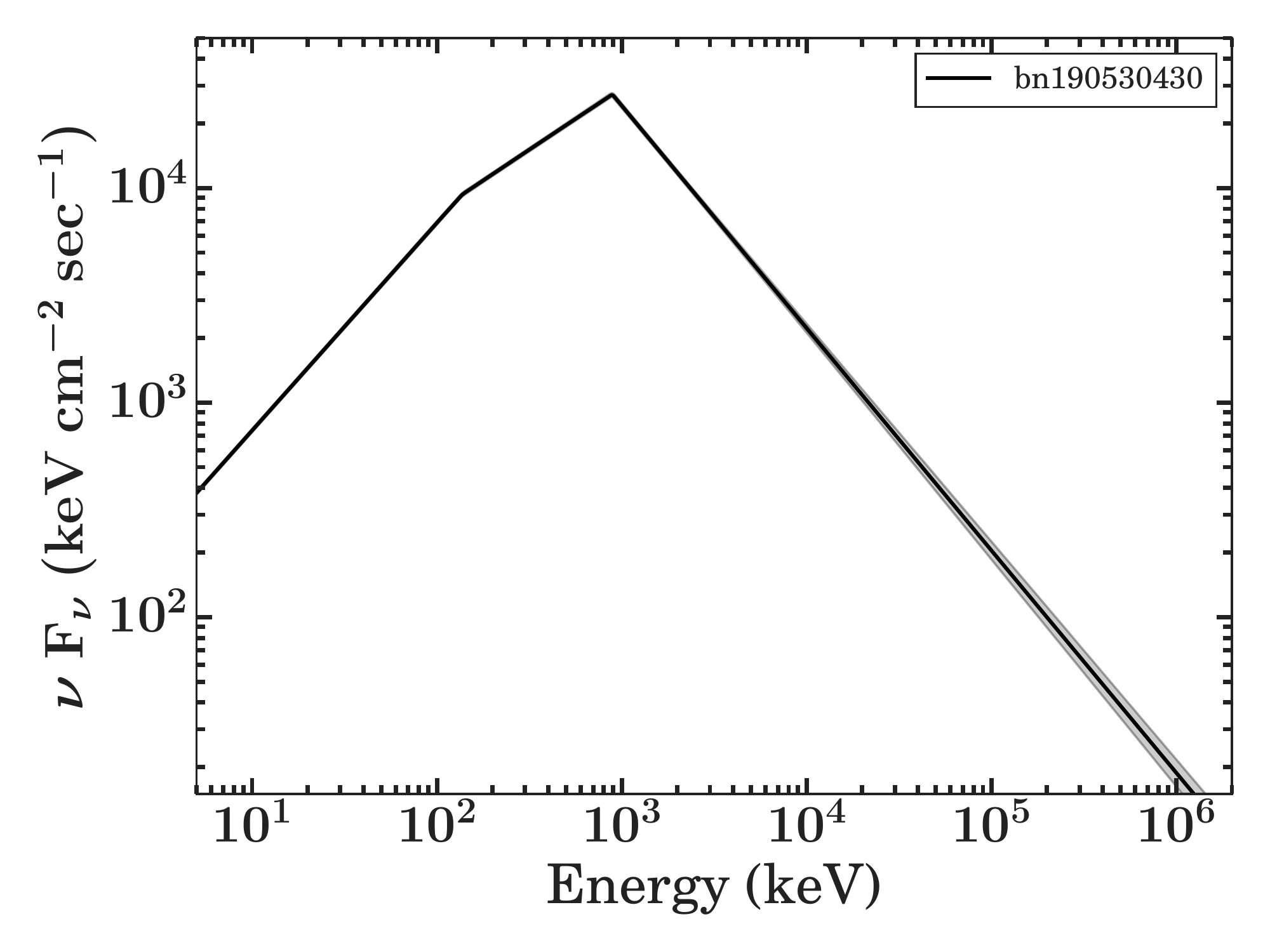}
\caption{The time-integrated best-fit energy spectrum of \thisgrb in model space modelled with a \sw{bkn2pow} model, a broken power-law model with two sharp breaks for an interval of 25 s (from \fermiT to \fermiT + 25 s) using joint spectral analysis of \fermi GBM and LAT data. The shaded grey region shows the 1 $\sigma$ uncertainty region. The legend indicates the \fermi trigger name of \thisgrb.}
\label{TAS_bkn2pow}
\end{figure} 

The best-fit spectral parameters of the joint analysis are presented in the appendix. We found that of all of the eight models used, the \sw{bkn2pow} model, a broken power-law model with two sharp breaks has the lowest BIC value. Therefore, we conclude that the time-averaged spectrum of \thisgrb is best described with \sw{bkn2pow} function with $\alpha_{1}$= 1.03$^{+0.01}_{-0.01}$, $\alpha_{2}$=1.42$^{+0.01}_{-0.01}$, $\it \beta_{\rm pt}$= 3.04$^{+0.02}_{-0.02}$, low-energy spectral break ($E_{\rm break, 1}$) = 136.65$^{+2.90}_{-2.88}$, and high-energy spectral break or peak energy ($E_{\rm break, 2}$) = 888.36$^{+12.71}_{-11.94}$. We noticed that the values of $\alpha_{1, 2}$ are consistent with the power-law indices expected for synchrotron emission. The best-fit time averaged spectra in model space is shown in Figure \ref{TAS_bkn2pow}. Next, we perform a detailed time-resolved analysis to search the low-energy spectral break with two different (coarser and finer) bin sizes.

\subsubsection{\bf {Time-resolved Spectroscopy of \thisgrb and Spectral Parameters Evolution}}
\label{trs_section}

The mechanisms producing the GRB prompt emission is still an open question \citep{2015AdAst2015E..22P}. The emission can be equally well described by a non-thermal synchrotron model \citep{2020NatAs...4..174B} as well as a thermal photospheric model \citep{2007MNRAS.379...73G}. Time-resolved spectral analysis of prompt emission is a propitious method to study the possible radiation mechanisms and investigate correlations between different spectral parameters. There are several methods to bin the prompt emission light curve, such as constant cadence, signal-to-noise (S/N), Bayesian blocks, and Knuth bins. Of these methods, the Bayesian blocks algorithm is the best method to identify the intrinsic intensity change in the prompt emission light curve \citep{2014MNRAS.445.2589B}.

Initially, we rebinned the total emission interval (from \fermiT to \fermiT+25 s) based on the constant cadence method with a coarse bin size of 1 s to perform the time-resolved spectral analysis. This provides a total of 25 spectra; however, the last five seconds of binned spectra do not have significant counts to be modelled. We used \sw{gtburst} to produce the 25 spectra. We modelled each spectrum with a \sw{Band} function and included various other models (\sw{Black Body}, and \sw{bkn2pow} or their combinations with \sw{Band} function) as we did for time-averaged spectral analysis, if required. We find that out of twenty modelled spectra, four spectra (0-1 s, 8-9 s, 9-10 s, and 14-15 s) were best fit by \sw{bkn2pow} model, indicating the presence of a low-energy spectral break, and the rest of the temporal bins are well described with the \sw{Band} function only.\footnote{8-9 s bin is equally described with both functions.} For the best fit \sw{bkn2pow} model, the calculated mean values of the four spectra are $<\alpha_{1}>$ = 0.93 (with $\sigma$ = 0.03), $<\alpha_{2}>$ = 1.38 (with $\sigma$ = 0.03), and $E_{\rm break, 1}$ = 106.00 (with $\sigma$ = 3.14), where $\sigma$ denotes the standard deviation. When calculating mean values, we have excluded the first bin spectrum (0-1 s) as it has $E_{\rm break, 1}$ less than 20 \keV (close to the lower edge of GBM detector). The calculated mean values of $<\alpha_{1}>$ and $<\alpha_{2}>$ are consistent with the power-law indices expected for synchrotron emission. The spectral parameters and their associated errors are listed in Tables A3 and A4 of the appendix.

Furthermore, we rebinned the light curve for the detector with a maximum illumination (i.e. NaI 1) based on the Bayesian blocks algorithm integrated over the 8 - 900 \keV. This provides 53 spectra; however, some of the temporal bins do not have sufficient counts to be modelled. Therefore, we combined these intervals, resulting in a total of 41 spectra for time-resolved spectroscopy. We find that out of 41 modelled spectra, five spectra have significant requirement for a low-energy spectral break ($\Delta$BIC$_{\rm Band/Black Body-Bkn2pow} \ge$6), three spectra are equally fitted with \sw{Band}+ \sw{Black Body} or \sw{bkn2pow} models ($\Delta$BIC$_{\rm Bkn2pow-Band+Black Body} \leq$6), and six spectra are equally fitted with \sw{Band} or \sw{bkn2pow} models ($\Delta$BIC$_{\rm Bkn2pow-Band} \leq$6).
The rest of the temporal bins are well described with the \sw{Band} function only. For the bins with signature of low-energy spectral break, the calculated mean values of the fourteen spectra are $<\alpha_{1}>$ = 0.84 (with $\sigma$ = 0.04), $<\alpha_{2}>$ = 1.43 (with $\sigma$ = 0.06), and $E_{\rm break, 1}$ = 79.51 (with $\sigma$ = 11.57). In this case, also, the calculated mean values are $<\alpha_{1}>$ and $<\alpha_{2}>$ are consistent with the power-law indices expected for synchrotron emission in a marginally fast cooling spectral regime. We also calculated the various spectral parameters such as \Ep, $\alpha_{\rm pt}$, and $\beta_{\rm pt}$ by modelling each spectra using \sw{3ML} software. The spectral parameters and their associated errors are listed in Tables A5 and A6 of the appendix. Figure \ref{TRS_FIG_190530A} shows the evolution of spectral parameters such as \Ep, $\alpha_{\rm pt}$, and $\beta_{\rm pt}$ along with the light curve for three brightest NaI detector in 8 - 900 \keV energy ranges. The value of \Ep changes throughout the burst. The \Ep evolution follows an intensity tracking trend throughout the emission episodes. The evolution of $\alpha_{\rm pt}$ also follows an intensity-tracking trend, and it is within the synchrotron fast cooling and line of death for synchrotron slow cooling (though, in the case of the third pulse, in some of the bins, $\alpha_{\rm pt}$ becomes shallower and exceeds the line of death for synchrotron slow cooling); therefore, the emission of \thisgrb may have a synchrotron origin for the first two-pulses. 

\begin{figure}
\centering
\includegraphics[scale=0.3]{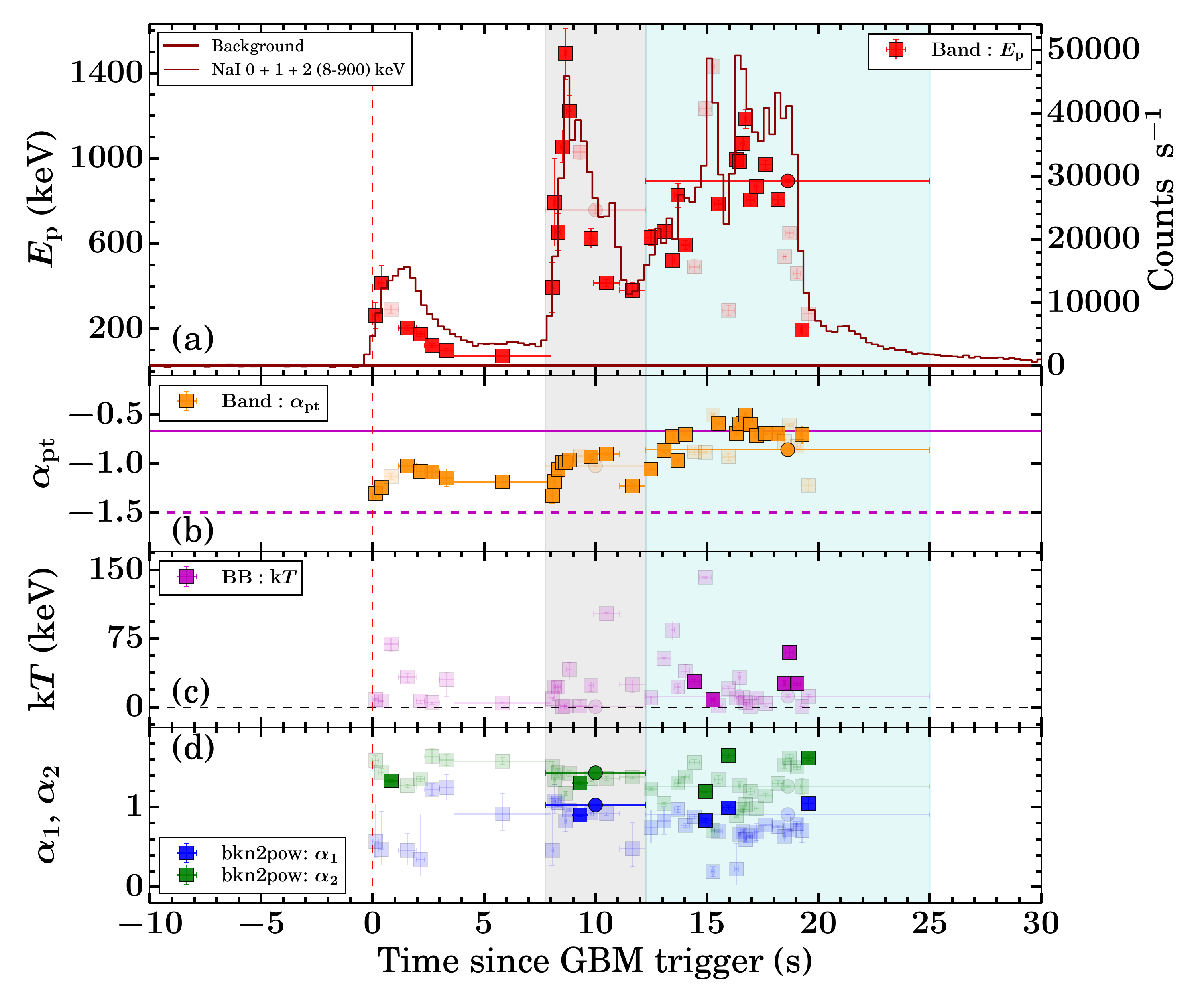}
\caption{{\bf Evolution of the spectral parameters using \fermi GBM data.} (a) The peak energy evolves with time and follows intensity tracking trends through the main emission episode. (b) The low energy spectral index also evolves and follows intensity tracking behaviour. The two horizontal lines are the line of death for synchrotron fast cooling ($\alpha_{\rm pt}$ = -3/2, magenta dashed line) and the line of synchrotron slow cooling ($\alpha_{\rm pt}$ = -2/3, magenta solid line). (c) The evolution of k$T$ (\keV) is obtained from the \sw{Black Body} component. The horizontal black dashed line shows k$T$ = 0 \keV. (d) Evolution of the photon indices ($\alpha_{1}$ and $\alpha_{2}$) for the \sw{bkn2pow} model. In respective panels, if a particular model is best fitted to a particular bin, data points are highlighted with dark colour, otherwise shown with light colour. The temporal binning has been performed based on the Bayesian block algorithm. The red dashed line shows the \fermiT. The vertical grey (best fit with the \sw{bkn2pow} model) and cyan (best fit with the \sw{Band} model) shaded regions show the intervals used for time-resolved polarization measurements using CZTI data, respectively. For these bins, spectral parameters are shown with circles in respective sub-panel (a,b,c, and d).}
\label{TRS_FIG_190530A}
\end{figure}

\subsection{\bf \AstroSat-Cadmium Zinc Telluride Imager}

\thisgrb detection had been confirmed from the ground analysis of the data of \AstroSat CZTI. The CZTI light curve observed multiple pulses of emission likewise followed by \fermi GBM prompt emission (see Figure \ref{promptlc}). The substantial peak was detected at 101:19:25.5 UT, having a count rate of 2745 counts per second of the combined data of all the four quadrants of the CZTI above the background \citep{2019GCN.24694....1G}. We calculated \tninty duration 23.9 s using the cumulative count rate. We found that 1246 Compton events are associated with this burst within the time-integrated duration. In addition, the CsI anticoincidence (Veto) detector working in the energy range of 100-500 \keV also detected this burst.

\subsubsection{\bf Prompt Emission Polarization Measurements}

During its ground calibration, the \AstroSat CZTI was shown to be a sensitive on-axis GRB polarimeter in the 100 - 350 \keV energy range \citep{2014ExA....37..555C, 2015A&A...578A..73V}. The azimuthal angle distribution of the Compton scattering events between the CZTI pixels is used to estimate the polarization. The detection of the polarization in Crab pulsar and nebula in the energy range of 100-380 \keV provided the first onboard verification of its X-ray polarimetry capability \citep{2018NatAs...2...50V}. CZTI later reported the measurement of polarization for a sample of 11 bright GRBs from the first year \AstroSat GRB polarization catalogue \citep{2019ApJ...884..123C}. The availability of simultaneous background before and after the GRB's prompt emission and the significantly higher signal to background contrast for GRBs compared to the persistent X-ray sources makes CZTI sensitive for polarimetry measurements even for the moderately bright GRBs.
To estimate the polarization fraction and to correct for the azimuthal angle distribution for the inherent asymmetry of the CZTI pixel geometry \citep{2014ExA....37..555C}, polarization analysis with CZTI for GRBs (see below) involves a Geant4 simulation of the \AstroSat mass model. Recently, a detailed study was carried on a large GRB sample covering the full sky based on imaging and spectroscopic analysis to validate the mass model (see \citet{2021JApA...42...93M, 2021JApA...42...82C}). The results are encouraging and boost confidence in the GRB polarization analysis. \citet{2019ApJ...884..123C} discusses the GRB polarimetry methodology in detail. Here we only give a brief description of the steps involved for polarization analysis for \thisgrb.  

\begin{enumerate}

\item The polarization analysis procedure begins by selecting the valid Compton events that are first identified as double-pixel events occurring within the 20 $\mu$s time window. The double-pixel events are further filtered against several Compton kinematics conditions like the energy of the events and distance between the hit pixels \citep{2014ExA....37..555C, 2019ApJ...884..123C}.

\item The above step is applied on both the burst region obtained from the light curve of \thisgrb (see \S~\ref{Prompt emission Polarization}) and at least 300 seconds of pre and post-burst background interval. The raw azimuthal angle distribution from the valid event list for the background region is subtracted from the GRB region.
    
\item The background-subtracted prompt emission azimuthal distribution is then normalized by an unpolarized raw azimuthal angle distribution to correct for the CZTI detector pixel geometry induced anisotropy seen in the distribution \citep{2014ExA....37..555C}. The unpolarized distribution is obtained from the \AstroSat mass model by simulating 10$^9$ unpolarized photons in Geant4 with the incident photon energy distribution the same as the GRB spectral distribution (modelled as Band function) and for the same orientation with respect to the spacecraft.
    
\item A sinusoidal function fits the corrected azimuthal angle distribution to calculate the modulation amplitude ($\mu$) and polarization angle in the CZTI plane using Markov chain Monte Carlo (MCMC) simulation.

\item To determine that the GRB is polarized, we calculate the Bayes factor for the sinusoidal and a constant model representing the polarized and unpolarized radiation \citep[see section 2.5.3 of][]{2019ApJ...884..123C}. Suppose the Bayes factor is found to be greater than 2. In that case, we estimate the polarization fraction by normalizing $\mu$ with $\mu_{100}$ (where $\mu_{100}$ is the modulation factor for 100 \% polarized photons obtained from Geant4 simulation of the \AstroSat mass model for 100 \% polarized radiation (10$^9$ photons) for the same GRB spectral distribution and orientation). For a GRB with Bayes factor $<$ 2, an upper limit of polarization is computed \citep[see][for the details of the upper limit of calculation]{2019ApJ...884..123C}. 

\end{enumerate}

\subsection{\bf Soft X-ray observations}
We obtained the X-ray afterglow data (both the light curve and spectrum) products from the \swift XRT online repository \footnote{\url{https://www.swift.ac.uk/xrt_curves/}  \url{https://www.swift.ac.uk/xrt_spectra/}} hosted by the University of Leicester \citep{eva07, eva09}. 

\subsubsection{\bf \swift X-ray Telescope}

The {\em Neil Gehrels Swift observatory} \citep[henceforth \swift;][]{2004ApJ...611.1005G} initiated a ToO observation to search for the X-ray and UV/optical counterparts of the \fermi GBM and LAT detected \thisgrb $\sim$ 33.8 ks after the GBM trigger \citep{2019GCN.24681....1E}. The X-ray telescope \citep[XRT;][]{2005SSRv..120..165B} onboard \swift observed 4.9 ks of Photon Counting (PC) mode data starting from \fermiT + 33.8 ks in soft X-rays (0.3 - 10 \keV). The XRT discovered four new uncatalogued X-ray objects. Of these four sources, only one is detected above the RASS limit, and therefore, was considered to be likely X-ray afterglow. The location of this source was coincident with the position of the optical afterglow candidate reported by MASTER \citep{2019GCN.24680....1L}. The \swift XRT enhanced position for this source, obtained using the alignment of XRT-UVOT data, is at RA, DEC = 120.53242, +35.47947 degrees (J2000) with an uncertainty radius of $\rm 1.4^"$ (90 \% confidence level). This location is 11.3 arcmin from the \fermi LAT position. The X-ray afterglow candidate was monitored until \fermiT+ $\sim$ 2.2 $\times$ 10$^{5}$ $\rm s$ \citep{2019GCN.24689....1M}.

\begin{figure}
\centering
\includegraphics[scale=0.32]{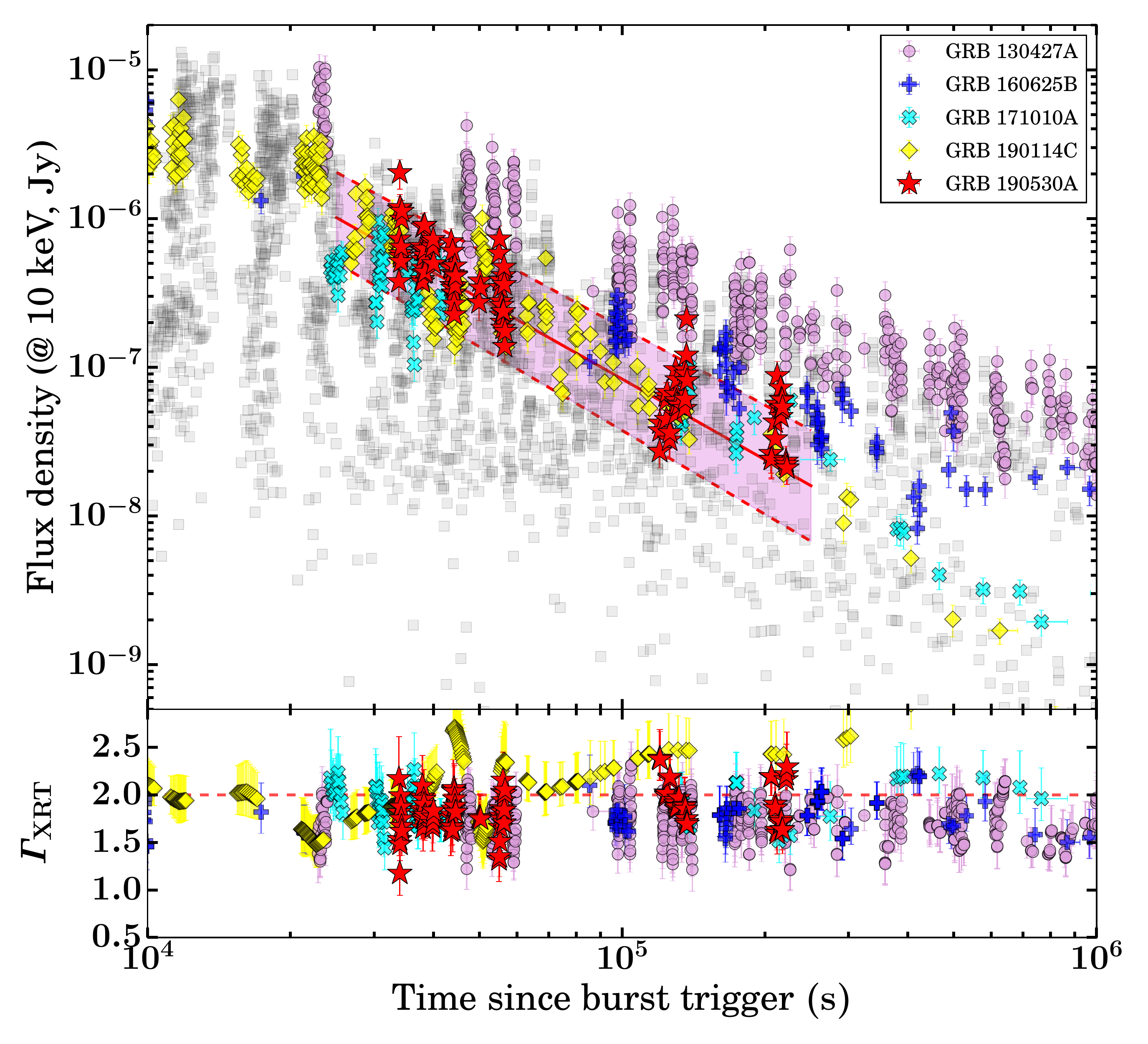}
\caption{{\bf X-ray afterglow of \thisgrb:} {\it Top panel:} The X-ray afterglow light curve of \thisgrb (shown with red stars) is best described with a simple power-law function. The magenta shaded region indicates the uncertainty in the index with a 90\% confidence range. For the comparison, the X-ray afterglow light curves of bursts brighter than \thisgrb are also shown. The grey squares show the XRT light curves (@ 10 \keV) for all the \fermi LAT detected GRBs. {\it Bottom panel:} The evolution of XRT photon indies for \thisgrb and other GRBs brighter than \thisgrb. The red horizontal dashed line shows the photon index value equal to 2.}
\label{Xray_afterglow_190530A}
\end{figure}

The XRT light curve is shown in Figure \ref{Xray_afterglow_190530A}. We fitted the XRT light curve using a power-law and a broken power-law function as expected from the external forward shock model of the afterglow. The X-ray flux light curve can be best explained with a simple power-law model. The temporal decay index ($\alpha_{\rm x}$) is $-1.80 \pm 0.07$ (see Table \ref{lcfits}).

\begin{table}
\centering
 \caption{The best fit models describing the X-ray and single filter optical/UV light curves. We have only fitted the optical light curves when at least nine data points are available for a particular filter. The g-band data are fitted after excluding the first data point as it was only marginally detected with a significant error (see \S~\ref{MASTER observations}).}
 \begin{tabular}{cccc}
 \hline
 \bf Wavelength & \bf Model &$\bf \alpha$ & $\bf \chi^2/dof$ \\
  \hline
 X-ray (@ 10 \keV)  &  power-law    & $-1.80 \pm 0.07$  &  308.3/100 \\
 \hline   
UVOT (@ U-band)       & power-law           & $-1.70 \pm 0.10$  &  10.30/7 \\ 
Optical (@ B-band)       & power-law           & $-1.57 \pm 0.11$  &  8.75/11 \\ 
Optical (@ V-band)       & power-law           & $-1.85 \pm 0.12$  &  1.42/7 \\ 
Optical (@ R-band)        &   power-law  & $-1.81 \pm 0.08$ &  5.59/8   \\
Optical (@ g-band)       &   power-law  & $-1.59 \pm 0.08$ &  38.01/14   \\
\hline
\end{tabular}
\label{lcfits}
\end{table}

For the analysis of the \swift XRT spectra, we used the X-Ray Spectral Fitting Package \citep[\sw{XSPEC};][]{1996ASPC..101...17A} version 12.10.1 of \sw{heasoft-6.25}. We modelled the XRT spectrum using an absorbed power-law model in the 0.3 - 10 \keV energy band. This model includes two absorption components (one for our Galaxy \sw{phabs} with a fixed galactic column density, and another for the host galaxy \sw{zphabs} with a free intrinsic hydrogen column density at the source redshift) together with a power-law component for the X-ray afterglow. We fixed the galactic hydrogen column density at $\rm NH_{\rm Gal}=5.07 \times 10^{20}{\rm cm}^{-2}$ \citep{2013MNRAS.431..394W}. We used \sw{C-Stat} statistics for optimization of the XRT data. The results of the spectral analysis are listed in Table \ref{tab:xrtspec}. The time-averaged spectral analysis using all the available PC mode observations is well described with a simple absorbed power-law model showing significant excess over the galactic hydrogen column density.

\begin{table}
\centering
\caption{The best-fit spectral modelling result for the X-ray afterglow of \thisgrb.}
\begin{tabular}{cccccc}
\hline
\bf Time (s) &$\rm \bf Photon ~index$ & $\rm \bf NH_z$ ($ \bf \times 10^{22} \bf cm^{-2}$)  &\bf Mode \\
\hline
33828-56758 & $ 1.75 \pm 0.09$ & 0.32 $ \pm 0.11$ & PC \\
\hline
\end{tabular}
\label{tab:xrtspec}
\end{table}

\subsection{\bf UV/Optical observations}
\label{optical observations}

\begin{figure}
\centering
\includegraphics[scale=0.31]{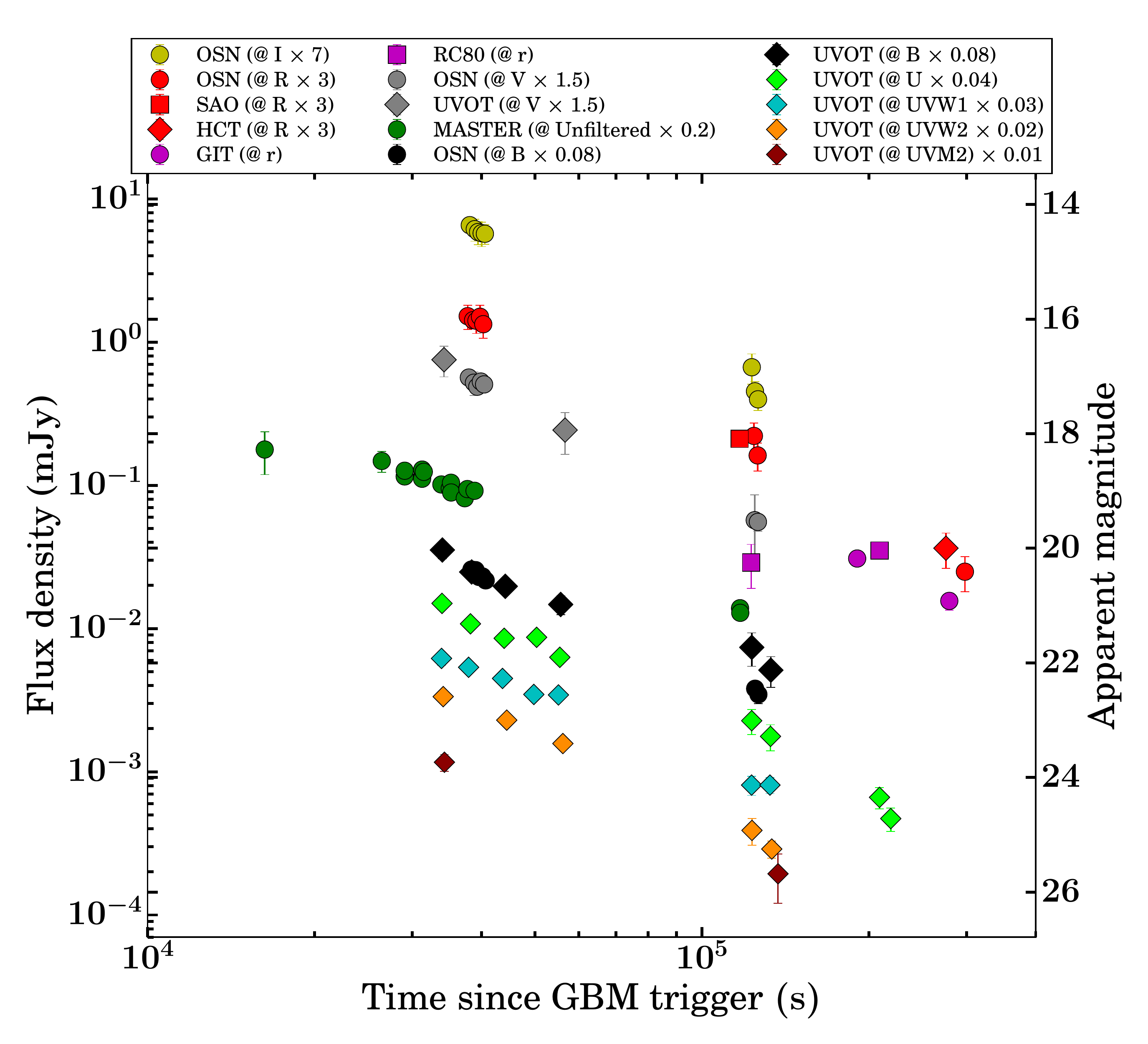}
\caption{{\bf Optical afterglow of \thisgrb:} The multi-band optical light curves of \thisgrb obtained from various telescopes. The observed magnitudes are corrected for galactic extinction and scaled with respect to the $r$-band, with the scaling factor given in the legend.}
\label{Optical_afterglow_190530A}
\end{figure}

Due to the large uncertainty on the \fermi position, an optical counterpart for \thisgrb was only detected at \fermiT+6hrs. The optical counterpart was discovered by MASTER Global Robotic Net \citep{2010AdAst2010E..30L} auto-detection system (AT2019gdw / MASTER OT J080207.73+352847.7) at 18:12:10 UT on May 30, 2019 \citep{2019GCN.24680....1L}. We obtained observations with MASTER, 1.5m OSN, RC80 robotic, 0.7m GIT, 2m HCT, 2.2m CAHA, and 10.4m GTC telescopes. We also obtained the Ultra-violet data taken with the \swift-UVOT telescope. We give the details of these observations and reduction methods below in their respective sub-sections. Multi-band light curves using our data are shown in Figure \ref{Optical_afterglow_190530A}. 

\subsubsection{\bf MASTER  optical observation}
\label{MASTER observations}

MASTER Global Robotic Net \citep{2010AdAst2010E..30L, 2019ARep...63..293L} is a network of identical twin wide-field 40 cm fully robotic telescopes on a high-speed mount of up to 30 deg per second. Each telescope has its own photometers with BVRI and two orthogonally oriented polarization filters \citep{2012ExA....33..173K} and its auto-detection system. Eight telescopes with a field of view 4-8 degrees up to 20 mag are distributed on the Earth, designed specially to discover and investigate optical counterparts of GRBs \citep{2016MNRAS.455..712L, 2018ApJ...861...48S, 2020ARep...64..126E}, GWs \citep{2017ApJ...850L...1L} and other high energy astrophysics sources in large error-fields. The automatic strategy (MASTER central planner) of optical follow-up to triggers with large error-fields (\fermi and others) depends on the error-box and coverage of the maximum probability region in open mode (8 square degrees), taking into account the altitude of the current and neighbouring square at present and the following time, the limit of received images and the possibility to observe squares at nearby MASTER observatories. If there is BALROG (BAyesian Location Reconstruction Of GRBs) localization \citep{2018MNRAS.476.1427B}, MASTER planner uses this position to observe all other things being equal. When \fermi-LAT detects the source, its position has the highest priority for follow-up. In the case of \thisgrb, LAT coordinates were known later, and by then, MASTER had discovered the OT inside the BALROG localization \citep{2019GCN.24677....1B}. 

The optical afterglow AT2019gdw / MASTER OT J080207.73+352847.7 of \thisgrb was discovered by MASTER auto-detection system \citep{2010AdAst2010E..30L, 2019ARep...63..293L} at the location RA, DEC = 120.532208 35.479917 degree (J2000) at 18:12:10 UT on May 30, 2019. It was the first ground-based telescope to report the optical afterglow \citep{2019GCN.24680....1L, 2019ATel12824....1V}. The afterglow was detected by all five stations of MASTER Global Robotic Net (see Table A8 in the appendix). At the time of the alert, all MASTER observatories were in the daytime. Observations of \thisgrb started ($\sim$ \fermiT+ 16 ks) at MASTER-Amur near Blagoveschensk with an exposure time of 180 s, and a transient source of brightness 16.68 $\pm$ 0.36 mag was marginally detected because these observations were carried out just after the sunset (very cloudy weather at the location of MASTER-Amur) and close to the horizon (error-box altitude 13$^\circ$, $\rm sun_{\rm alt}$ =-17$^\circ$). Observations continued at MASTER-Tunka (near Baykal Lake), MASTER-SAAO (started at sunset, at 16:34:41 UT on May 30, 2019, with very cloudy at all horizon, 13.5 degrees error-box altitude, without OT detection up to unfiltered m$_{lim}$=14.5 mag), MASTER-Kislovodsk (automatic detection), MASTER-Tavrida, and MASTER-IAC \citep{2019GCN.24693....1L}.

We performed the photometry using the standard method and averaged the nearby images for each telescope of twin ones of MASTER-Kislovodsk, Tavrida, IAC, and Tunka. We have listed the log of observations and photometry of MASTER data in Table A8 of the appendix, where T$_{\rm mid}$ is the middle of exposure in seconds, exp is the exposure duration in seconds, unfiltered magnitudes with error and MASTER observatory, which made observations. The reported magnitudes are taken in the clear filter and calibrated using Gaia $g$ mag with 30 reference field stars. The error of the photometry is calculated with the following formula:
$\Delta{m}=\sqrt{\frac{\sum_{i=1}^N(\overline{m_i}-m_{ij})^2}{N}}$
where $\overline{m_i}$ is the mean magnitude of check star i during the observation time, $m_{ij}$ is the magnitude of check star i on frame j, N is the number of check stars.

The first points at Kislovodsk($\sim$ \fermiT+ 29 ks) were observed with a polarization filter (oriented as 90 degrees for MASTER-Kislovodsk-east telescope (222 camera) and as zero degrees for MASTER-Kislovodsk-west telescope (223 camera)). The primary assumption of MASTER polarimetry is of zero polarization of the background stars. Therefore, it is necessary to calibrate the average polarization degree to zero. The error in the polarimetry will consist of the deviation from zero of the polarization of stars with a brightness comparable to the object. A similar procedure was performed for each pair of frames. This afterglow was recorded with MASTER telescopes in two perpendicularly oriented polarizing filters (0 and 90 relative to RA). This is not enough to measure the polarization of an object \citep{2016MNRAS.455.3312G}.
Nevertheless, it is possible to estimate the low limit of the polarization degree for \thisgrb. From the MASTER polarization measurements (average $P_{lowlim} < 1.3\%$), it is clear that the optical afterglow polarization should either be near zero or have an orientation close to 45 or 135 degrees
\citep{2012MNRAS.421.1874G}.

\subsubsection{\bf \swift UVOT data}

The \swift Ultra-Violet and Optical telescope (UVOT) started observing the field of \thisgrb 33.8 ks after the \fermi trigger \citep{2019GCN.24681....1E, 2019GCN.24703....1S}. A fading optical/UV afterglow candidate, at RA, DEC = 120.53209, +35.47968 degree (J2000) (with an uncertainty radius of $\rm 0.49^"$, 90\% confidence), was discovered in the initial UVOT observations. This location was consistent with the optical afterglow position first reported by \cite{2019GCN.24680....1L} using MASTER robotic telescope observations. The source was detected in the six UV/optical filters of \swift UVOT, and observations were carried out in image and event mode. We downloaded the \swift UVOT observation data using the online \swift data archive page\footnote{\url{http://swift.gsfc.nasa.gov/docs/swift/archive/}}. For the analysis, we utilized the \sw{heasoft} software with the latest calibration release. Initially, we carried out the astrometric corrections for the UVOT event data following the methodology described in \cite{oates09}. We extracted the object counts using a region of three arcsec radius. Then, the count rates were corrected to five arcsec using the curve of growth contained in the calibration files using standard methods to be consistent with the \swift UVOT calibration. Background counts were estimated considering a circular region of radius twenty arcsec from an empty area of the sky near the object. The count rates were retrieved using the event and image lists utilizing the \swift standard tools \sw{uvotevtlc} and \sw{uvotsource}, respectively\footnote{\url{https://www.swift.ac.uk/analysis/uvot/}}. The count rates were converted to magnitudes using the UVOT photometric zero points \citep{2011AIPC.1358..373B}. The UVOT afterglow photometry is given in AB magnitudes and has not been corrected for galactic (E(B-V)= 0.05 \citep{sch11}) and host extinction in the direction of the GRB. All the upper limits on magnitudes are given with a three-sigma level. The UVOT photometry is given in Table A7 in the appendix.

\subsubsection{\bf 1.5m OSN Telescope}

Following the trigger of \thisgrb, the 1.5 m telescope of Sierra Nevada Observatory (OSN, Granada, south Spain)\footnote{\url{http://www.osn.iaa.es/}} started to observe the source position at 20:47:52 UT on May 30, 2019 (10.5 h after \fermiT). It was observed again on another four nights, i.e., May 30, May 31, June 2 and June 3, 2019. A series of images were obtained in Johnson-Cousins broadband filters: B, V, R, and I with exposures of 120 s and 360 s during the first two epochs of observations. The afterglow counterpart was initially detected in a single frame. A series of R-band images with significant exposures were taken during the late (third) epoch to obtain a deep field image. The afterglow is still detectable in the combined image of the third epoch. The photometric data were derived using aperture photometry through standard procedures after bias-subtraction and flat-field correction. The magnitudes were calibrated against nearby reference stars in the field of view listed in USNO-B1, GSC 2.3 catalogue \citep{2003AJ....125..984M, 2008AJ....136..735L}, see Table A8 in the appendix.

\subsubsection{\bf RC80 Robotic Telescope}

The 0.8m Ritchy-Chrétien (RC80) robotic telescope at Piszk\'estet\H\o station of Konkoly Observatory detected \thisgrb on two epochs: 2019-05-31.85 UT, and 2019-06-01.85 UT, 1.42, and 2.42 days after burst, respectively. The total exposure time was 60 minutes per night, and the observations were made with the Sloan r-band filter. The frames, after bias, and flat field corrections, were co-added using standard procedures in IRAF by the RC80 automatic pipeline. Forced aperture photometry on the co-added frames was applied at the position of \thisgrb. The optical afterglow was clearly detected at both the epochs. To get the final photometry for \thisgrb, the fluxes from the nearby contaminating source ($r_{PS1} = 20.947 \pm 0.056$ AB mag) were subtracted from the fluxes taken within the fixed aperture, then converted to AB magnitudes based on 19 local tertiary comparison stars from the PanSTARRS DR1 photometry catalog. The final AB magnitudes are listed in Table A8 in the appendix.


\begin{figure*}
\centering
\includegraphics[scale=0.26]{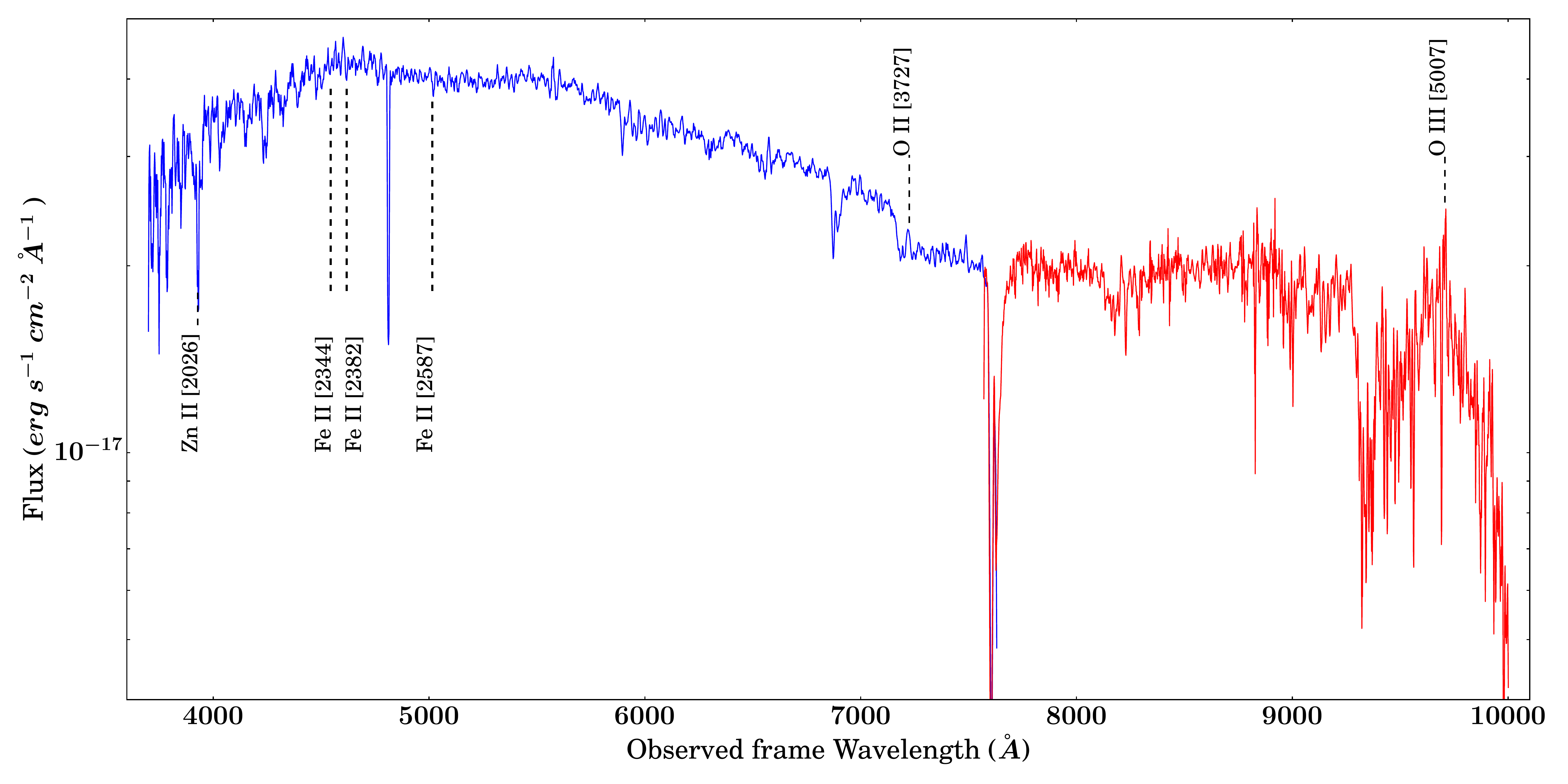}
\caption{{\bf Redshift determination:} The 10.4m GTC optical spectrum in the range $3,700-1,000$ \AA provides the redshift of \thisgrb. The Fe II and Zn II lines are shown in absorption and the O\,III and O\,II emission lines of the underlying host galaxy at the same redshift ($z$ = 0.9386). The measured redshift is consistent with the value obtained from a joint spectral energy distribution with the XRT and UVOT data.}
\label{redshiftSpectrum}
\end{figure*}

\subsubsection{\bf 0.7m GIT Telescope}

We triggered 0.7m GROWTH-India Telescope (GIT) on 1$^{st}$ June 2019 at 14:38:01 UT to observe \thisgrb. The telescope is located at Indian Astronomical Observatory (IAO), in Hanle, Ladakh (India), and is equipped with a 4K $\times$ 4K Andor iKon-XL 230 camera. We used $r^{\prime}$ band for all observations. The GRB afterglow candidate was followed up for two consecutive nights, i.e., on 1$^{st}$ and 2$^{nd}$ June 2019. We successfully detected the afterglow in our frames. However, the observation was undertaken at a very low altitude of $\sim$ 20~deg, which resulted in poor S/N for the detection leading to high uncertainty in the magnitude estimation. The data were downloaded and reduced in real-time via the automated GIT data processing pipeline. Images were calibrated using the bias and flat frames acquired on the same night of observation. The pipeline used the Astro-SCRAPPY~\citep{2019ascl.soft07032M} package to remove cosmic-rays streaks from the images. Using offline astrometry solve-field engine \citep{Lang_2010}, we obtained the transformation from image to sky coordinates. Photometry was performed using standard methods. To calculate the zero-point of the image, we cross-matched the Sextractor \citep{bertin11} generated catalogue to the PanSTARRS DR1 with the help of Vizier. The zero points we obtained were used to standardize the magnitudes. Photometric measurements for GIT data are listed in Table A8 of the appendix. 

\subsubsection{\bf 2m HCT Telescope}

We carried out observations of the field of \thisgrb \citep{2019GCN.24676....1F, 2019GCN.24689....1M} with the 2m Himalayan Chandra Telescope (HCT) located at the Indian Astronomical Observatory, Hanle, India. The follow-up observations started on 2019-06-02 14:50:43 UT, i.e., around 3.19 days post burst in Bessell R filter. We processed the images using IRAF routines \citep{1986SPIE..627..733T}. After cleaning the images, we stacked them and performed aperture photometry using DAOPHOT II packages\footnote{\url{http://www.star.bris.ac.uk/~mbt/daophot/}}. The optical afterglow first reported by \cite{2019GCN.24680....1L} is detected in the stacked image with a total exposure time of 15 min. We calculated the magnitude of the source as 21.3 $\pm$ 0.3 mag, calibrated using the field stars from the USNO-B1.0 catalogue \citep{2003AJ....125..984M}.

\subsubsection{\bf 2.2m CAHA Telescope}

The 2.2 m telescope at Centro Astron\'omico Hispano-Alem\'an (CAHA, Almer\'ia, south Spain)\footnote{\url{https://www.caha.es/}} which is equipped with the Calar Alto Faint Object Spectrograph (CAFOS) also observed \thisgrb on the night of June 12, 2019, starting at 20:26:08 UT (13.4 days after \fermiT) with the Cousins R filter. In the resulting 1950 s co-added image, the afterglow is not detectable, providing a 3$\sigma$ upper limit of 22.70 mag, which is calibrated against nearby reference stars in USNO-B1.0 catalogue \citep{2003AJ....125..984M}.

\subsection{\bf 10.4m GTC spectroscopy observations and Redshift determination}

\cite{2019GCN.24686....1H} performed spectroscopic observations of the optical afterglow using the 2.5-m Nordic Optical Telescope (NOT). They carried out the spectroscopy observations for a sum of 2 $\times $ 600 s, in a wavelength range of 3650 - 9450 \text{\AA}. They found a blue continuum from 3900 \text{\AA} to 9000 \text{\AA} and were unable to detect any significant absorption or emission lines in their low-resolution spectrum. However, they report an upper limit on the redshift as $z$ $<$ 2.2 based on the observed continuum. In the present section, we report the redshift determination of \thisgrb using our observations.
 
We performed spectroscopic observations of \thisgrb using OSIRIS mounted on the 10.4m Gran Telescopio Canarias (GTC; Canary Island, Spain). We obtained a spectrum in the wavelength range 3,700 to 10,000 \text{\AA}, 35.1 hours post-burst (in the observer-frame). We carried out the standard calibration using IRAF routines. The two-dimensional raw spectroscopic frames were corrected for bias, divided by a normalized flat-field, corrected for cosmic rays (using the L. A. Cosmic algorithm, \cite{2001PASP..113.1420V}), extracted across the spatial direction after having interpolated the background below the GRB with a low-order polynomial fit on the surrounding regions and calibrated in wavelength against NeArHg arcs. Then the extracted one-dimensional spectra were calibrated in flux using a spectrophotometric standard star. 

The reduced spectrum has a low signal-to-noise ratio (SNR). To improve the SNR, we have smoothed the spectrum. We have used both the R1000B and R2500I grisms obtained on 31 May 2019 using 10.4m GTC to constrain the redshift. We identified the Fe II \& Zn II absorption lines (2344, 2382, 2587, and 2026 \text{\AA}) in the observed spectrum (see Figure \ref{redshiftSpectrum}) at a common redshift $z$ = 0.9386. We also identified O\,III (5007 \AA{}) and O\,II emission lines (3727 \text{\AA}) of the underlying galaxy at the same redshift. Therefore, we report that $z$ = 0.9386 is the redshift of \thisgrb.

\subsection{\bf Spectral Energy Distribution of the afterglow}
\label{SED}

A spectral energy distribution (SED) is a useful tool to constrain the spectral regime of the broadband afterglow emission. We created a SED during the first orbit of \swift XRT and UVOT observations ($\sim$ 30.6-60.7 ks) following \cite{depas07}, which is based on the methodology of \cite{schady07}. We performed the joint optical and X-ray data modelling using XSPEC \citep{1996ASPC..101...17A} software. We used two models, a power-law model and a broken-power law model, according to the expectation of the external forward shock model. In the case of the broken power-law model, the difference between the indices (before and after the spectral break) was fixed at 0.5, consistent with the synchrotron cooling break \citep[e.g.][]{zhang04}. In each model, we include a Galactic and intrinsic absorber using the XSPEC models \sw{phabs} and \sw{zphabs}. The Galactic absorption is fixed to $\rm NH_{\rm Gal}=5.07 \times 10^{20}{\rm cm}^{-2}$ \citep{2013MNRAS.431..394W}. We also include Galactic and intrinsic dust components using the XSPEC model \sw{zdust}, one at redshift $z$= 0, and the other fixed at $z$ = 0.9386. The Galactic reddening was fixed at E(B-V)= 0.04 mag according to the map of \cite{sch11}. For the extinction at the redshift of the burst, we test Milky Way, Large and Small Magellanic Clouds (MW, LMC, and SMC) extinction laws \citep{pei92}. The results of SED fitting are presented in \S~\ref{sedEL}.

\subsection{\bf Low frequency data}
In addition to above mentioned optical data as part of the present work, we also used low-frequency afterglow observations, helpful to constrain the self-absorption frequency. \cite{2019GCN.24978} started observing the field of \thisgrb with Northern Extended Millimeter Array (NOEMA) telescope at a frequency between 76 and 150 GHz from \fermiT + 1.17  to \fermiT + 16.44 days. The mm afterglow was detected on their first epoch with a flux density of 1.0 mJy at 92 GHz. Subsequently, the afterglow declined consistently in flux density until it was no longer detected in their last observation (flux density of 0.066 mJy at 92 GHz). 

\subsection{Host Galaxy Search}

We performed late time photometric observations using the 4Kx4K CCD Imager \citep{2018BSRSL..87...42P, Kumar2022} mounted at the axial port of the 3.6m DOT in UBVRI filters to search for the host galaxy of \thisgrb. The data reduction was conducted using the standard procedure as discussed in \cite{2021MNRAS.502.1678K}. We detected a bright source around 3.8 arcsec away from the location of optical afterglow. This source is also visible in PanSTARRS images. However, our analysis indicates that this source does not have a similar profile and colour to a typical galaxy. Therefore, we performed deeper observations using 10.4m GTC in the griz filters to search for the host galaxy. However, no new source was found consistent with the position of \thisgrb. This indicates that the host could be a very faint galaxy (the host is substantially fainter than the known GRB host luminosity distribution). A log of photometric observations is given in Table A9 of the appendix.

\section{\bf Results}
\label{results}
In this section, we present the results based on analysis of multiwavelength prompt emission and afterglow observations of \thisgrb obtained from various space and ground-based facilities (see \S\ref{multiwaveength observations and data reduction}).

\subsection{\bf Prompt emission characteristics}

Prompt emission properties of \thisgrb using \fermi and \AstroSat data are discussed and compared the observed properties using other well-studied samples of long GRBs.

\subsubsection{\bf Prompt emission Polarization}
\label{Prompt emission Polarization}

\thisgrb with a total duration of around 25 seconds (since \fermiT) registers $\sim$ 1250 Compton counts in \AstroSat CZTI. This corresponds to a high level of polarimetry sensitivity, making this GRB suitable for polarization analysis \citep[see section 2.5.3 of][]{2019ApJ...884..123C}. 

\begin{figure*}
\centering
    \includegraphics[width=0.3\linewidth]{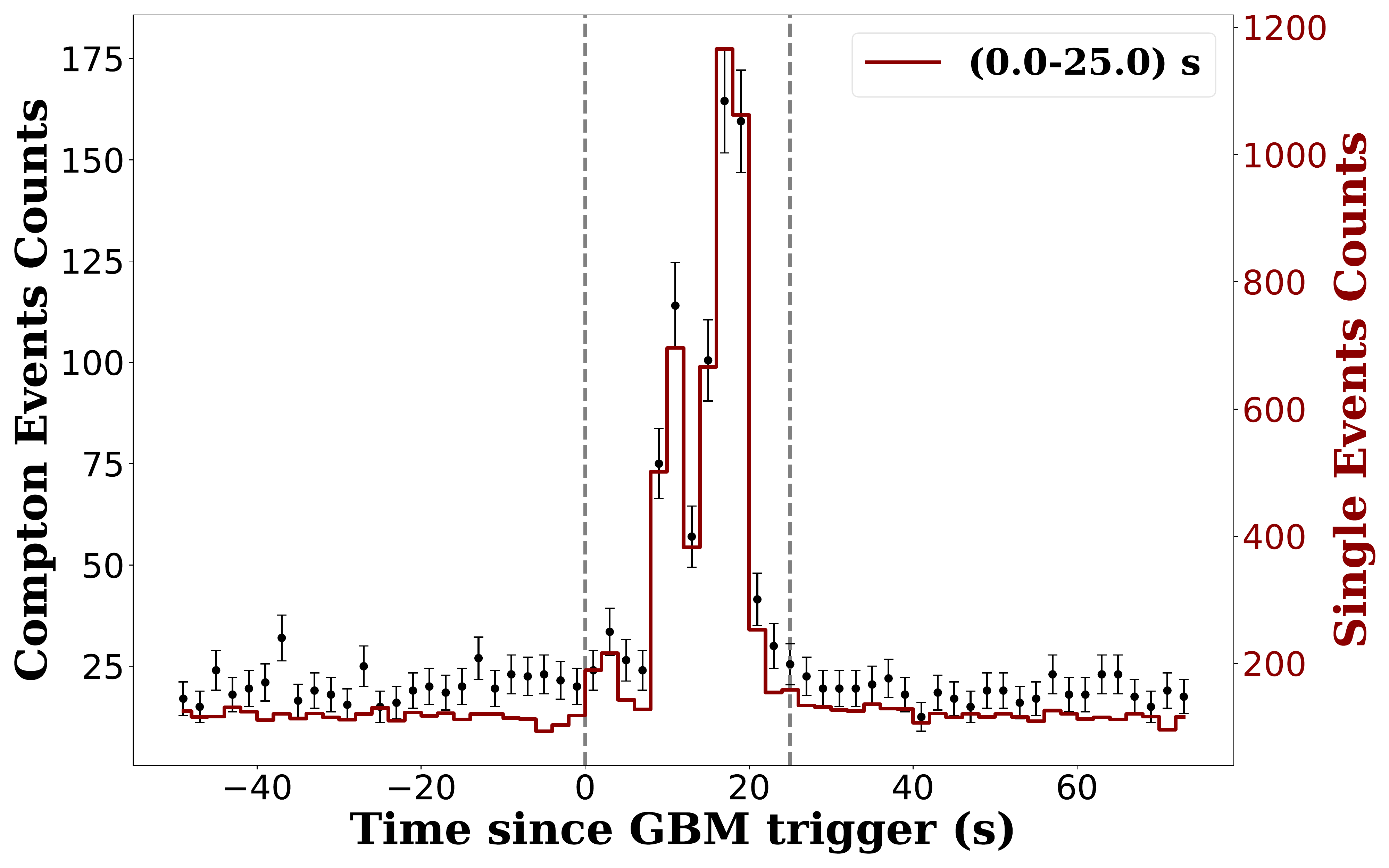}
    \includegraphics[width=0.3\linewidth]{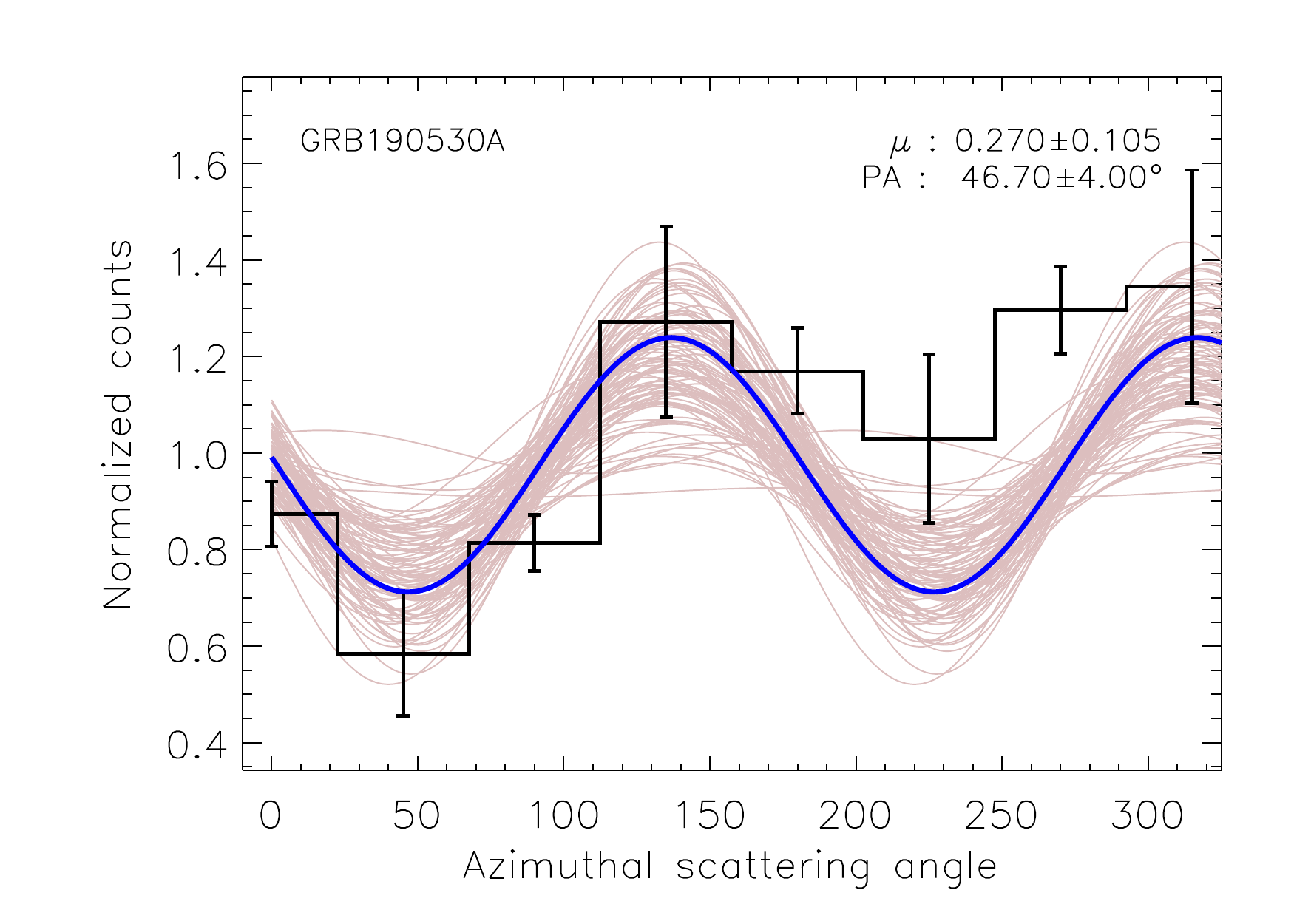}
    \includegraphics[width=0.3\linewidth]{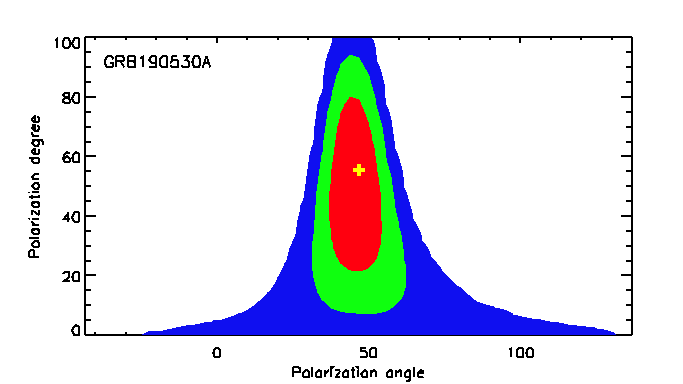}
    \includegraphics[width=0.3\linewidth]{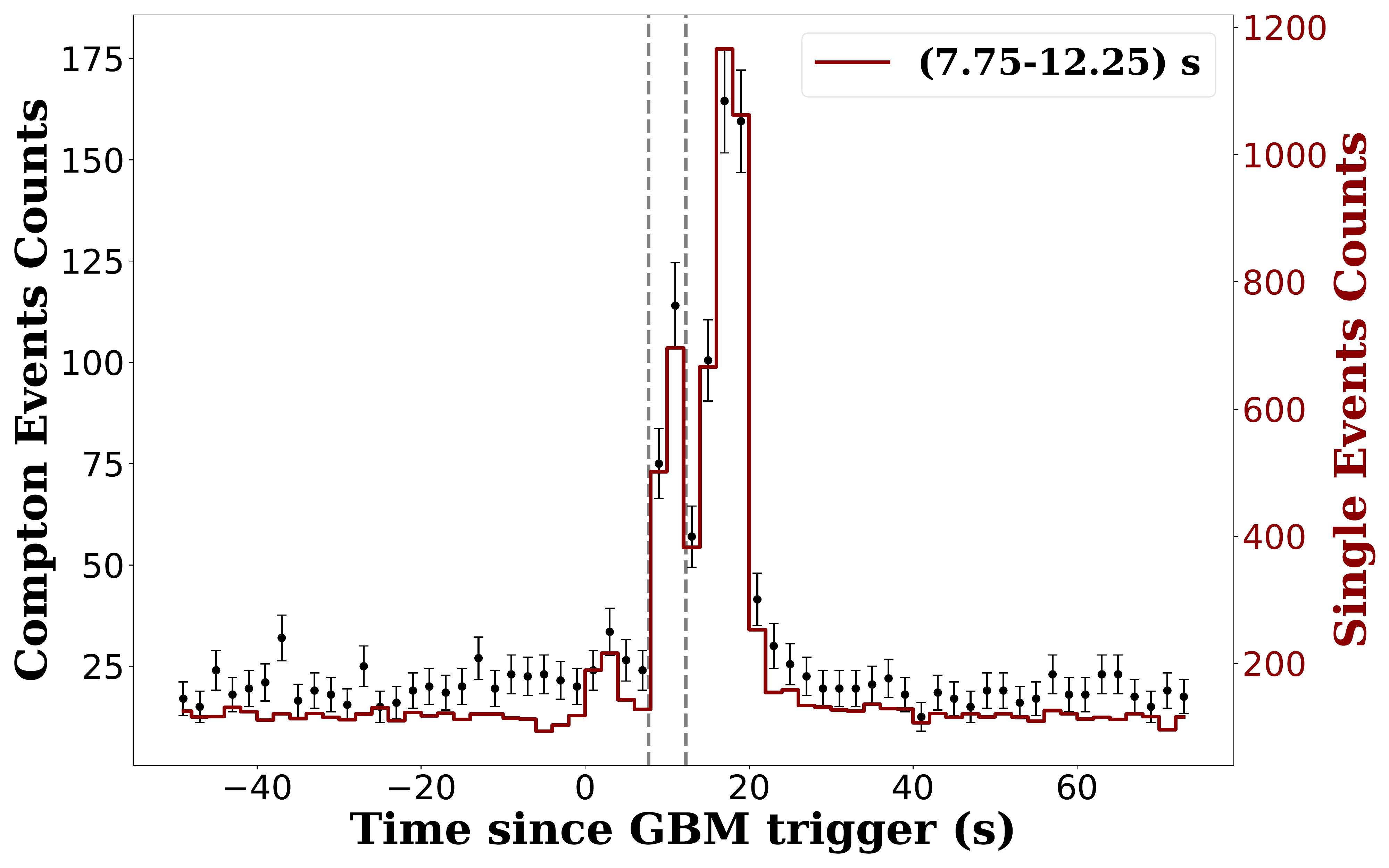}
    \includegraphics[width=0.3\linewidth]{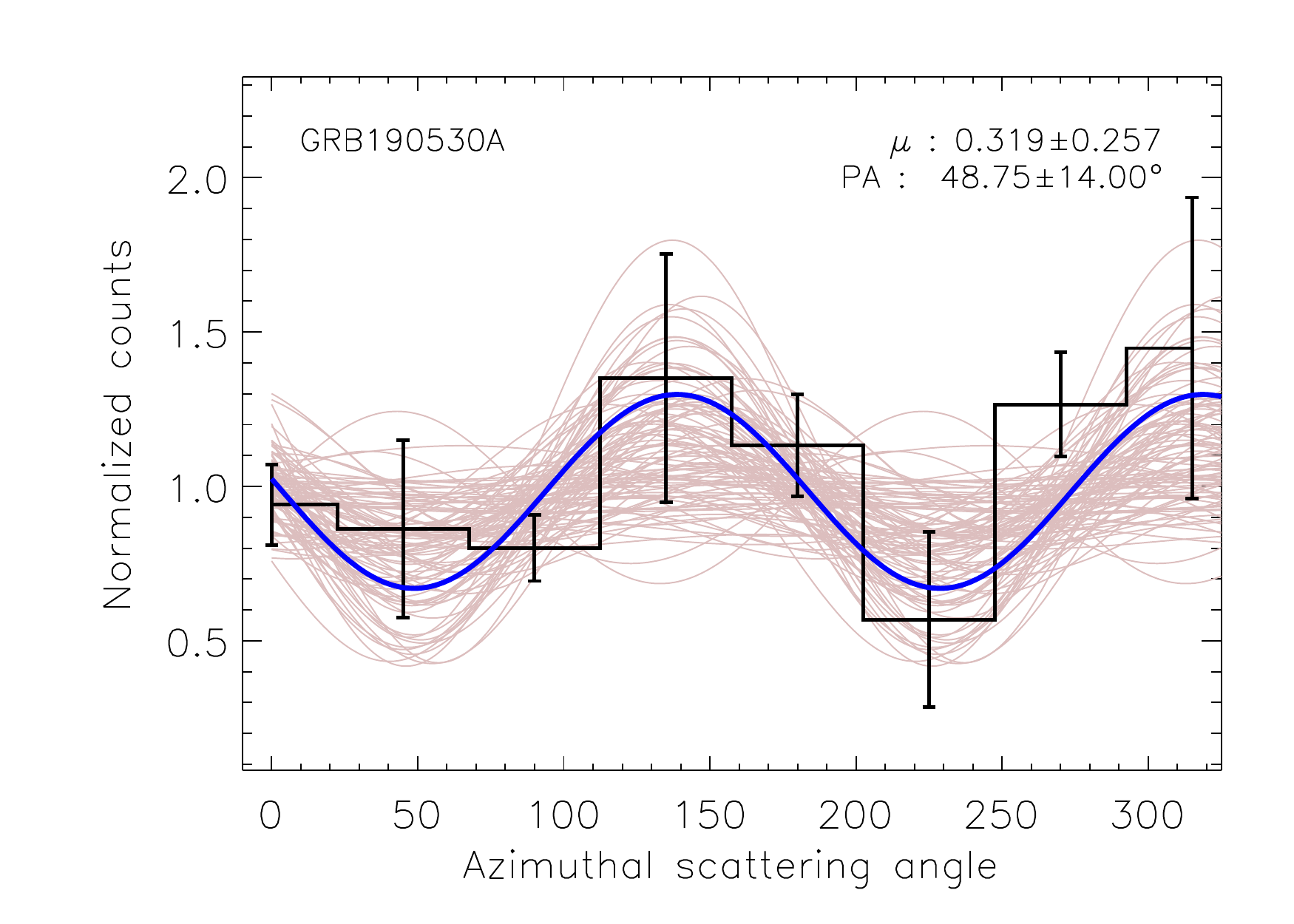}
    \includegraphics[width=0.3\linewidth]{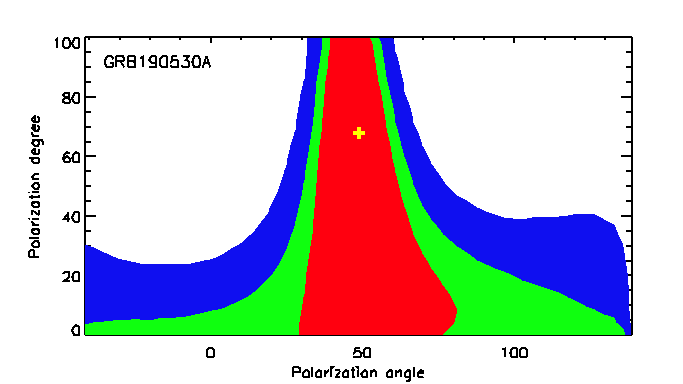}
    \includegraphics[width=0.3\linewidth]{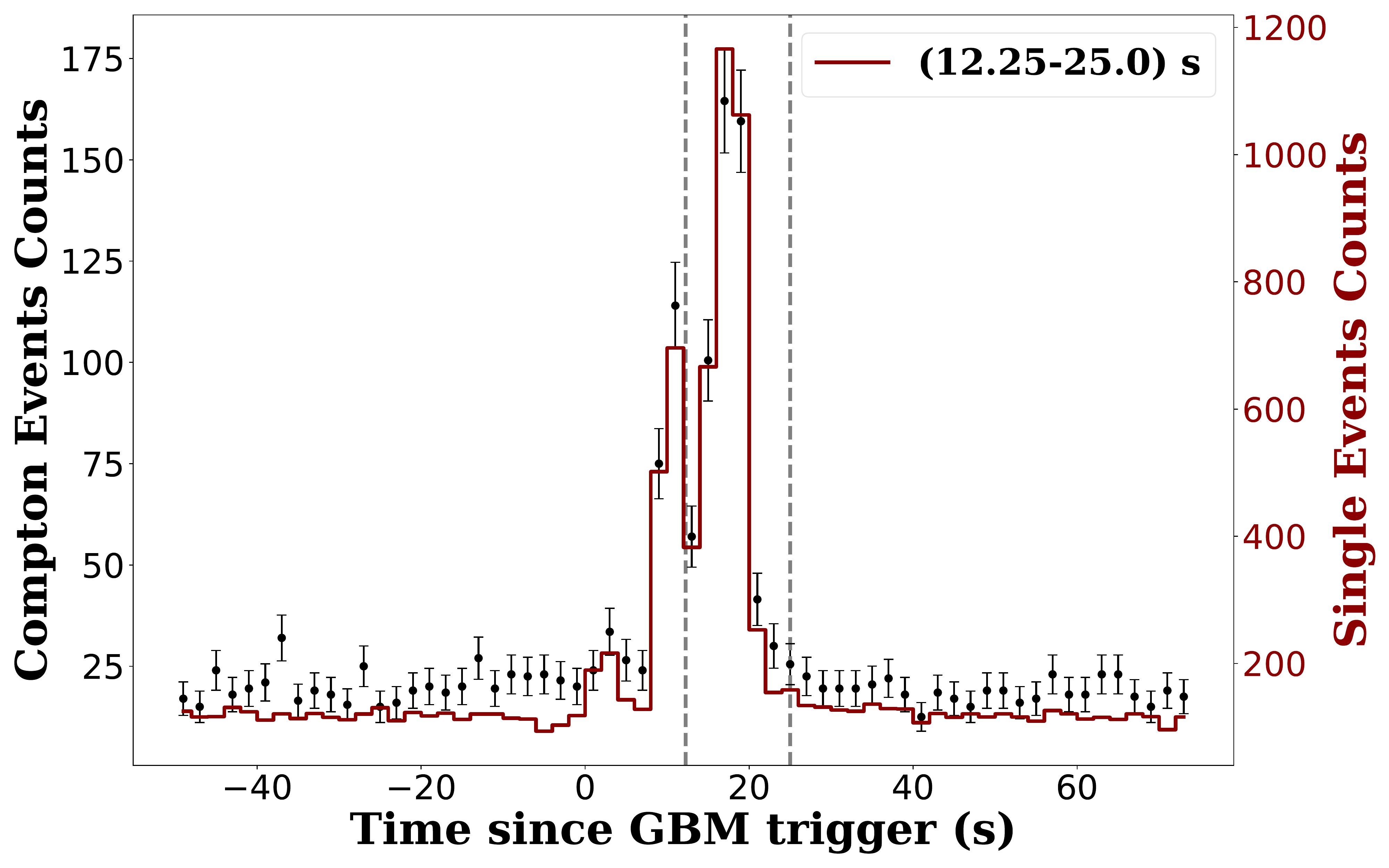}
    \includegraphics[width=0.3\linewidth]{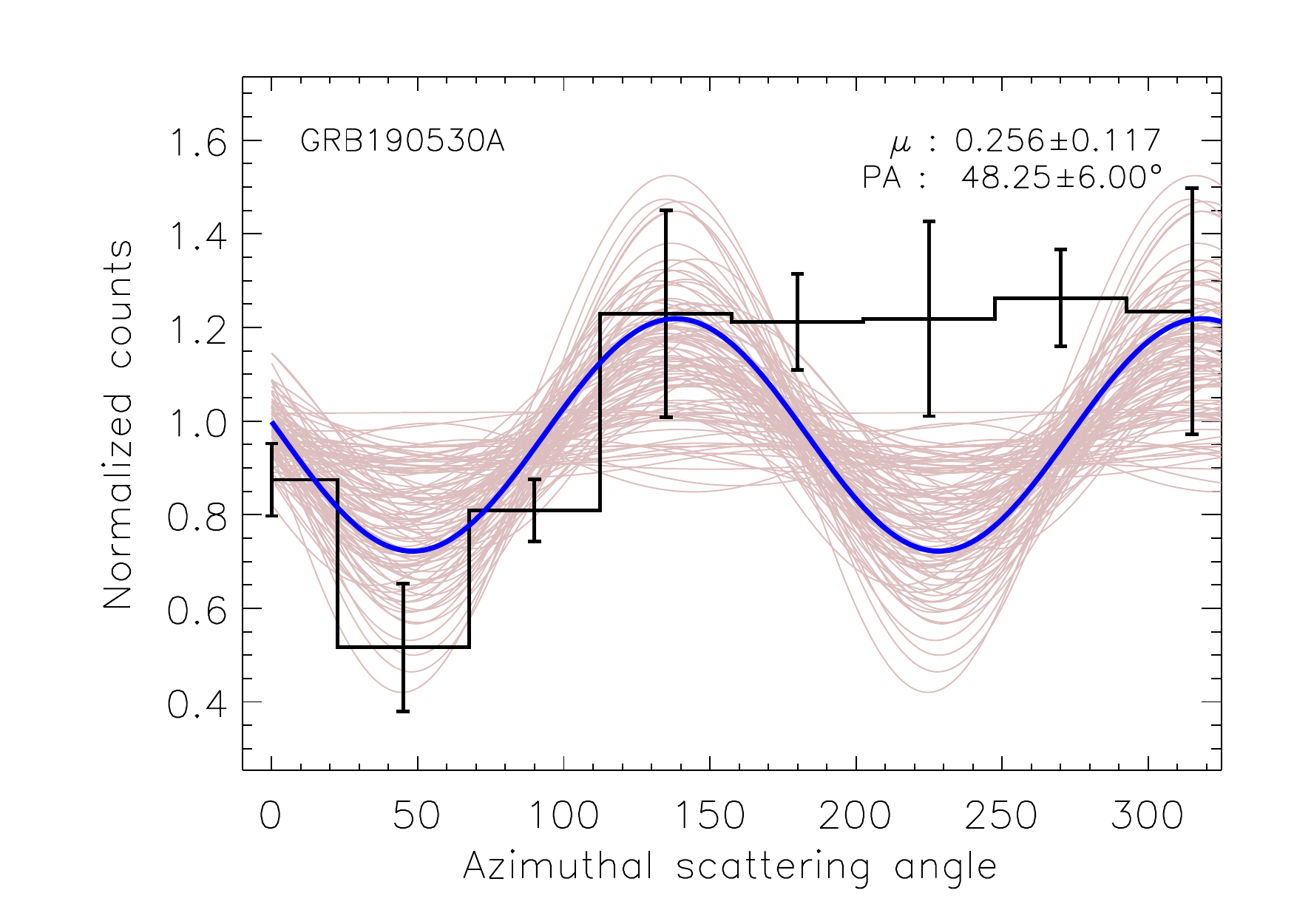}
    \includegraphics[width=0.3\linewidth]{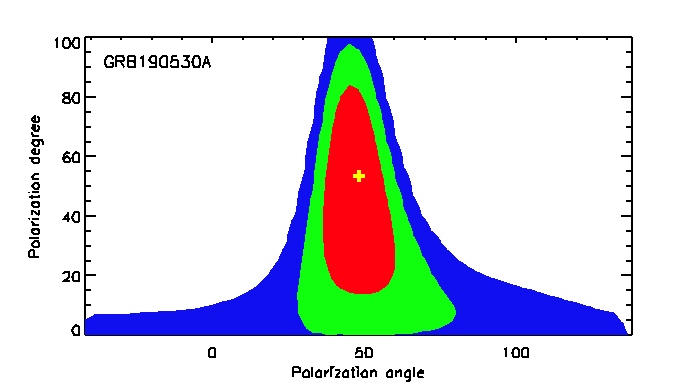}
    \includegraphics[width=0.3\linewidth]{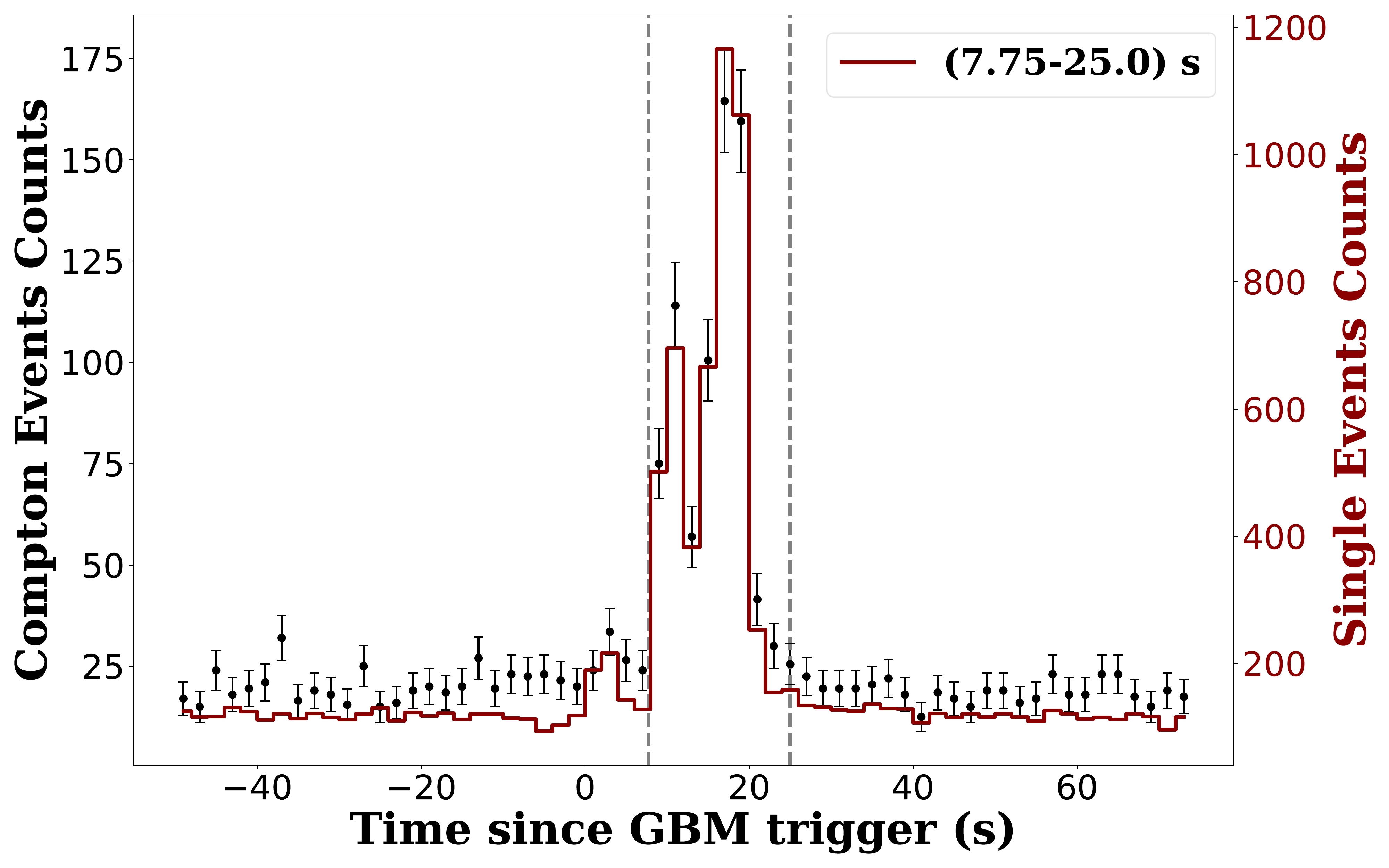}
    \includegraphics[width=0.3\linewidth]{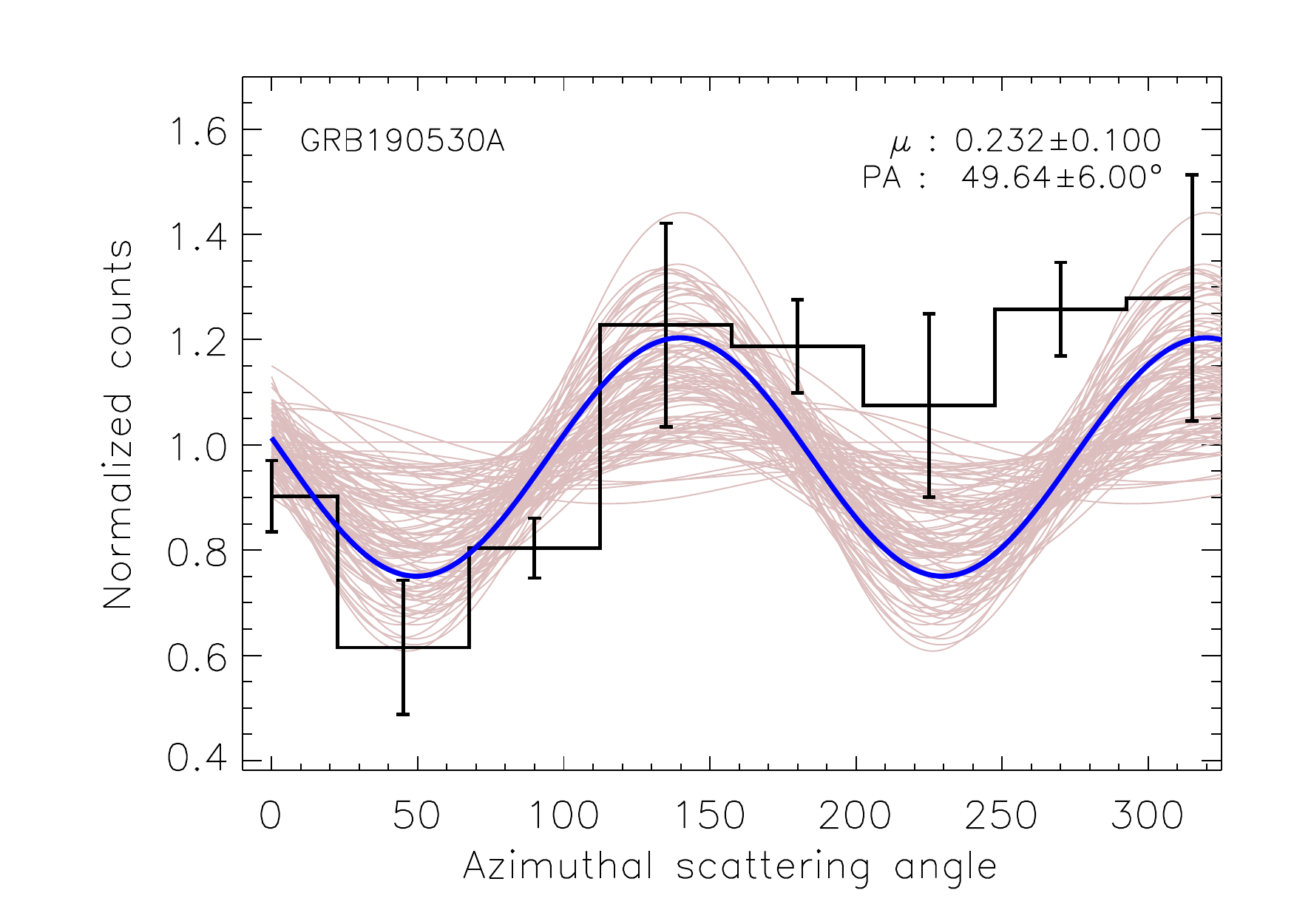}
    \includegraphics[width=0.3\linewidth]{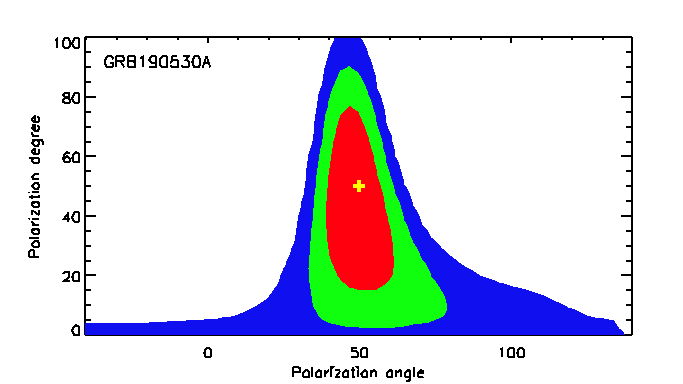}
    \includegraphics[width=0.3\linewidth]{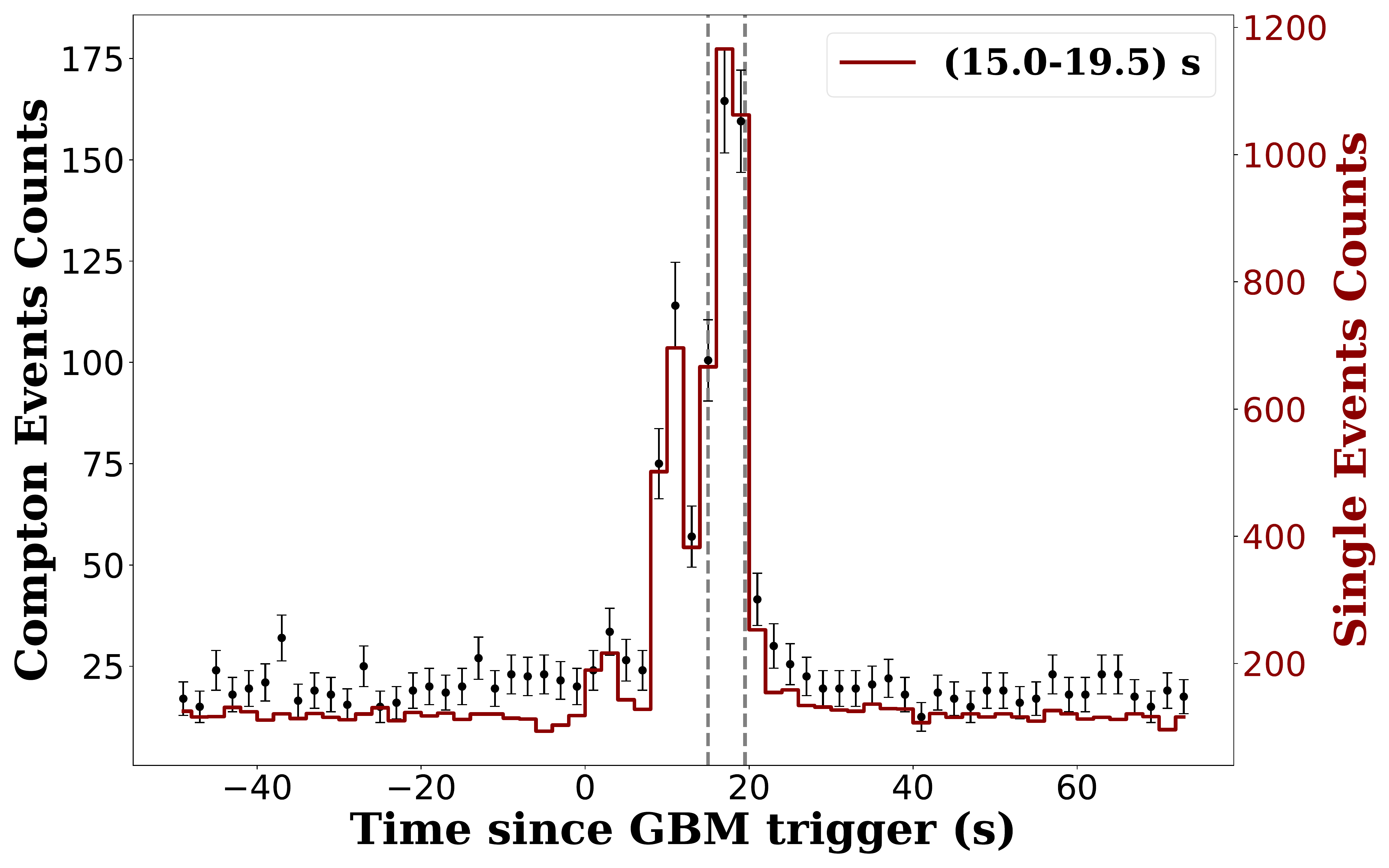}
    \includegraphics[width=0.3\linewidth]{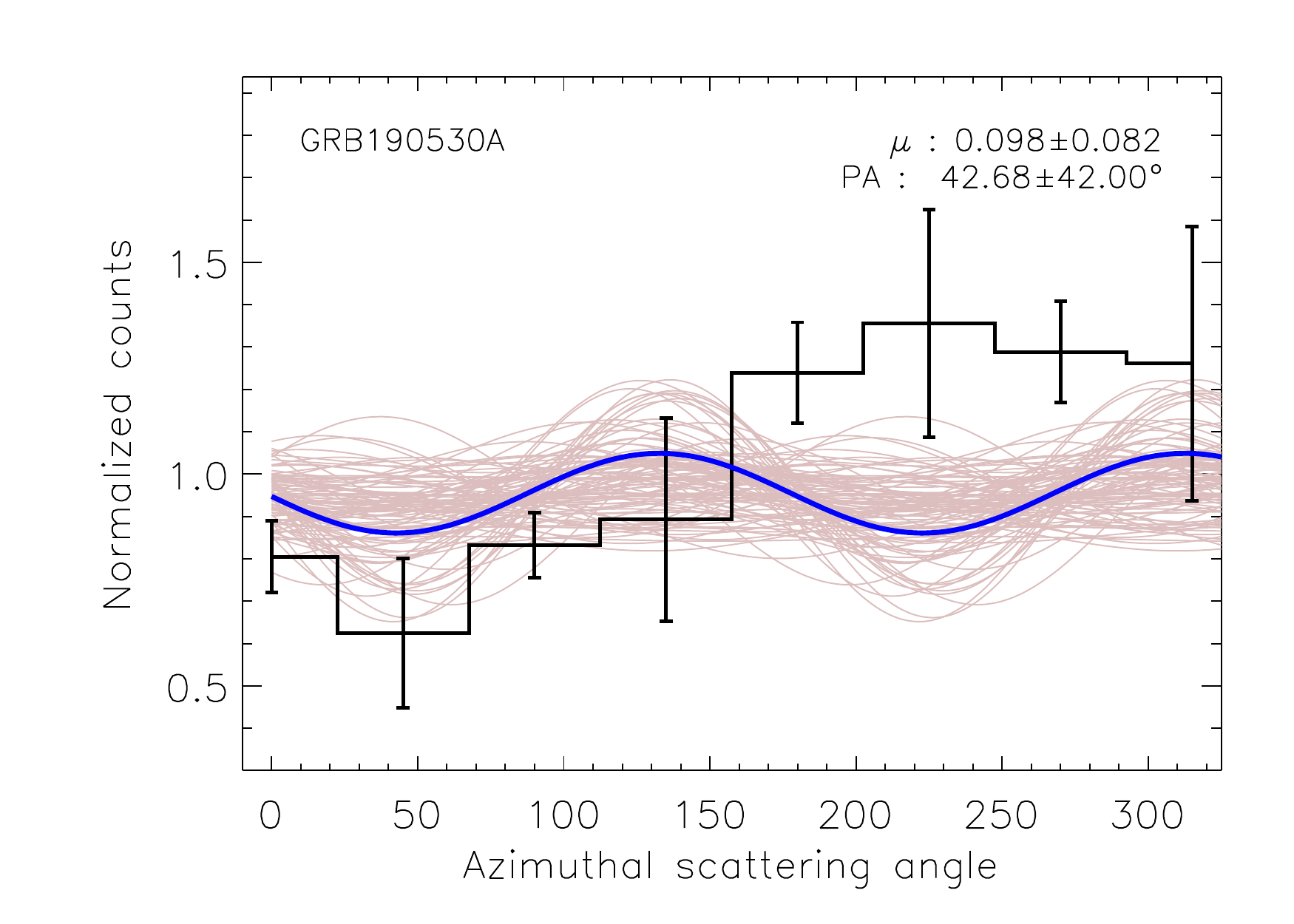}
    \includegraphics[width=0.3\linewidth]{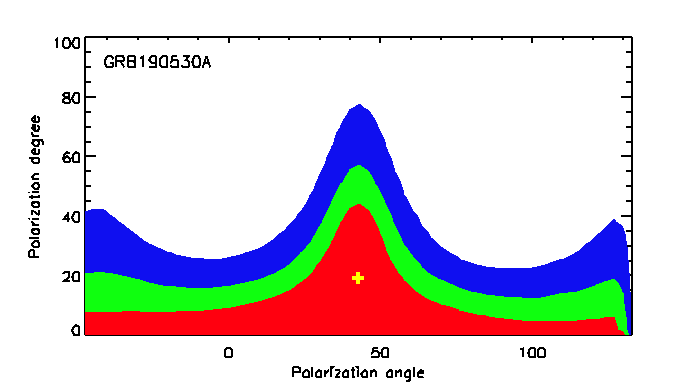}
\caption{The left panels depict the light curve of \thisgrb for the single-pixel (marked in red), and double pixel (marked in black) counts in 100-300 \keV energy range with a temporal bin size of 2 s, obtained using \AstroSat CZTI data. The vertical grey dashed lines indicate the time intervals used for the time-integrated (in the top row) and time-resolved polarization (in the last four rows) measurements within the burst. The middle and right panels depict the contour plots of polarization fraction and angle and modulation curves for the corresponding intervals. Detailed info about the figure has been discussed in \S~\ref{Prompt emission Polarization}.}
\label{pol_190530}
\end{figure*}

\begin{table*}
\caption{The \AstroSat CZTI polarimetry results of \thisgrb in the time-integrated and time-resolved temporal window in 100-300 \keV energy range.}
\label{polarization_table} 
\begin{tabular}{ccccccc}
\hline
\bf Burst interval & \bf Energy & \bf No. of Compton events  &\bf Modulation amplitude &\bf Polarization angle  &\bf Bayes factor & \bf Polarization Fraction\\
\bf (s)& \bf (\keV)&   &\bf ($\mu$) & \bf ($\bf ^\circ$) &  &  \\
\hline
0.0 $-$ 25.0&100-300&1246 &0.27 $\pm$ 0.10&46.74 $\pm$ 4.0&3.51& 55.43 \% $\pm$ 21.30\% \\\\
7.75$-$12.25 &100-300&319&0.32 $\pm$ 0.26&--&1.08& $<$64.40\% (95\%) \\\\
12.25$-$25.0 &100-300&870&0.26 $\pm$ 0.12&48.17 $\pm$ 6.0&2.11& 53.95 \% $\pm$ 24.13\%\\\\
7.75$-$25.0 &100-300&1189&0.23 $\pm$ 0.10&49.61 $\pm$ 6.0&2.52& 49.99 \% $\pm$ 21.80\%\\\\
15.0 $-$ 19.5&100-300&577 &0.09 $\pm$ 0.08&--&0.71& $<$65.29\% (95\%)  \\
\hline
\end{tabular}
\end{table*}

Figure \ref{pol_190530} left panels show the CZTI light curves of \thisgrb in 100-300 \keV energy range for both 1-pixel (marked in red) and 2-pixel Compton events (marked in black). This shows that the burst can be seen in Compton events. The grey dashed lines show the time intervals used for polarization analysis. The mean background registered is around 20 counts per second. The middle panels show respective azimuthal angle distributions for the burst obtained during the same time intervals. As shown in Figure \ref{pol_190530}, for \thisgrb, polarization analysis has been performed for the full burst (\fermiT to \fermiT+25 s, see top panel of Figure \ref{pol_190530}) as well as for the two brightest emission episodes (referred as 2$^{nd}$ and 3$^{rd}$ emission episodes, respectively, see panels in second row and panels in third row of Figure \ref{pol_190530}). The 2$^{nd}$ episode (time interval: \fermiT+7.75 to \fermiT+12.25 s) and 3$^{rd}$ emission episode (time interval: \fermiT+12.25 to \fermiT+25 s) recorded around 319 and 870 Compton events, respectively in CZTI detector. For the complete burst, we estimate a polarization fraction of 55.43 $\pm$ 21.30 \% with a Bayes factor around 3.5 (see the panel in the first row of Figure \ref{pol_190530}). We also see the polarized signature in the azimuthal angle distribution for the 2$^{nd}$ episode (see the panel in the second row of Figure \ref{pol_190530}). However, the polarization could not be constrained because of the small Compton events (Bayes factor of 1.08). We estimated the 2$\sigma$ upper limit on polarization fraction around 64 \% for this episode (see Table \ref{polarization_table}). On the other hand, the 3$^{rd}$ episode (see the panel in the third row of Figure \ref{pol_190530}) with a relatively more significant number of Compton events yields a hint of polarization in this region, having a polarization fraction around 53.95 $\pm$ 24.13\% with a Bayes factor value around 2. The panel in the fourth row light curve and azimuthal angle distribution in Figure \ref{pol_190530} is the combined analysis of the 2$^{nd}$ and 3$^{rd}$ episodes which yield a hint of polarization fraction of 49.99 $\pm$ 21.80\% with a Bayes factor around 2.5. Furthermore, we attempted to measure polarization for a temporal window (see the panel in the last row of Figure \ref{pol_190530}) where the low energy spectral index is found to be harder. However, we could only constrain the limits during this window due to a low number of Compton events. A hint of high polarization signature for both  time-integrated (with a Bayes factor of around 3.5) as well as time-resolved analysis confirms that polarization properties remain independent across the burst. This can be further verified because the polarization angles obtained for different burst intervals are within their error bar, indicating no significant change in the polarization properties with burst evolution.

\subsubsection {\bf Hardness Ratio, Minimum Variability Time Scale, and Spectral lag}

GRBs have traditionally been classified based on the \tninty-spectral hardness distribution plane. LGRBs are softer in comparison to SGRBs. In the case of \thisgrb, we obtained the \tninty duration in 50-300 \keV from the \fermi GBM catalogue, and its value is consistent with LGRBs in the bimodal duration distribution of GRBs. Furthermore, we measured the spectral hardness of this burst using the three brightest NaI detectors. To calculate the HR, we divided the observed counts in soft (10 - 50 \keV) and hard (50 -300 \keV) energy channels for these detectors (see Table \ref{tab:prompt_properties}). We placed \thisgrb in \tninty-spectral hardness distribution plane along with other data points (\fermi detected GRBs) published in \cite{2017ApJ...848L..14G}. The top panel of Figure A1 in the appendix displays the results of \tninty-spectral hardness distribution for \thisgrb (shown with a red star). The probabilities of a burst classified as a short or long burst from the Gaussian mixture model are given using a logarithmic colour bar scaling (obtained from \citealt{2017ApJ...848L..14G}).

GRB's prompt emission light curves are highly variable \citep{1990SvAL...16..129M}, as a result of internal shocks and central engine activities. The minimum variability time scale \citep[][\mvts]{2013MNRAS.432..857M} is an important parameter to constrain the central engine, the source emission radius ($\rm R_{\rm c}$) and the minimum Lorentz factor \citep[][\lmin]{2015ApJ...805...86S} of GRBs. The values of \mvts for LGRBs is larger than that of SGRBs, suggesting that SGRBs have a more compact central engine. In the case of \thisgrb, we measured the minimum variability time scale using continuous wavelet transforms\footnote{\url{https://github.com/giacomov/mvts}} presented in \cite{2018ApJ...864..163V}. We determine \mvts $\sim$ 0.5 s for this GRB. Furthermore, we place \thisgrb in \tninty-\mvts distribution (see the position of the red star in the bottom panel of Figure A1 in the appendix) along with other data points studied by \citep{2015ApJ...811...93G}.

Using the calculated value of \mvts, we measured $\rm \Gamma_{\rm min}$ and $\rm R_{\rm c}$ using the following relations taken from \cite{2015ApJ...811...93G}:

\begin{equation}
\rm \Gamma_{\rm min} \gtrsim 110 \, \left (\frac{L_{\rm iso}}{10^{51} \, \rm erg/sec} \, \frac{1+z}{\rm t_{\rm mvts} / 0.1 \, \rm s } \right )^{1/5}
\label{gamma_min}
\end{equation}
\begin{equation}
\rm R_c \simeq 7.3 {\times} 10^{13} \, \left (\frac{L_{\rm iso}}{10^{51} \, \rm erg/s} \right )^{2/5} \left (\frac{ t_{\rm mvts} / 0.1 \, \rm s }{1+z} \right )^{3/5} \, \rm cm.
\label{minimum_source}
\end{equation}

We find $\rm \Gamma_{\rm min}$ $\gtrsim$ 330 and $\rm R_{\rm c}$ $\simeq$ 1.69 $\times$ $10^{15}$ cm for \thisgrb. The lower limit of the Lorentz factor is consistent with the value of the Lorentz factor found using $\Gamma_{0}$-$E_{\gamma, \rm iso}$ correlation in \S~\ref{pem}.

The spectral evolution of GRBs can be measured by a spectral lag -- a relative shift between the prompt emission light curves in different energy ranges. The lag is defined as positive if the hard light curve is forward of the soft one, and it could be significant (up to a few seconds) for long GRBs. To investigate the spectral lag for \thisgrb, we applied the cross-correlation method as described in \citet{min12, min14} for the \fermi GBM data. The prompt emission light curves were constructed using the TTE data of the brightest detectors NaI 0, NaI 3, NaI 5, BGO 0, and BGO 1 of the \fermi GBM experiment. Ten NaI-based light curves cover the energy range of (5, 850) \keV, while five BGO based light curves cover the range of (0.2, 10) MeV. The NaI-based energy channel (90, 120) \keV is used as the reference to cross-correlate the data of the other channels.

We performed the spectral lag analysis in four-time intervals, covering the total emission interval (time interval: (-1, 20) s relative to GBM trigger), it first (time interval (-1, 5) s), second (time interval (7, 12) s) and third (time interval (12, 20) s) episodes. The results of the cross-correlation analysis are presented in Figure A2 of the appendix. Although the burst is expected to have significant lag as a long GRB \citep[primarily based on empirical fact, there are many long GRBs consistent with zero lag][]{2015MNRAS.446.1129B}, it demonstrates overall the lag between the reference light curve and that in different energy ranges is small $<$ 0.1 s; however, there is a significant positive trend such the lag time increases when the reference light curve is compared to those of increasing energy (top left at Figure A2 of the appendix), well fitted by logarithmic model $lag$ $\propto$ $A\lg{E}$ in range (20, 300) \keV with the spectral lag index of $A = 0.025 \pm 0.002$. At energies above 300 \keV lag -- energy dependence becomes flat. The lag -- energy dependence of the first emission episode is well fitted by logarithmic function with spectral lag index $A = 0.27 \pm 0.03$ in range (100, 1000) \keV (top right at Figure A2 of the appendix). The lag -- energy dependence of the second emission episode is well fitted by logarithmic function with spectral lag index $A = 0.067 \pm 0.004$ in range (40, 1000) \keV (bottom left at Figure A2 of the appendix). The lag -- energy dependence of the third emission episode is well fitted by logarithmic function with spectral lag index $A = 0.025 \pm 0.003$ in range (5, 400) \keV (bottom right at Figure A2 of the appendix).

Non-monotonic behaviour (breaks in lag -- energy dependence) found for all analyzed emission episodes of \thisgrb can be explained by the superposition effect: each episode consists of several overlapping pulses, having unique spectral-temporal properties \citep{min14}. The first emission episode demonstrates the most pronounced spectral lag, typical for long bursts, while it has the smoothest time profile (longer pulses) and the softest energy spectrum. It is in agreement with other published works, showing that longer pulses have, in general, softer spectrum and more significant spectral lag (see, e.g., \citet{nor05, hakk11, min14}).

\begin{figure}
\centering
\includegraphics[scale=0.35]{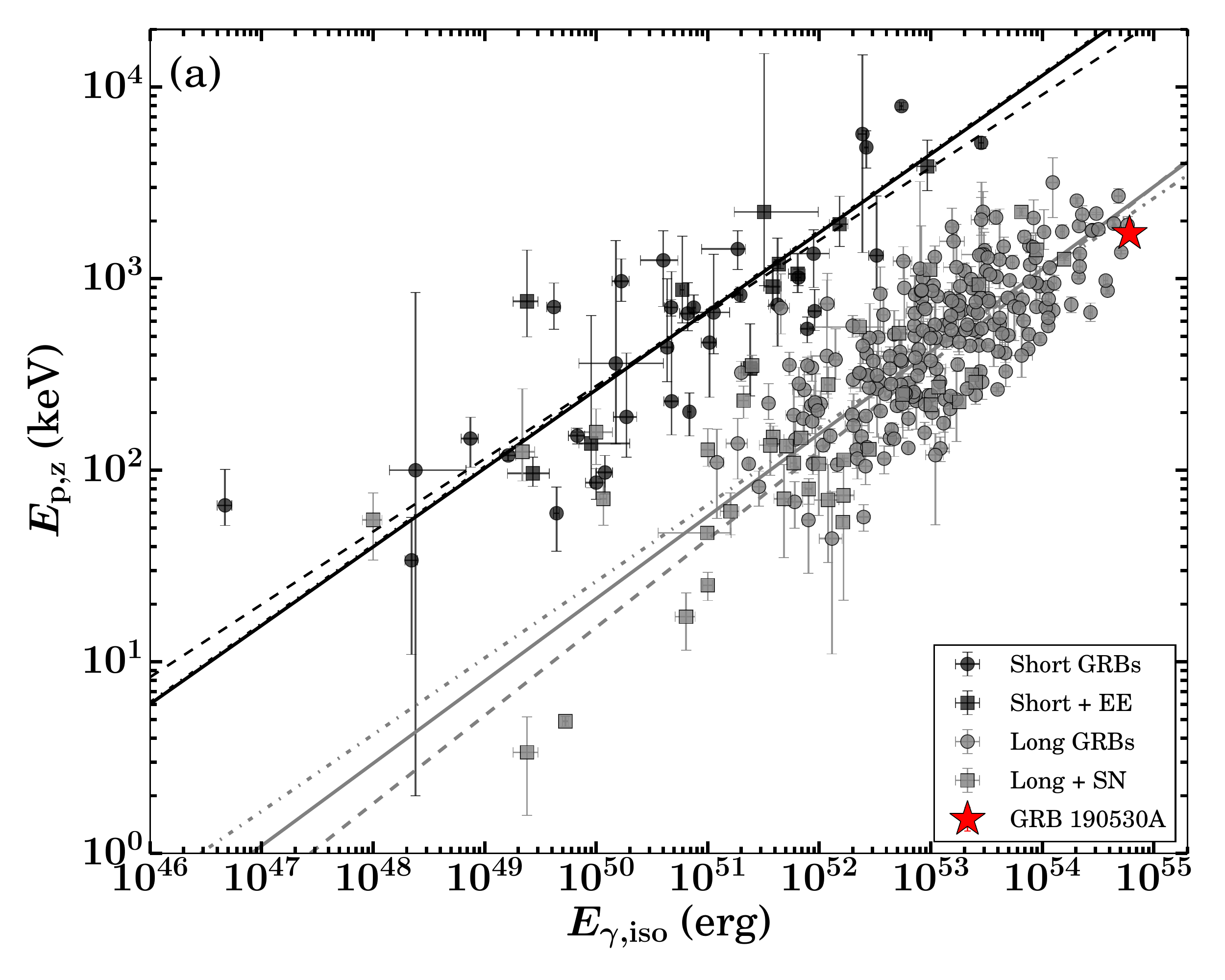}
 \includegraphics[scale=0.35]{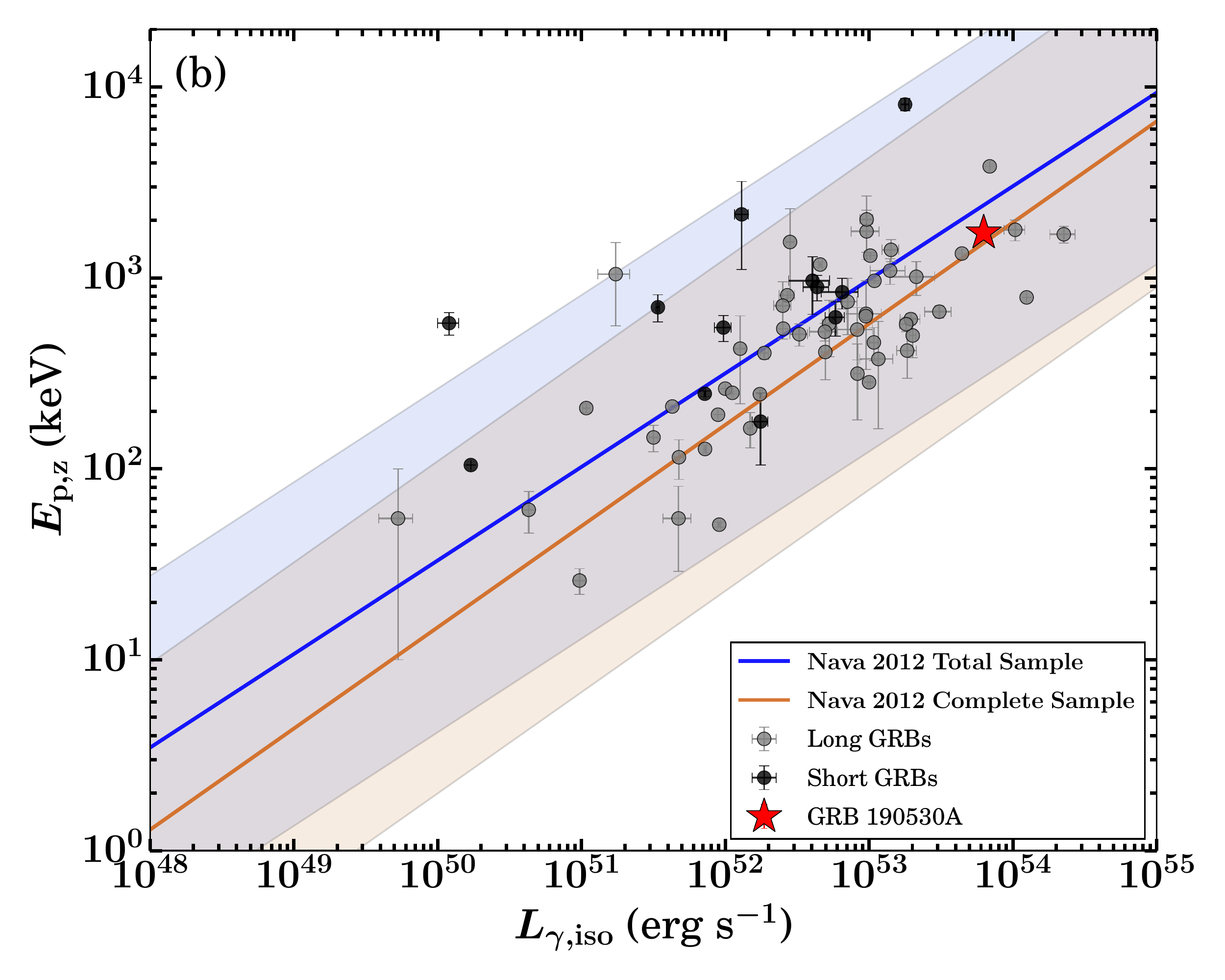}
\caption{{\bf Prompt emission characteristics of \thisgrb (shown with a red star):}  (a) Amati correlation:  \thisgrb along with the data points for long (grey circles for typical LGRBs and grey squares for LGRBs with associated supernovae) and short GRBs (black circles for typical SGRBs and black squares for SGRBs with extended emission) published in \protect\cite{2020MNRAS.492.1919M}. Grey colour solid, dashed, and dashed-dotted lines correspond to the best-fit lines for the complete sample of LGRBs, for LGRBs with and without associated supernovae, respectively. Similarly, black colour solid, dashed, and dashed-dotted lines correspond to the best-fit lines for the complete sample of SGRBs, for SGRBs with and without extended emission, respectively. (b) Yonetoku correlation: \thisgrb along with the data points for long (grey circles) and short GRBs (black circles) published in \protect\cite{2012MNRAS.421.1256N}. The coloured solid lines indicate the best-fit and shaded region represents the 3$\sigma$ scatter of the correlations \citep{2012MNRAS.421.1256N}.}
\label{fig:prompt_properties_amati_yonetoku}
\end{figure}

\subsubsection{\bf Amati and Yonetoku correlations}

The Amati correlation \citep{2006MNRAS.372..233A} is a correlation between the time-integrated peak energy in the source frame ($E_{\rm p,z}$) and isotropic equivalent $\gamma$-ray energy ($ E_{\gamma,\rm iso}$) of GRBs. The $E_{\gamma,\rm iso}$ depends on time-integrated bolometric (1-10,000 \keV in the rest frame) energy fluence. In the case of \thisgrb, we calculated the rest frame peak energy and $E_{\gamma,\rm iso}$ using the joint GBM and LAT spectral analysis (\fermiT to \fermiT + 25 s). The calculated values of these parameters are listed in Table \ref{tab:prompt_properties}) and shown in Figure \ref{fig:prompt_properties_amati_yonetoku} (a) along with other data points for long and short bursts published in \cite{2020MNRAS.492.1919M}. We noticed that \thisgrb lies towards the upper right edge and is consistent with the Amati correlation of long bursts. We compared the energetic of \thisgrb with a large sample of GBM detected GRBs with a measured redshift \citep{2021ApJ...908L...2S}. We noticed that \thisgrb is one of the most energetic GRBs ever detected, with only GRB 140423A and GRB 160625B reported as more energetic.

Furthermore, we also examined the location of \thisgrb on the Yonetoku correlation \citep{2004ApJ...609..935Y}. This is a correlation between the time-integrated peak energy in the source frame ($E_{\rm p,z}$) and isotropic peak luminosity ($L_{\gamma,\rm iso}$). To calculate the value of $L_{\gamma,\rm iso}$, we measured the peak flux in 1-10,000 \keV energy range for \thisgrb. The calculated value of $L_{\gamma,\rm iso}$ is listed in Table \ref{tab:prompt_properties}). The position of \thisgrb on the Yonetoku relation is given in Figure \ref{fig:prompt_properties_amati_yonetoku} (b) together with data points for other short and long GRBs, published in \cite{2012MNRAS.421.1256N}. In this plane, \thisgrb lies within the 3 $\sigma$ scatter of the total and complete samples\footnote{\url{https://www.mpe.mpg.de/events/GRB2012/pdfs/talks/GRB2012_Nava.pdf}} of GRBs studied by \cite{2012MNRAS.421.1256N}. \thisgrb is one of the bursts with the largest $L_{\gamma,\rm iso}$.

\subsection{\bf Correlation between spectral parameters}

The prompt emission spectral parameter correlations play an important role in investigating the intrinsic behaviour of GRBs. In the case of \thisgrb, we investigated the correlation between \Ep - flux, $\alpha_{\rm pt}$-flux, and \Ep-$\alpha_{\rm pt}$ obtained using the \sw{Band} function based on time-resolved analysis of the GBM data (for each bin obtained from the Bayesian Block binning algorithm). We noticed a strong correlation between the \Ep and the flux in 8 \keV - 30 MeV energy range with a Pearson coefficient (r) and \sw{p-value} of 0.82 and 4.00 $\times$ $10^{-11}$, respectively. We also noticed a strong correlation between $\alpha_{\rm pt}$ and flux with r and \sw{p-value} of 0.76 and 9.91 $\times$ $10^{-9}$. As \Ep and $\alpha_{\rm pt}$ show a strong correlation with flux, we investigated the correlation between \Ep and $\alpha_{\rm pt}$. They also show a moderate correlation with r and \sw{p-value} of 0.52 and 4.92 $\times$ $10^{-4}$. Therefore, \thisgrb is consistent with being a ``Double tracking'' GRB (Both $\alpha_{\rm pt}$ and the \Ep follow the “intensity-tracking” trend) similar to GRB 131231A \citep{2019ApJ...884..109L} and GRB 140102A \citep{2021MNRAS.505.4086G}. The correlation results are shown in Figure A3 in the appendix.

\subsection{\bf Nature of the afterglow}
\label{afterglow}

The early X-ray and optical afterglows of this GBM localized burst were missed, and we could not get early phase data till the MASTER network of telescopes provided the precise localization \citep{2019GCN.24676....1F, 2019GCN.24680....1L}. In the following section, we present the results of afterglow closure relations and multiwavelength modelling of the afterglow of \thisgrb. 

\subsubsection{\bf {Spectral Energy Distribution: Extinction law}}
\label{sedEL}

\begin{figure}
\centering
\includegraphics[scale=0.36]{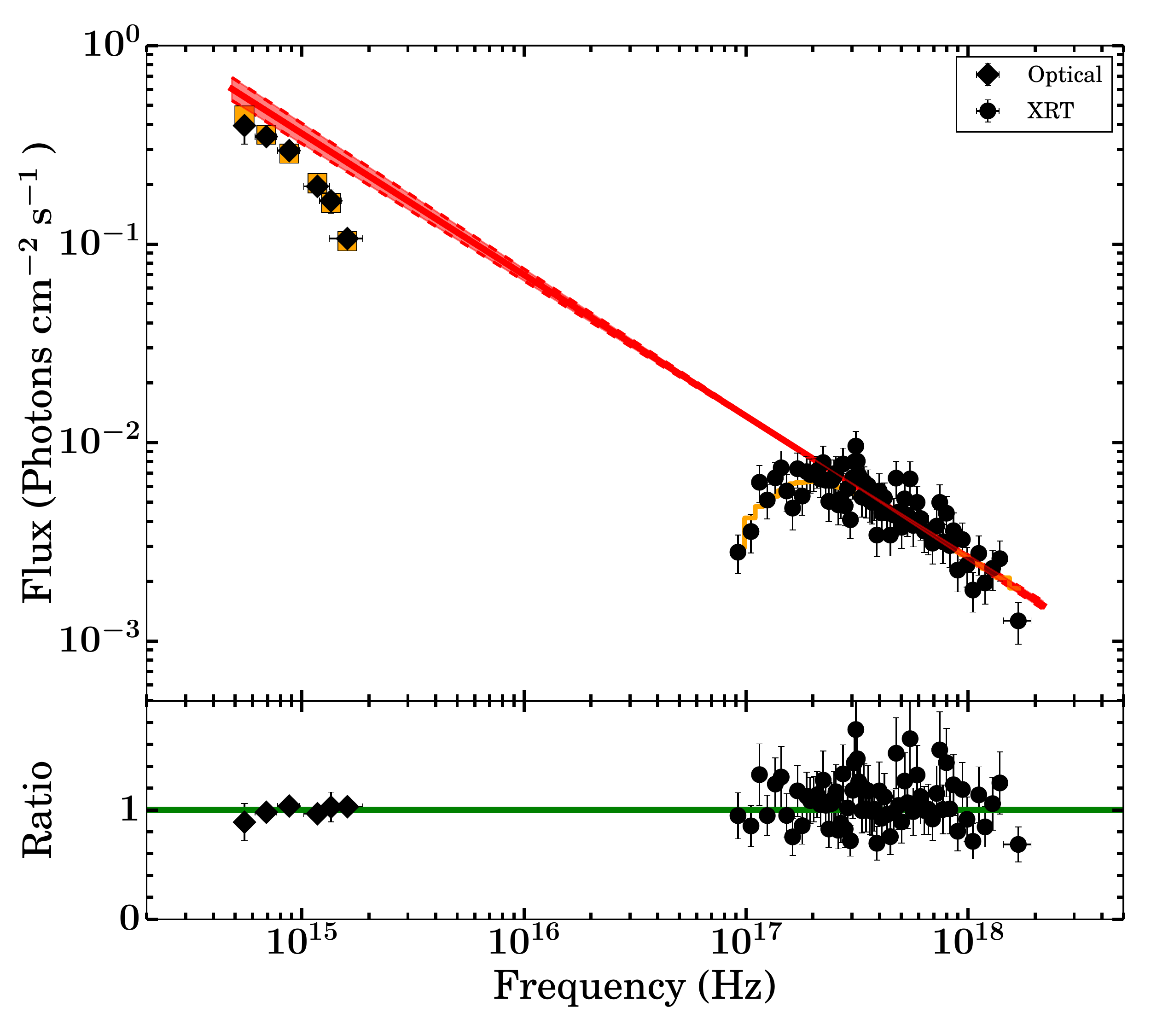}
\caption{{\bf Spectral Energy Distribution}: The best fit SED for the LMC extinction law with a power-law fit was obtained using joint XRT and UVOT data analysis during the temporal window from 30.6 ks to 60.7 ks. The red line shows the best fit unabsorbed spectral index, and the shaded region shows the 1 $\sigma$ associated uncertainty. The orange line and squares show the best fit model for the observed X-ray, and optical SEDs, respectively. The bottom panel indicates the ratio of data (observed) to the model. The horizontal green solid line corresponds to the ratio equal to one.}
\label{SED_fig}
\end{figure}

Following the methodology discussed in \S~\ref{SED}, we created the SED using \swift XRT and UVOT observations between 30.6 and 60.7 ks. During this temporal window, there is no break in the X-ray light curve; also, no spectral evolution is observed. The evolution of the X-ray photon index ($\it \Gamma_{\rm XRT}$) measured during this time window is consistent with not changing (see Figure \ref{Xray_afterglow_190530A}). We fit the SED with the simplest model, a power-law, which we find to be practically indistinguishable with \sw{$\chi^2$} values for the LMC and SMC 68.96 and 69.08 respectively for 76 degrees of freedom. However, for the MW model, we find slightly larger \sw{$\chi^2$} (70.43) for the same number of degrees of freedom. Therefore, for a power-law model, the LMC model has the lowest reduced \sw{$\chi^2$} value (0.91), however, the reduced \sw{$\chi^2$} values for all the considered extinction curves do not have much difference. Further, we fit the SED with the broken power-law model. For the MW, LMC, and SMC, all the fits are statistically acceptable with \sw{$\chi^2$} of 66.76, 65.20, 65.03, respectively, for 75 degrees of freedom. Further, we performed the \sw{F-test}\footnote{\url{https://heasarc.gsfc.nasa.gov/xanadu/xspec/manual/node83.html}} to find the best fit model among the different combinations of power-law and broken power-law models. We find the F statistic values (probability) as 4.12 (0.05) for MW, 4.33 (0.04) for LMC, and 4.67 (0.03) for SMC model, respectively. This suggests that the LMC model for single power-law is the best fit model with the lowest \sw{$\chi^2$}/dof for the SED. However, other two extinction laws also provide an acceptable to the data. We calculated the host extinction ($0.02 \pm 0.01$ mag). The SED is shown in Figure \ref{SED_fig}. All the results of SED are listed in Table \ref{SED_table}.

\begin{table}
\begin{center}
\caption{The best fit spectral parameters and spectral regime were obtained from the joint XRT and UVOT afterglow SED. The parameter $p$ denotes the mean value of the electron distribution index obtained from the observed value of the temporal and spectral index for the best spectral regime. Uncertainty in the calculation of $p$ is obtained with a confidence level of 95 \%. $\chi_r^{2}$ notify the reduced chi-square values.} 
\label{SED_table}
\begin{scriptsize}
\begin{tabular}{cccc}
\hline
\textbf{Time interval} & \textbf{$\bf \beta_{\rm \bf X-ray/opt}$}  & \textbf{$p$} & \textbf{$\bf \chi_r^{2}$} \\ 
\textbf{(ks)} &  & \textbf{(Spectral regime)} &  \\ \hline
30.6 - 60.7& 0.71$^{+0.02}_{-0.02}$ & \begin{tabular}[c]{@{}c@{}}2.84 $\pm$ 0.36 \\ ($\nu_{\rm opt}$ $<$ $\nu_{\rm x-ray}$ $<$ $\nu_{\rm c}$)  \end{tabular} & 0.90 \\ \hline
\end{tabular}
\end{scriptsize}
\end{center}
\end{table}

\subsubsection{\bf Origin of X-ray and Optical afterglows}
\label{closure_relation}

The optical and X-ray temporal slopes obtained using the simple power-law fits are consistent with each other. To understand the origin of the X-ray and optical afterglow data, we produced the spectral energy distribution (from \fermiT+30.6 to \fermiT+60.7 ks) using joint UVOT and XRT data. We explain the joint spectral analysis method in \S~\ref{SED} and present the results in Figure \ref{SED_fig}. Considering the slow cooling and constant medium case without energy injection, using the external shock model for $\nu_c<\nu_{opt}<\nu_{x-ray}$, $\nu_{opt}<\nu_c<\nu_{x-ray}$, and $\nu_{opt}<\nu_{x-ray}<\nu_c$ spectral regimes of FS \citep{2013NewAR..57..141G}, we can calculate the power law index of the shocked electrons by using the closure relations for these spectral regimes. We found that temporal decay $\alpha_{opt} = 1.59 \pm 0.08$, and $\alpha_{x-ray}$ = $1.80 \pm 0.07$. The value of spectral indices $\beta_{opt/x-ray}$ = 0.71 $\pm$ 0.02. We used the observed values of $\alpha_{\rm opt}-\beta_{\rm opt}$, $\alpha_{\rm x-ray}-\beta_{\rm x-ray}$ to constrain the $p$ value and position of the cooling-break frequency ($\nu_c$). We found that the $p$ value is most consistent for $\nu_{\rm opt}$ $<$ $\nu_{\rm x-ray}$ $<$ $\nu_{\rm c}$ spectral regime during the given segment of SED of \thisgrb. 
We calculated the $p$ value using observed value of $\alpha_{\rm opt}-\beta_{\rm opt}$, $\alpha_{\rm x-ray}-\beta_{\rm x-ray}$ and find $p$ = 2.84 $\pm$ 0.36, this is consistent with that calculated from afterglow modelling (see \S~\ref{optical_afterglow_modelling}). Hence, we can conclude that the afterglow of \thisgrb is formed in an external forward shock consistent with the slow cooling ISM medium case.

\subsubsection{\bf {Broadband afterglow light curve modelling}}
\label{optical_afterglow_modelling}

The X-ray light curve of \thisgrb declines from the beginning of observations as power-law and shows no superimposed features, such as steep and shallow decays phases or any flaring activity (see Figure \ref{Xray_afterglow_190530A}). The multi-band optical/UV afterglow light curves also follow a simple power-law decay behaviour. The best fit temporal flux decay indices and statistics used are listed in Table \ref{lcfits}. The optical and X-ray decay indices are consistent at $2\sigma$ and can thus be explained as originating from a single component model.

\begin{figure}
\centering
\includegraphics[scale=0.29]{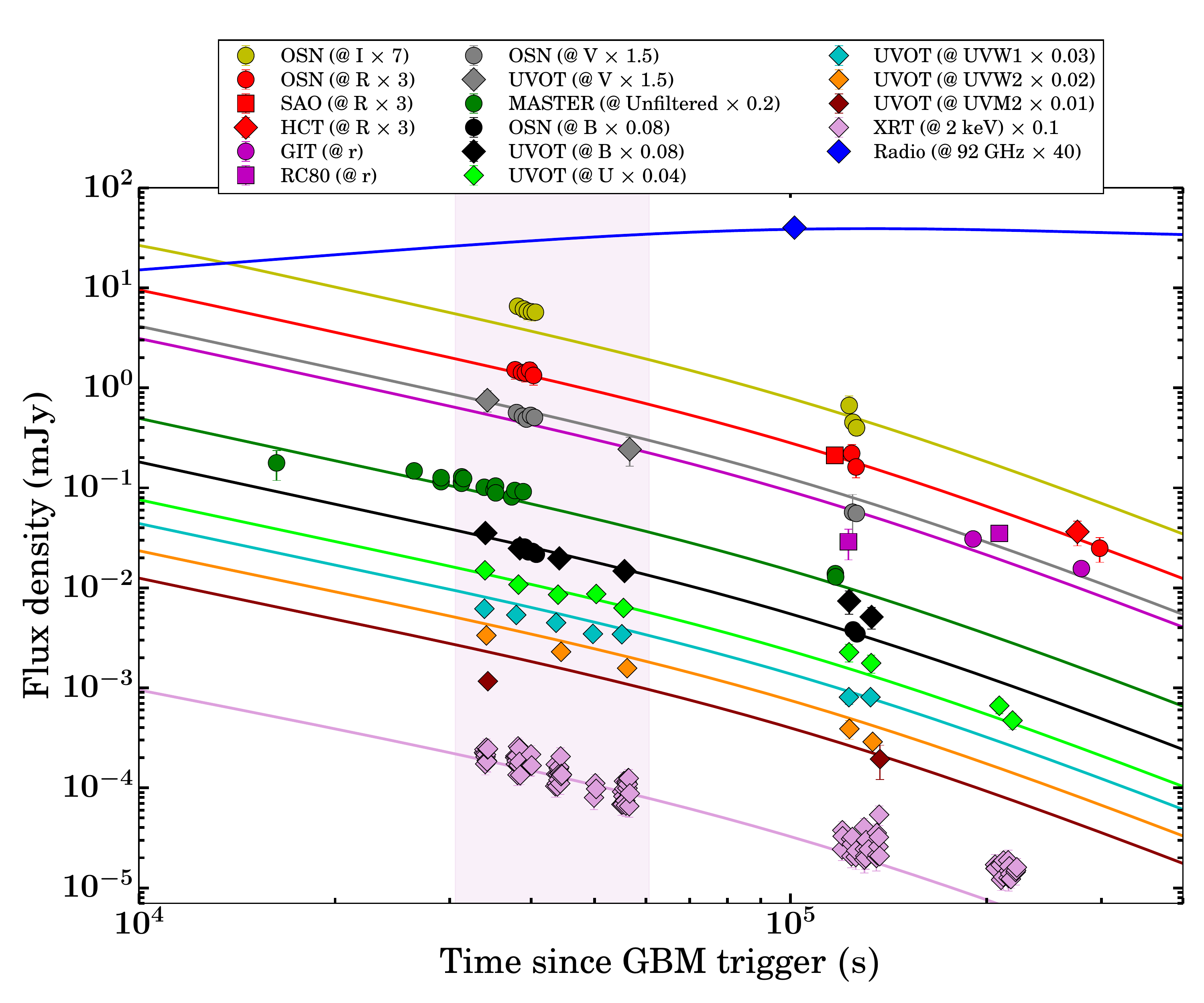}
\includegraphics[scale=0.28]{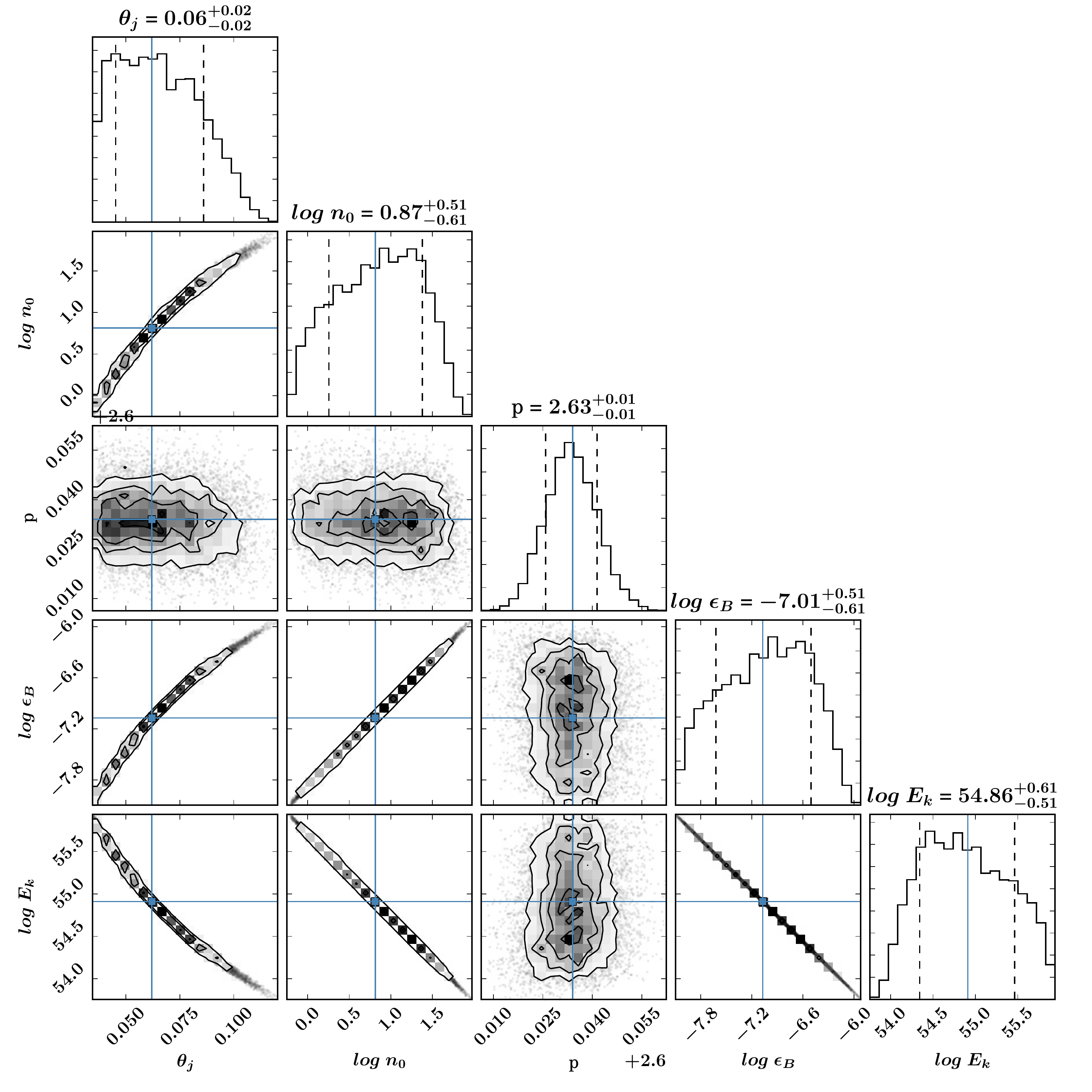}
\caption{{\bf Broadband afterglow modelling of \thisgrb:} {\em Top panel:} The multiwavelength afterglow light curves and the best fit model for each frequency are given by the corresponding coloured line. The vertical plum shaded region indicates the epoch used for the spectral energy distribution analysis. The legends for the optical data are similar as given in Figure \ref{Optical_afterglow_190530A}. {\em Bottom panel:} Corner plots obtained using simulation of afterglow data using external forward shock model. The best fit model parameters are provided at the top of each column.}
\label{afterglow_190530A}
\end{figure}

Presently, the external fireball forward shock model (synchrotron emission up to the X-ray wavelengths and synchrotron self Compton (SSC) emission for GeV-TeV photons) is the most accepted model used to explain the observed broadband afterglow emission from GRBs. According to this model, the interaction of the relativistic ejecta with the external medium is responsible for the observed afterglow at different frequencies. The temporal ($\alpha$) and spectral ($\beta$) characteristics of the afterglows are explained by the closure relations \citep[][see the references for a comprehensive list]{sari98, mes97, sari99, racusin09}. The values of $\alpha$ and $\beta$ are connected to the electron energy distribution index $p$ (generally found to be between 2 and 3 for relativistic shocks) for different ambient media densities (ISM or wind-like) and evolving spectra with frequencies (in the synchrotron spectrum, mainly the synchrotron cooling frequency $\nu_{\rm c}$ and the synchrotron peak frequency $\nu_{\rm m}$. Another break frequency is the synchrotron self-absorption frequency, though it mainly influences the low-frequency data) as a function of micro-physical parameters. 

To model the observed data, we consider a constant density external medium and adiabatic external shock without energy injection (as suggested by the closure relations, see \S\ref{closure_relation}). In this case, the peak synchrotron flux is defined as $f_{\rm max}^f \propto t^{0}$ for $t<t_j$ and after the jet break time it follows as $t^{-1}$, where $f$ and $t_j$ denote the forward shock and jet break time, respectively. The synchrotron peak frequency also evolves with time as $\nu_m^f \propto t^{-3/2}$ for $ t< t_j$ and $ \nu_m^f \propto t^{-2}$ for $ t>t_j$. The cooling frequency $\nu_c^f \propto t^{-1/2}$ for $t< t_j$ and $\nu_c^f \propto t^{0}$ after $t_j$. We apply this model to the optical, X-ray and radio data, using the \sw{PyMultiNest} python module to perform the fitting and parameter estimation\footnote{\url{https://johannesbuchner.github.io/PyMultiNest/}}. The best fit model light curves are given in the top panel of Figure \ref{afterglow_190530A}. However, we observed a late-time discrepancy from the best fit model for the X-ray data, which might be due to the significant scattering at late epochs. In addition, we also noticed a discrepancy for the early optical data of I and UVM2 filters, which might be due to the unavailability of continuous observations in these bands. We plot the two-dimensional posteriors in the bottom panel of Figure \ref{afterglow_190530A} and provide the best fit values at the top of each column. The parameters determined are: the electron energy index $p$, micro-physical parameters $\epsilon_e$ and $\epsilon_B$.

\section{\bf Discussion}
\label{discussion}

The radiation mechanism of the prompt emission of GRBs is still an open question. Besides temporal and spectral properties, a polarization measurement is a powerful tool for investigating the radiation mechanisms in the prompt emission.

\subsection{\bf Prompt Emission mechanism of \thisgrb}
\label{pem}

The different emission processes invoked to explain the prompt emission of GRBs is associated with unique polarization signatures. In the case of \thisgrb, we found a hint of high polarization fraction in both the time-integrated and time-resolved polarization measurements. We do not notice any significant variation in polarization fraction and polarization angle in our time-resolved polarization analysis, supporting the synchrotron emission model, an ordered magnetic field produced in shocks \citep{2003ApJ...597..998L} for the first two pulses. Such high polarization ($\sim$ 40-70 \%) cloud also be produced using synchrotron emission with a random magnetic field, in the case of a narrow jetted emission ($\Gamma_{0}$ $\theta_{\rm j}$ $\sim$ 1, where $\Gamma_{0}$ is the bulk Lorentz factor and $\theta_{\rm j}$ is the jet opening angle) and seen along the edge. To verify both possibilities, we calculated bulk Lorentz factor $\Gamma_{0}$ of the fireball ejecta and $\theta_{\rm j}$ (see \S\ref{optical_afterglow_modelling}). There are several methods to calculate $\Gamma_{0}$ using both prompt emission and afterglow properties \citep{2018A&A...609A.112G}. We calculated the value of the Lorentz factor using the prompt emission correlation between $\Gamma_{0}$-$E_{\gamma, \rm iso}$\footnote{$\Gamma_{0}$ $\approx$ 182 $\times$ $E_{\gamma, \rm iso, 52}^{0.25 \pm 0.03}$} \citep{2010ApJ...725.2209L} as $\Gamma_{0}$ decreases towards afterglow phase. The calculated value of $\Gamma_{0}$ is 902.63$^{+191.23}_{-157.80}$ using the normalization and slope of the $\Gamma_{0}$-$E_{\gamma, \rm iso}$ correlation. The calculated value of $\theta_{j}$ is 0.062 radian (3.55$^\circ$) derived from the broadband afterglow modelling (see \S \ref{optical_afterglow_modelling}). We obtained $\Gamma_{0}$ $\theta_{\rm j}$ equal to $\sim$ 56, which supports the synchrotron emission model with an ordered magnetic field \citep{2009ApJ...698.1042T}. We also calculated the beaming angle ($\theta_{\rm beam}$) of the emission equal to 0.001 radian (0.06$^\circ$) using the relation between Lorentz factor and $\theta_{\rm beam}$, i.e., $\theta_{\rm beam}$= 1/$\Gamma_{0}$. Thus, \thisgrb had a wider jetted emission with a narrow beaming angle.

In addition to polarization results, our time-resolved spectral analysis indicates that the low-energy spectral indices of the \sw{Band} function are consistent with the prediction of synchrotron emission for the first two pulses. Moreover, the presence of a low-energy spectral break in the time-integrated and time-resolved spectra with power-law indices consistent with the prediction of synchrotron emission model confirms synchrotron emission as the mechanism dominating during the first two pulses of \thisgrb. However, during the third pulse, the low-energy spectral indices become harder and exceed the synchrotron death line in few bins. During this window, we find a signature of the thermal component along with the synchrotron component in our time-resolved spectral analysis, suggesting some contribution to the emission from the photosphere.

\subsection{\bf Afterglow origin of LAT GeV Photons}
\label{lat_TRS}

In the case of \thisgrb, the extended GeV emission becomes harder and slightly brighter (consistent with statistical fluctuation) after the end of prompt \keV-MeV emission. This indicates that the LAT high energy emission started later than the prompt \keV-MeV emission and is from a different spatial region (originated due the external shock). This section will study the possible origin and emission mechanism of the GeV photons detected by \fermi LAT. For this purpose, we measure the maximum photon energy emitted by synchrotron radiation in an adiabatic external forward shock during the decelerating phase in a constant ambient medium. In this case, we use equation 4 from \cite{2010ApJ...718L..63P}.

We consider $\rm n_0$ = 7.41 (see \S~\ref{optical_afterglow_modelling}) for the present analysis (see Figure \ref{afterglow_190530A}). We noticed that one of the late time photons (source association probability $>$ 90 $\%$) lies slightly above the maximum synchrotron energy line, which indicates that this photon could be from a SSC process, as observed in the recent VHE detected GRBs \citep{2019Natur.575..455M, 2021RMxAC..53..113G}. 

Time-resolved \fermi LAT analysis shows that the high energy emission could be momentary increasing ($0 - 8$ s \&  $8 - 11$ s), peaked in the third temporal bin ($11 - 13$ s) and then decreasing with time up to fifth bin ($13 - 15$ s \& $15 - 22$ s) in both energy and photon fluxes during the prompt emission phase. The prompt phase of \thisgrb ends $\sim$ 25 s after the trigger, which can also be seen in the time-resolved spectra, which show substantial temporal variation in the photon index in the first five bins. After the prompt phase, the \fermi LAT photon flux light curve shows temporal variation as a power-law with an index $0.33 \pm 0.24$. The energy flux light curve shows temporal variation as a power-law an index of $0.10 \pm 0.30$ (nearly flat). The time-resolved spectra do not show substantial temporal variation in the photon index in the last four bins after the prompt phase. The observed flattening could be explained using SSC emission. The \fermi LAT (GeV) light curve may flatten and the \fermi LAT spectrum to harden when the peak of the SSC component passes through the LAT energy range \citep{2014Sci...343...42A}.

\subsubsection{\bf Comparison with \fermi-LAT catalogue}
\label{Comparison with LAT catalogue}

We compared the high energy properties of \thisgrb observed with \fermi-LAT instrument with other LAT detected GRBs \citep[the second GRB LAT catalogue (2FLGC;][]{2019ApJ...878...52A}. We compared the energy fluence values in GBM (10-1000 \keV) and LAT (0.1-100 GeV) energy ranges during a temporal window of 18.4 s (\tninty duration) since \fermiT. In this time interval, we calculated the LAT energy fluence value equal to 2.41 $\times$ $10^{-7}$ erg $\rm cm^{-2}$ $\rm s^{-1}$ in 0.1 - 100 GeV energy range and compared with the GBM fluence value for \thisgrb. \thisgrb lies on the line for which GBM fluence is 100 times brighter than LAT fluence for all the LAT detected samples (see Figure A4 (a) in the appendix). 

For this burst, \fermi LAT observed many high energy GeV photons for an extended duration. The detection significance is computed by Test Statistics (TS) value calculated from the Likelihood Ratio Test. Likelihood Ratio Test analyzes two different models; the first one regards only the background, the second model includes an extra presumed GRB as a point object. The ratio of the likelihoods by fitting these two models provide a TS value; a higher TS value suggests a high detection significance for a given object, TS equal to 36 corresponds to approximately six sigma. Here we carried out the \fermi LAT data likelihood analysis using \sw{gtburst} software, see \S\ref{section:LAT} for more details. The TS value we calculated for the \tninty from 0-18.4 s is 189, which is among the highest TS values for GRBs observed using \fermi LAT. Usually, the TS value is less than 150 if LAT GeV photons are detected, as presented in Figure A4 (b) of the appendix. We also calculate the TS value for a time window of 25 s (\fermiT - \fermiT+25 s; during the prompt emission phase), TS = 240.

In the case of \thisgrb, the highest energy photon observed with \fermi LAT is at 8.7 GeV and was detected 96 s after the \fermiT. In Figure A4 (c) of the appendix, we have shown the maximum photon energy of the highest-energy photon as a function of arrival time for \thisgrb along with other data points taken from 2FLGC. For \thisgrb, the highest-energy photon arrives after the GBM \tninty duration, consistent with a large fraction of LAT detected GRBs. We also calculated isotropic $\gamma$-ray energy in the 100 MeV -10 GeV rest frame ($E_{\rm LAT, iso}$) using \fermi LAT observations for \thisgrb.  We have shown the distribution of $E_{\rm LAT, iso}$ as a function of redshift ($z$) for \thisgrb along with data points for the 34 \fermi LAT detected bursts with a measured redshift from the 2FLGC (see Figure A4 in the appendix). We notice that \thisgrb is one of the most energetic \fermi LAT detected GRBs below z $<$ 1, with the highest-energy photon of a 16.87 GeV photon in the rest frame.

\subsection{\bf Central Engine}

\begin{figure}
\centering
\includegraphics[scale=0.35]{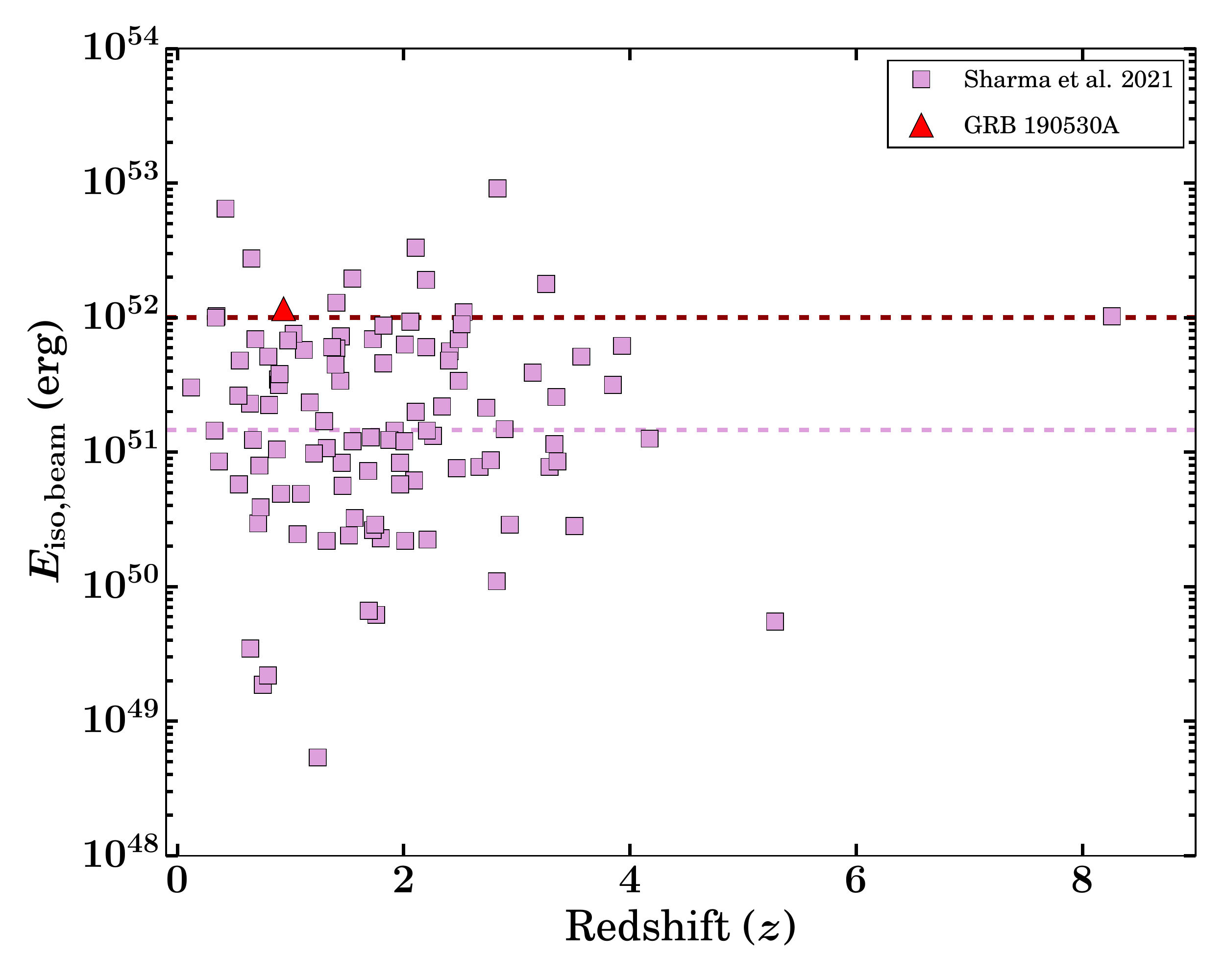}
\caption{{\bf The central engine of \thisgrb:} Redshift distribution as a function of beaming corrected isotropic $\gamma$-ray energy for the \fermi detected bursts, data points taken from \protect \cite{2021ApJ...908L...2S}. \thisgrb is shown with a red triangle. The horizontal red and pink dashed lines indicate the maximum possible energy budget of the magnetars and the median value of beaming corrected isotropic $\gamma$-ray energy of the sample studied by \protect \cite{2021ApJ...908L...2S}, respectively.}
\label{central engine}
\end{figure}

Based on the properties of prompt and afterglow emission, e.g. variability in the gamma-ray light curves and X-ray flares and plateau \citep{2012A&A...539A...3B, 2020ApJ...896...42Z}, there are two types of the object thought to be powering the central engine: a stellar-mass black hole (BH), and a rapidly spinning, highly magnetized 'magnetar.' Recently, \cite{2018ApJS..236...26L} studied the X-ray light curves sample of 101 bursts with a plateau phase and measured redshift. They calculated the isotropic kinetic energies and the isotropic X-ray energies for each burst. They compared them with the maximum possible rotational energy budget of the magnetar ($10^{52}$ ergs). They found only $\sim$ 20 \% of GRBs were consistent with having a magnetar central engine. The rest of the bursts were consistent with having a BH as the central engine. More recently, \cite{2021ApJ...908L...2S} also identified GRBs with BH central engines based on the maximum rotational energy of the magnetar that powers the GRB, i.e., the upper limit of rotational energy of magnetar. They analyzed the sample of \fermi detected GRBs with a measured redshift. They calculated the beaming corrected isotropic gamma-ray energies and compared them with the magnetars' maximum possible energy budget. In the case of \thisgrb, we could not follow the methodology discussed by \cite{2018ApJS..236...26L} due to the absence of plateau phase in the X-ray light curves; therefore, we follow the methods discussed by \cite{2021ApJ...908L...2S}. We calculated the beaming corrected energy assuming the fraction of forward shock energy into the electric field $\epsilon_{e}$ = 0.1. We performed broadband afterglow modelling to constrain the limiting value of the jet opening angle (see \S\ref{afterglow}). We find beaming corrected energy for \thisgrb equal to 1.16 $\times$ $10^{52}$ ergs, and this value is well above the mean energy of the sample studied by \cite{2021ApJ...908L...2S}, see also Figure \ref{central engine}. In addition, this value is also higher than the maximum possible energy budget of the magnetar. No flares, plateau features are present in the X-ray light curve. 
We proposed that BH could be the possible central engine of \thisgrb. However, the magnetar option could also be feasible as beaming corrected energy for \thisgrb is close to the upper limit of magnetar's rotational energy, and no early X-ray observations of the X-ray afterglow of this burst are available.

\section{\bf Summary \& Conclusion}
\label{conclusion}

We studied the temporal, spectral, and polarization characteristics of the prompt emission of \thisgrb using \fermi and \AstroSat CZTI observations. \thisgrb (the sixth brightest burst ever observed by GBM) consists of three peaks with increasing hardness ratio. We noticed that the time-averaged spectrum (\fermiT to \fermiT+25 s) has a peculiar low-energy break in addition to the typical \Ep break. Such a low-energy break in addition to \Ep has only been seen in a few of the brightest GBM detected long bursts \citep{2019A&A...625A..60R, 2017ApJ...846..137O, 2018A&A...616A.138O, 2019A&A...628A..59O}. We performed a time-resolved analysis based on coarse (constant cadence) and fine bins (Bayesian algorithm) techniques to study the spectral evolution and search for the low-energy spectral break. Low-energy breaks were detected in some of the time-resolved bins with mean photon indices $<\alpha_{1}>$ = 0.84 (with $\sigma$ = 0.04) and $<\alpha_{2}>$ = 1.43 (with $\sigma$ = 0.06), consistent with the power-law indices expected by synchrotron emission in a marginally fast cooling spectral regime. Taking the low-energy break as due to the synchrotron cooling frequency, we constrain a limit on the co-moving magnetic field (B) following equation 8 of \cite{2018A&A...613A..16R}. We calculated B $\leq$ 9 gauss for \thisgrb. However, this value is small and not consistent with the expected value for a typical emitting region located at $\sim$ 10$^{14}$ cm \citep{2018A&A...613A..16R}.

In addition, we also found interesting spectral evolution within the \sw{Band} spectral parameters obtained using the detailed time-resolved spectroscopy. The spectral evolution of \Ep tracks the intensity of the GBM light curve and exhibits a strong correlation. Usually, the $\alpha_{\rm pt}$ evolution does not have any particular trend, but for \thisgrb, we found that it also tracks the intensity of burst; therefore, \thisgrb exhibits characteristics of a double-tracking burst. So far, this tracking behaviour has only been found in a few GRBs, i.e. in GRB 131231A \citep{2019ApJ...884..109L} and GRB 140102A \citep{2021MNRAS.505.4086G}, the low energy spectral index remains in the synchrotron limit ($\alpha_{pt} =-2/3$). Similarly, in the case of \thisgrb, $\alpha_{\rm pt}$ values are within the synchrotron limits for the first two pulses. However, during the third and the brightest pulse, $\alpha_{\rm pt}$ values become harder and exceed the synchrotron line of death in a few bins. During this temporal window, we found a signature of a thermal component along with a synchrotron one in our time-resolved spectral analysis, suggesting an additional contribution from the photosphere.

For \thisgrb, we found a hint of high polarization fraction in our time-integrated (55.43 $\pm$ 21.30 \%; 2.60 $\sigma$) as well as time-resolved (53.95 $\pm$ 24.13 \%; 2.24 $\sigma$ for the third pulse) polarization measurements in the 100-300 \keV energy range, based on our observations with \AstroSat CZTI. 
We investigated the origin of a high degree of polarization fraction and found that a synchrotron model with an ordered magnetic field could explain such a high polarization fraction. Our time-resolved polarization analysis does not show any substantial variation in the polarization fraction or angles. Based on our detailed spectro-polarimetric analysis, we suggest that the first two pulses of \thisgrb have a synchrotron origin, and it lies within a small subset of long GRBs with the credible signature of a high degree of prompt emission polarization \citep{2019ApJ...884..123C, 2020A&A...644A.124K}.

Apart from the prompt emission hard X-ray polarization measurements, we also constrained optical afterglow polarization using MASTER telescope data, making the burst the first case where both prompt emission and afterglow polarization measurements are constrained. These observations were carried out at different times: the \AstroSat CZTI data refer to the active stage of the gamma-ray burst, and the MASTER optical observations to the afterglow. Relatively high polarization of the intrinsic prompt hard X-ray radiation demonstrates a high ordering of the magnetic field in the region close to the jet base. It is apparently associated with the radiation of colliding relativistic plasma flows under conditions of multiple internal shocks. The optical afterglow is formed behind the shock in the driven plasma of the progenitor stellar wind \citep{sari98, sari99, 2000ApJ...545..807K}. The absence of significant afterglow optical polarization of more than 1\% indicates that the jet's own magnetic field has decayed due to the expansion of the radiation region, and the raked up chaotic magnetic field averaged and the radiation ceased to be polarized for the same reason \citep{2004MNRAS.347L...1L}. Overall, \thisgrb provides a detailed insight into the prompt spectral evolution and emission polarization and challenges the traditionally used spectral model.

We also studied the multiwavelength afterglow behaviour of this \thisgrb, one of the brightest bursts observed. We included observations taken from various ground-based telescopes along with \swift XRT, UVOT, and radio data as part of this analysis. We performed the modelling of broadband afterglow data considering an ISM ambient medium \citep{sari98}. The broadband afterglow is explained using an external forward shock model in the case of slow cooling. The closure relations indicate that the optical and X-ray emission is consistent with $\nu_{\rm opt}$ $<$ $\nu_{\rm x-ray}$ $<$ $\nu_{\rm c}$ spectral regime with slow cooling and an ISM ambient medium. We calculated the jet opening angle and beaming angle and found that \thisgrb consists of a jet with a wider jet opening and narrower beaming angles. Late time observations using the 3.6m DOT and the 10.4m GTC do not find signatures of the host galaxy to deeper limits indicative of an optically faint galaxy (see Table A9 in the appendix).

We also investigated the nature of the central engine of \thisgrb using the methodology discussed by \cite{2021ApJ...908L...2S}. We find beaming corrected energy for \thisgrb equal to 1.16 $\times$ $10^{52}$ ergs, larger than the mean beaming energy of a sample of GRBs studied by \cite{2021ApJ...908L...2S}. This energy is higher than the maximum possible energy budget of a magnetar, and no flares/plateau features are present in the X-ray light curve. This possibly support a BH based central engine for this GRB. We also constrain the radiative gamma-ray efficiency using the formula $\eta$= $E_{\rm \gamma, iso}$/($E_{\rm \gamma, iso}$+ $E_{\rm k}$)), finding $\eta$ $<$ 0.45 for \thisgrb. We conclude that the prompt emission polarization analysis, along with spectral and temporal information, has a unique capability to solving the long debatable topic of the emission mechanisms of GRBs.

\section*{Acknowledgements}
We thank the anonymous referee for providing constructive and positive comments.
RG, DB, SBP, KM, and VB acknowledge BRICS grant {DST/IMRCD/BRICS/PilotCall1/ProFCheap/2017(G)} for the financial support. RG and SBP thank Prof. A. R. Rao and Prof. Soebur Razzaque for fruitful suggestions to improve the manuscript. RG is also thankful to Dr P. Veres for sharing data files related to Figure A1 (a). This publication uses data from the \AstroSat mission of the Indian Space Research Organisation (ISRO), archived at the Indian Space Science Data Centre (ISSDC). CZT-Imager is built by a consortium of institutes across India, including the Tata Institute of Fundamental Research (TIFR), Mumbai, the Vikram Sarabhai Space Centre, Thiruvananthapuram, ISRO Satellite Centre (ISAC), Bengaluru, Inter University Centre for Astronomy and Astro-physics, Pune, Physical Research Laboratory, Ahmedabad, Space Application Centre, Ahmedabad. This research also has used data obtained through the HEASARC Online Service, provided by the NASA-GSFC, in support of NASA High Energy Astrophysics Programs. This work is based on data from the OSN Public Archive at IAA (IAA-CSIC) and the CAHA Archive at CAB (INTA-CSIC). MASTER equipment is supported by Lomonosov MSU Development Program. VL and VK are supported by RFBR 19-29-11011 grant. The research group of JV is supported by the project ``Transient Astrophysical Objects" GINOP 2.3.2-15-2016-00033 of the National Research, Development and Innovation Office (NKFIH), Hungary, funded by the European Union. AA acknowledges funds and assistance provided by the Council of Scientific \& Industrial Research (CSIR), India with file number 09/948(0003)/2020-EMR-I. AP, SB, and PM acknowledge a support of the  RSCF  grant  18-12-00378. The GROWTH India Telescope (GIT) is a 70-cm telescope with a 0.7-degree field of view, set up by the Indian Institute of Astrophysics and the Indian Institute of Technology Bombay with support from the Indo-US Science and Technology Forum (IUSSTF) and the Science and Engineering Research Board (SERB) of the Department of Science and Technology (DST), Government of India (https://sites.google.com/view/growthindia/). It is located at the Indian Astronomical Observatory (Hanle), operated by the Indian Institute of Astrophysics (IIA). MCG acknowledges support from the Ram\'on y Cajal Fellowship RYC2019-026465-I. YDH acknowledges support under the additional funding from the RYC2019-026465-I. B.B.Z acknowledges support by the National Key Research and Development Programs of China (2018YFA0404204), the National Natural Science Foundation of China (Grant Nos. 11833003, U2038105, 12121003), the science research grants from the China Manned Space Project with NO.CMS-CSST-2021-B11, and the Program for Innovative Talents, Entrepreneur in Jiangsu. RSM acknowledge support under the CSIC-MURALES project with reference
20215AT009.

\section*{Data Availability}

The data presented in this work can be made available based on the individual request to the corresponding authors.



\bibliographystyle{mnras}
\bibliography{GRB190530A} 



\section*{Affiliations}
\small{$^{1}$ Aryabhatta Research Institute of Observational Sciences (ARIES), Manora Peak, Nainital-263002, India \\
$^{2}$ Department of Physics, Deen Dayal Upadhyaya Gorakhpur University, Gorakhpur-273009, India \\
$^{3}$ Department of Physics, Indian Institute of Technology Bombay, Powai, Mumbai-400076, India \\
$^{4}$ Inter-University Center for Astronomy and Astrophysics, Pune, Maharashtra-411007, India\\
$^{5}$ Kavli Institute of Particle Astrophysics and Cosmology, Stanford University, 452 Lomita Mall, Stanford, CA 94305, USA \\
$^{6}$Lomonosov Moscow State University, SAI, Physics Department, 13 Univeristetskij pr-t, Moscow 119991, Russia \\
$^{7}$ Instituto de Astrof\'isica de Andaluc\'ia (IAA-CSIC), Glorieta de la Astronom\'ia s/n, E-18008, Granada, Spain \\
$^{8}$ Departamento de Ingenier\'ia de Sistemas y Autom\'atica, Escuela de Ingenier\'ias, Universidad de M\'alaga, C\/. Dr. Ortiz Ramos s\/n, E-29071, M\'alaga, Spain \\
$^{9}$ School of Physics and Astronomy, University of Birmingham, Birmingham B15 2TT, UK \\
$^{10}$ School of Studies in Physics and Astrophysics, Pandit Ravishankar Shukla
University, Raipur, Chattisgarh-492010, India \\
$^{11}$ Universidad de Granada, Facultad de Ciencias Campus Fuentenueva s/n, E-18071 Granada, Spain \\
$^{12}$ Special Astrophysical Observatory of Russian Academy of Sciences, Nizhniy Arkhyz 369167, Russia \\
$^{13}$ Crimean Astrophysical Observatory, Russian Academy of Sciences, Nauchnyi 298409, Russia  \\
$^{14}$ Space Research Institute, Russian Academy of Sciences, Profsoyuznaya ul. 84/32,  Moscow, 117997, Russia \\
$^{15}$ Lebedev Physical Institute, Leninsky Avenue 53, Moscow, 119991, Russia \\
$^{16}$ Konkoly Observatory, Research Centre for Astronomy and Earth Sciences, Konkoly Thege ut 15-17, 1121 Budapest, Hungary \\ 
$^{17}$ ELTE E\"otv\"os Lor\'and University, Institute of Physics, P\'azm\'any P\'eter s\'et\'any 1/A, Budapest, 1117, Hungary \\
$^{18}$ Department of Optics \& Quantum Electronics, University of Szeged, D\'om t\'er 9, Szeged, 6720, Hungary \\
$^{19}$ Department of Physics, KTH Royal Institute of Technology, AlbaNova, 10691 Stockholm, Sweden \\
$^{20}$ Facultad de Ciencias, Universidad de M\'alaga, 29010 M\'alaga, Spain \\
$^{21}$ National Research University “Higher School of Economics”, Myasnitskaya 20, Moscow, 101000 Russia \\
$^{22}$ School of Astronomy and Space Science, Nanjing University, Nanjing 210093, China \\
$^{23}$ Key Laboratory of Modern Astronomy and Astrophysics (Nanjing University), Ministry of Education, Nanjing 210093, China \\
$^{24}$ Ioffe Institute, 26,  Politechnic str., St Petersburg, 194021 Russia \\
$^{25}$ Indian Institute of Astrophysics, II Block Koramangala, Bengaluru 560034, India \\
$^{26}$ Instituto de Astrofísica de Canarias (IAC), Calle Vía Láctea s/n, E-38200, La Laguna, Tenerife, Spain \\
$^{27}$ INAF, Istituto di Astrofisica e Planetologia Spaziali, via Fosso del Cavaliere 100, I-00133 Rome, Italy \\
$^{28}$ Nikolaev National University, Nikolska 24, Nikolaev 54030, Ukraine \\
$^{29}$ Nikolaev Astronomical Observatory, Nikolaev 54030, Ukraine \\
$^{30}$ Physical Research Laboratory, Navrangpura, Ahmedabad, Gujarat-380009, India \\
$^{31}$ Purple Mountain Observatory, Academia Sinica, Nanjing 210008, China
}

\appendix
\section{FIGURES and TABLES}

\renewcommand{\thefigure}{A\arabic{figure}}
\setcounter{figure}{0}

\begin{figure}
\centering
\includegraphics[scale=0.35]{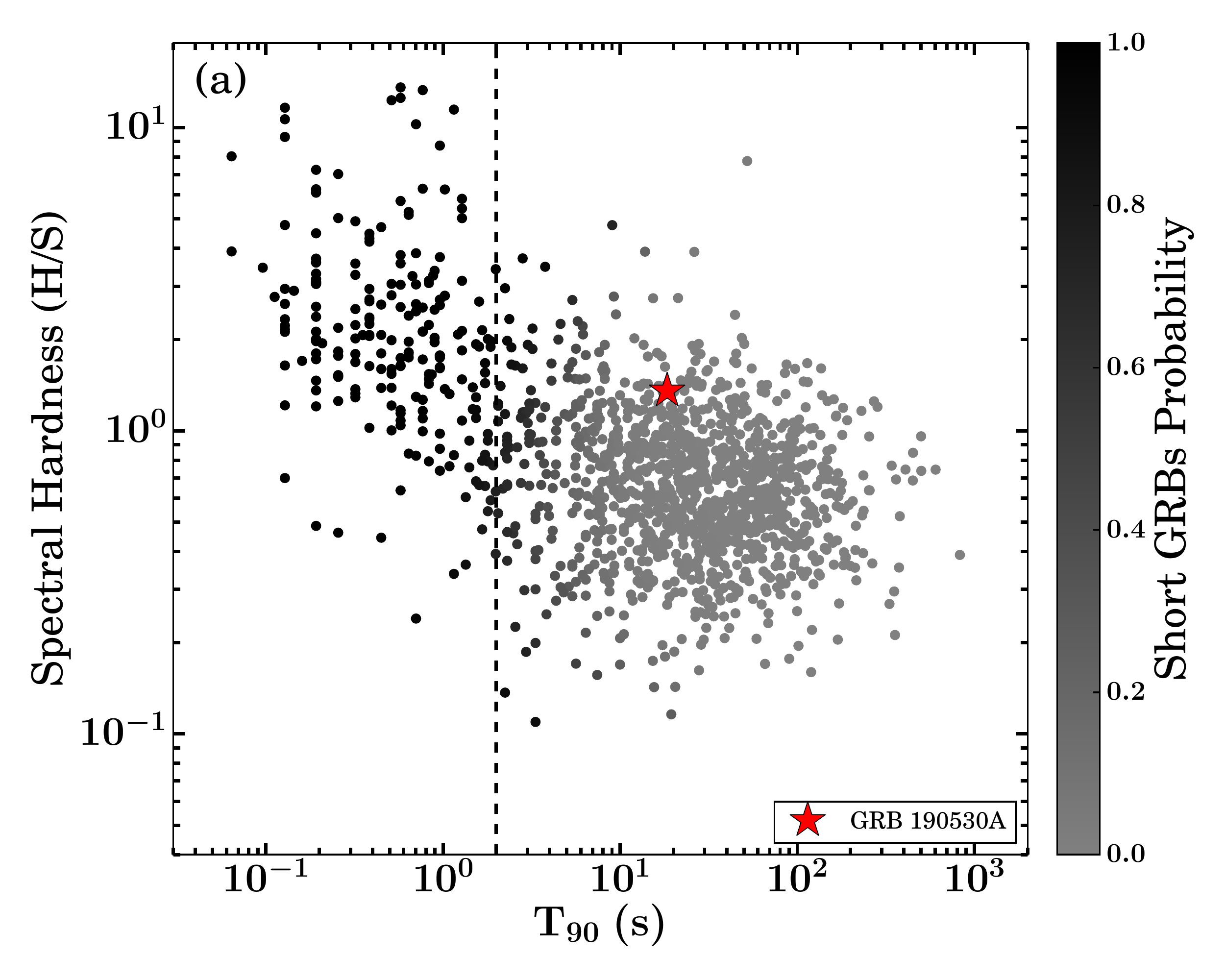}
\includegraphics[scale=0.35]{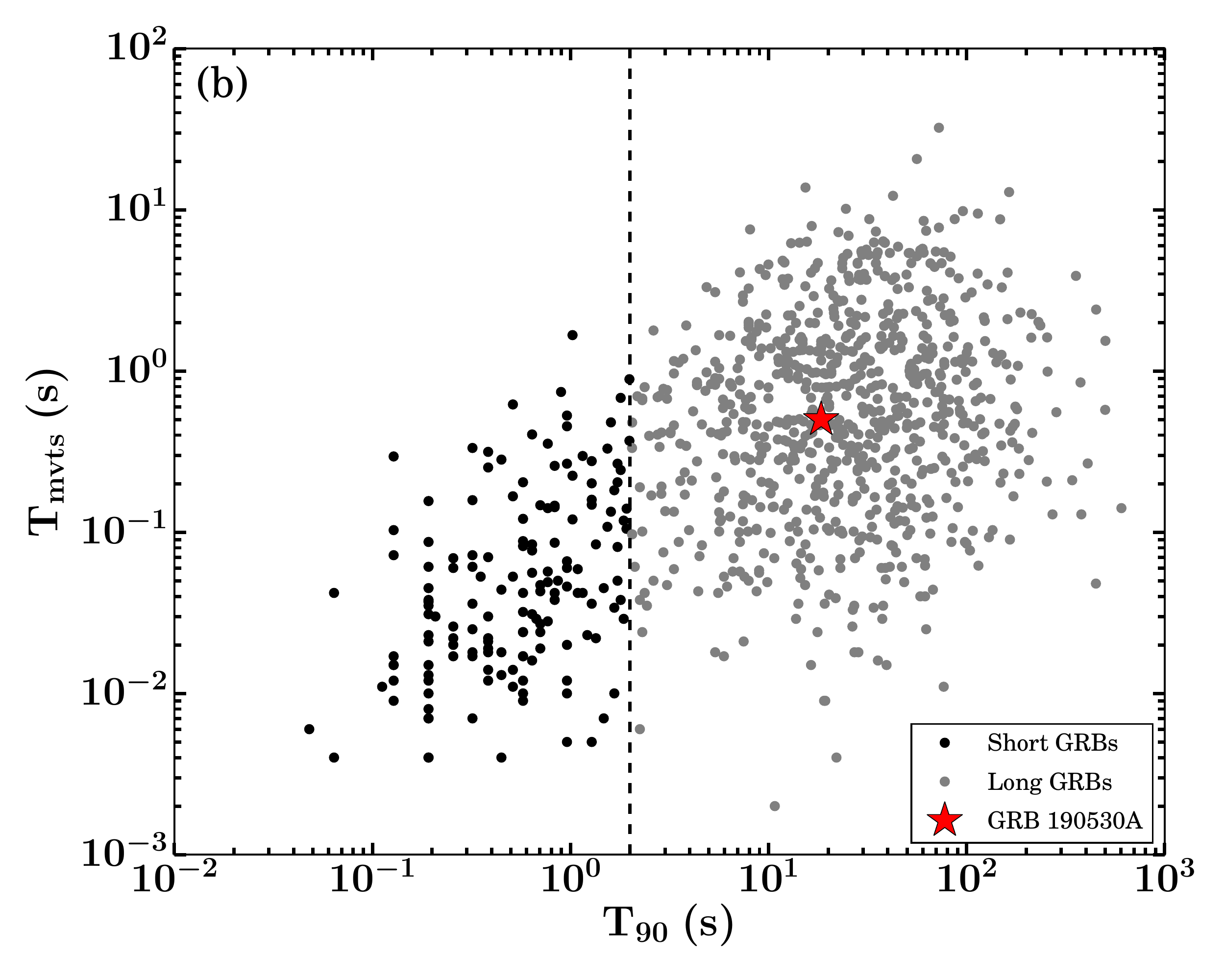}
\caption{{\bf Prompt emission characteristics of \thisgrb (shown with a red star):} (a) The spectral hardness as a function of \tninty duration for \thisgrb along with the data points for short (black circles) and long bursts (grey circles) used in Goldstein et al. (2017). The right side colour scale shows the probability of a GRB belonging to the short bursts class. The vertical dashed lines show the boundary between short and long GRBs. (b) Minimum variability time scale (\mvts) as a function of \tninty duration for \thisgrb along with the short and long GRBs sample studied by Golkhou et al. (2015).}
\label{fig:prompt_properties}
\end{figure}

\begin{figure*}
\includegraphics[width=2\columnwidth]{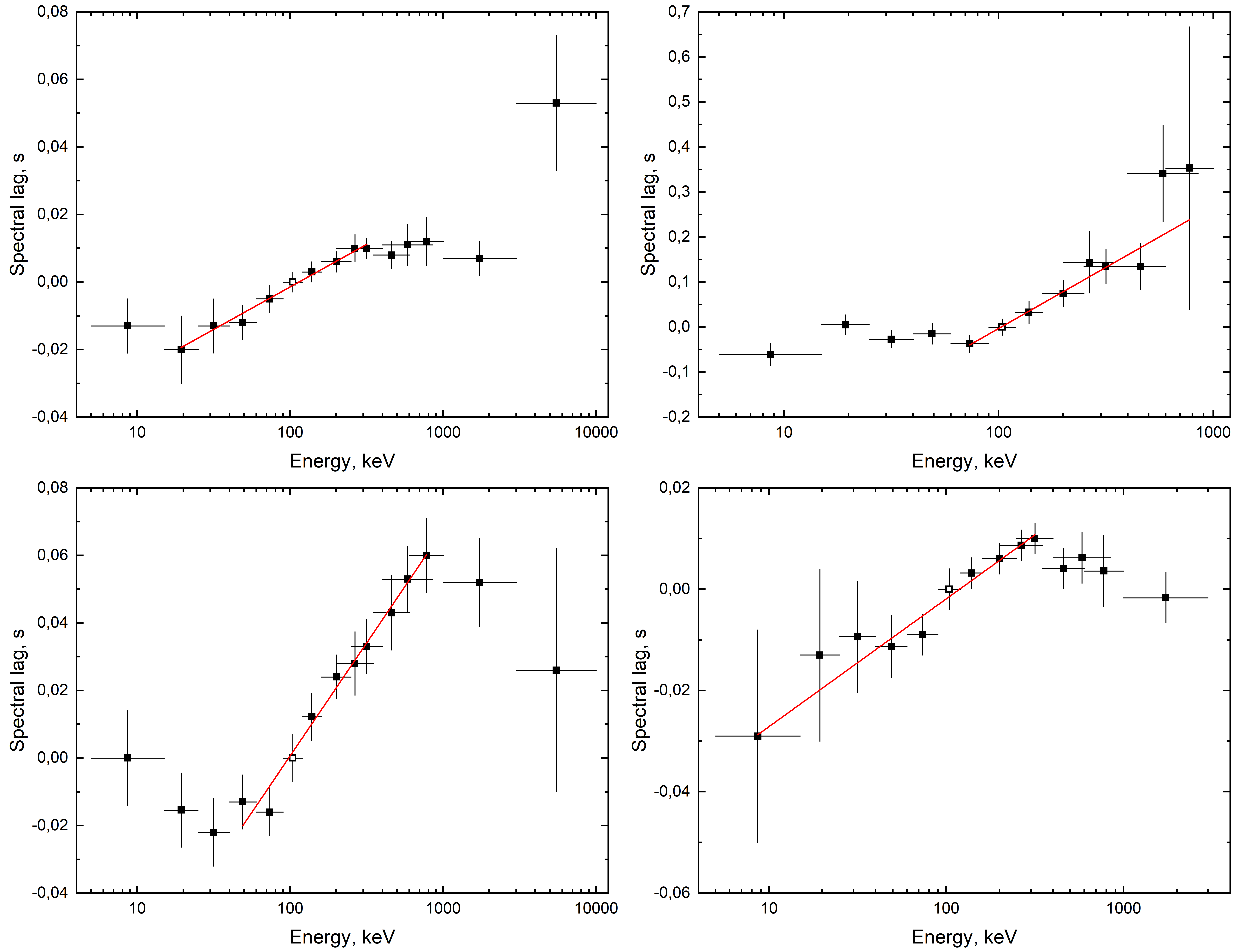}
\caption{Spectral evolution of the total emission episode of \thisgrb (top left), it first (top right), second (bottom left), and third (bottom right) episodes, based on \fermi GBM data. The horizontal axis -- the energy in units of \keV, the vertical axis -- the spectral lag in units of seconds relative to the (90, 120) \keV channel, shown by the unfilled symbol. Red lines represent logarithmic function fits.}
\label{fig:lags}
\end{figure*}

\begin{figure}
\centering
\includegraphics[scale=0.33]{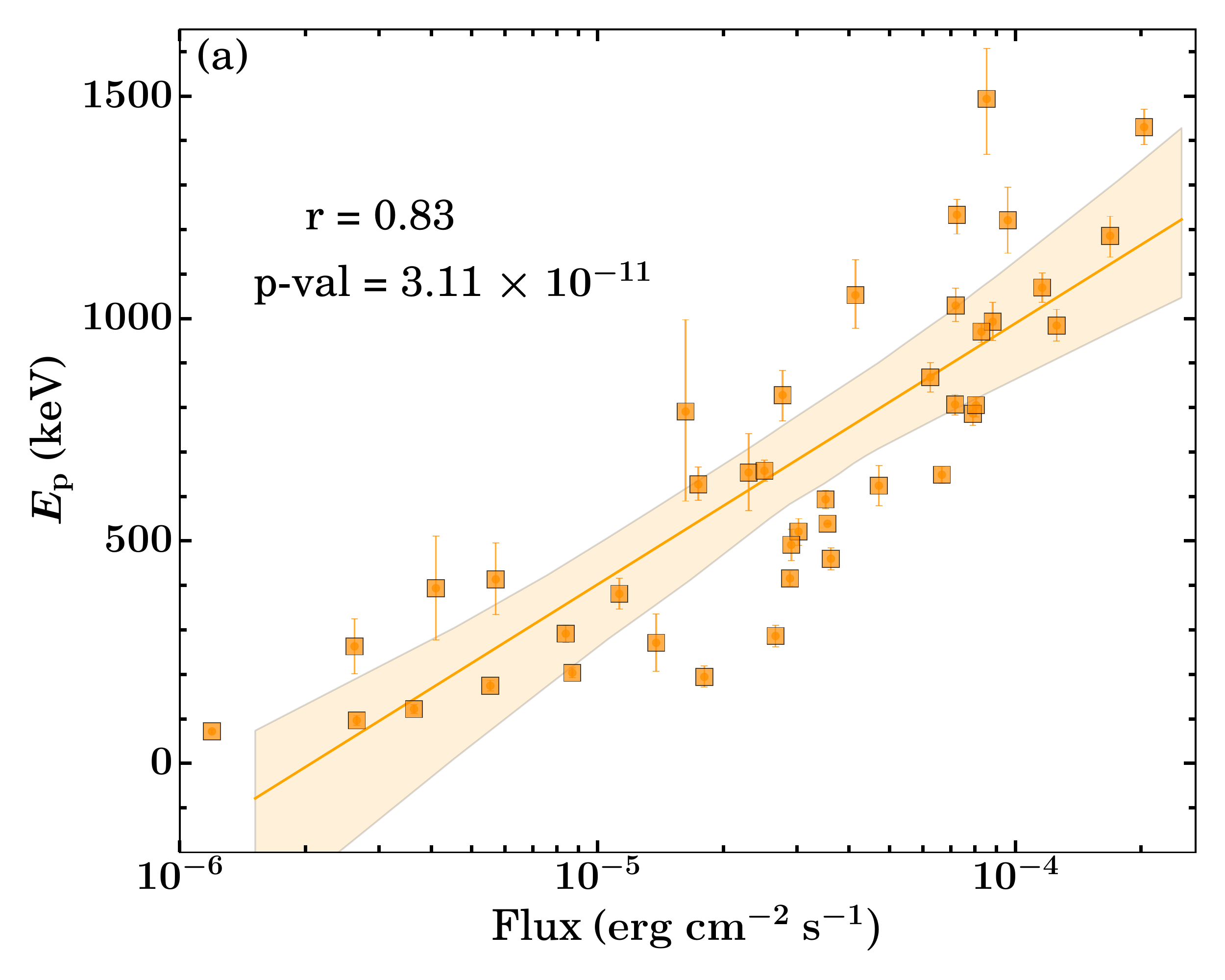}
\includegraphics[scale=0.33]{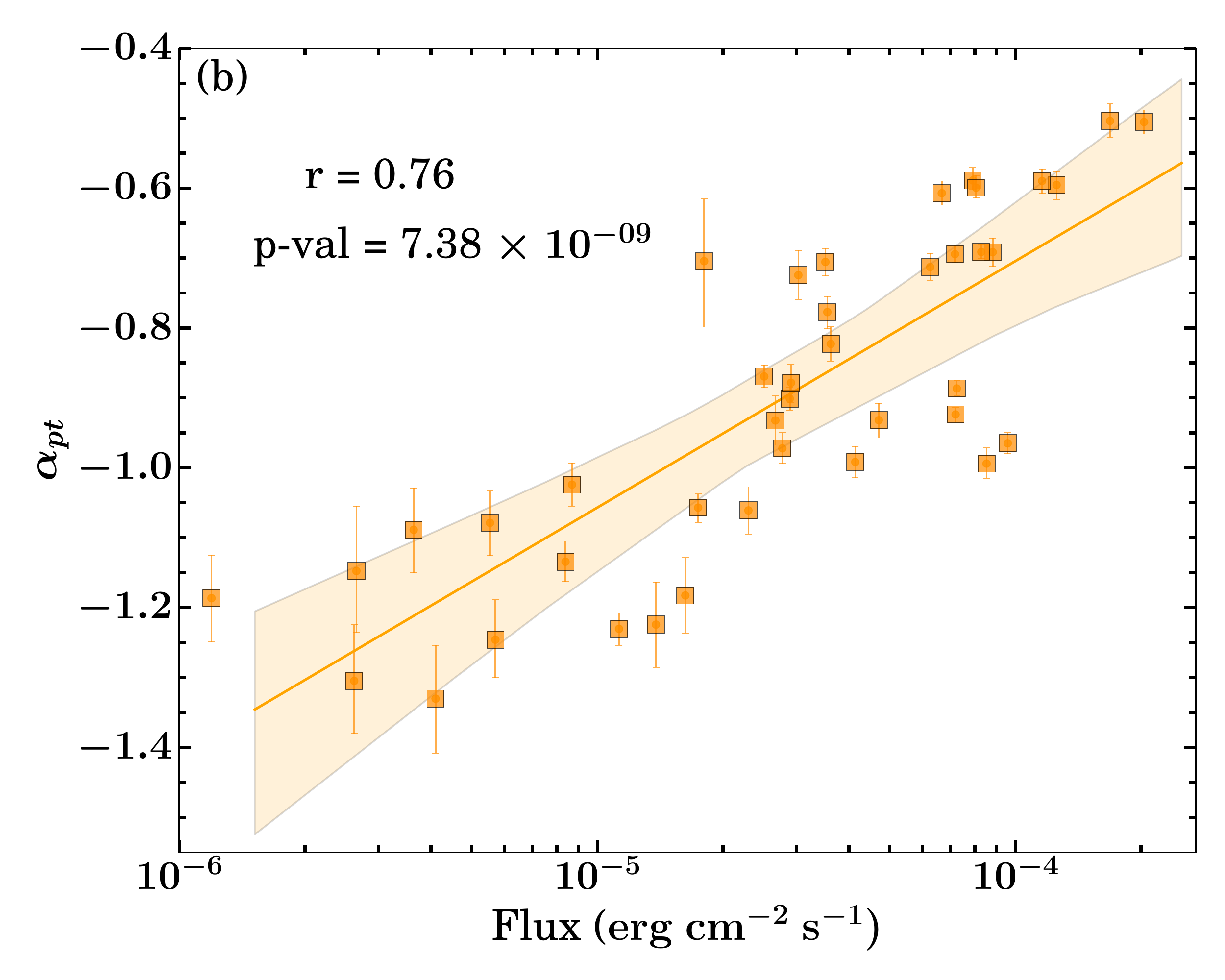}
\includegraphics[scale=0.33]{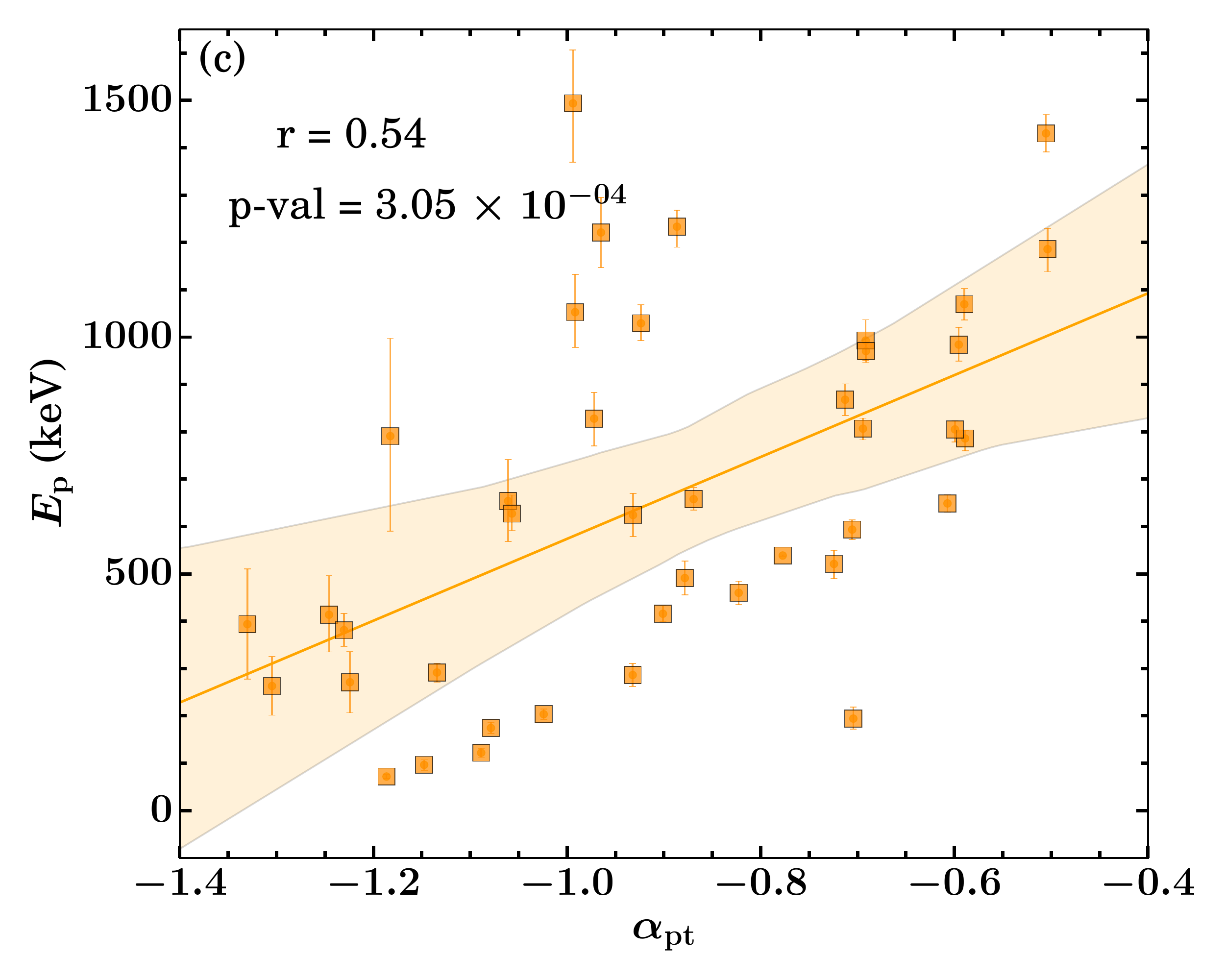}
\caption{{\bf Correlation between spectral parameters obtained from time-revolved spectral analysis using the \fermi GBM data.} (a) Peak energy (\Ep) of \sw{Band} function versus and flux, (b) low-energy spectral index ($\alpha_{\rm pt}$) of \sw{Band} function versus flux, (c) Peak energy (\Ep) as a function of low-energy spectral index ($\alpha_{\rm pt}$). Correlation shown in (a), (b), and (c) are obtained using \fermi GBM observations and modelling with \sw{Band} function. The best fit lines are shown with orange solid lines and shaded grey region show the 2 $\sigma$ confidence interval of the correlations.}
\label{spc}
\end{figure}

\begin{figure*}
\centering
\includegraphics[scale=0.35]{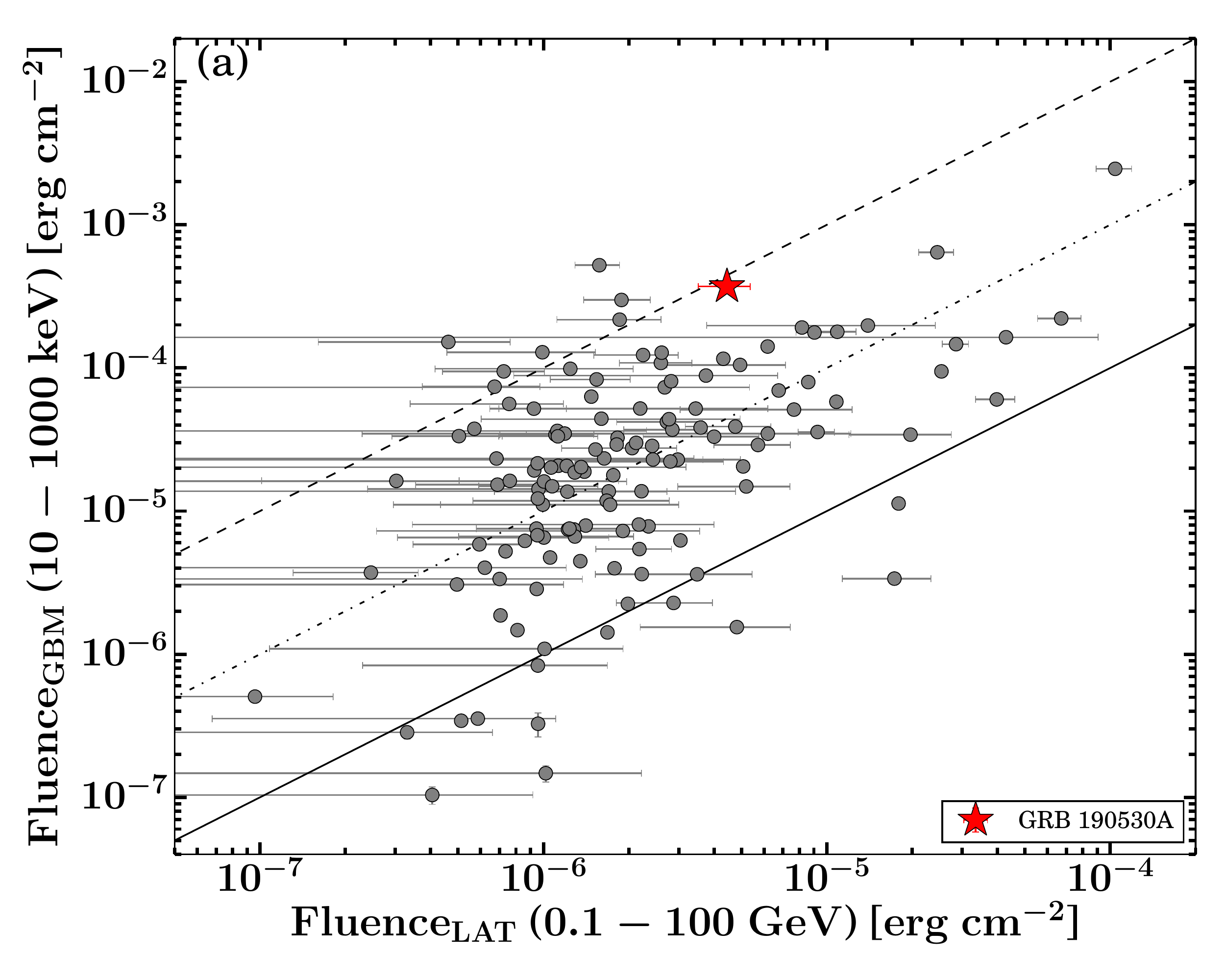}
\includegraphics[scale=0.35]{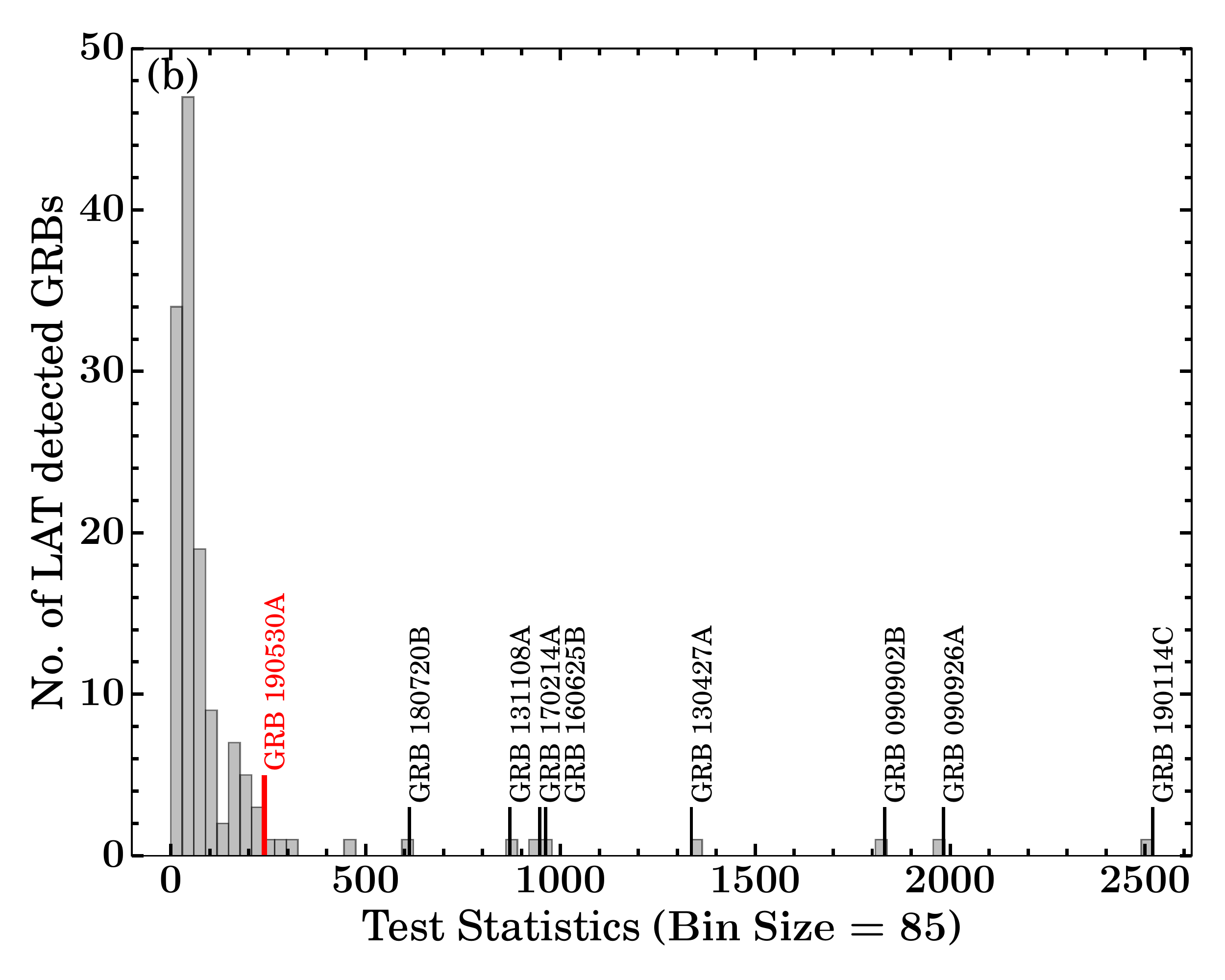}
\includegraphics[scale=0.35]{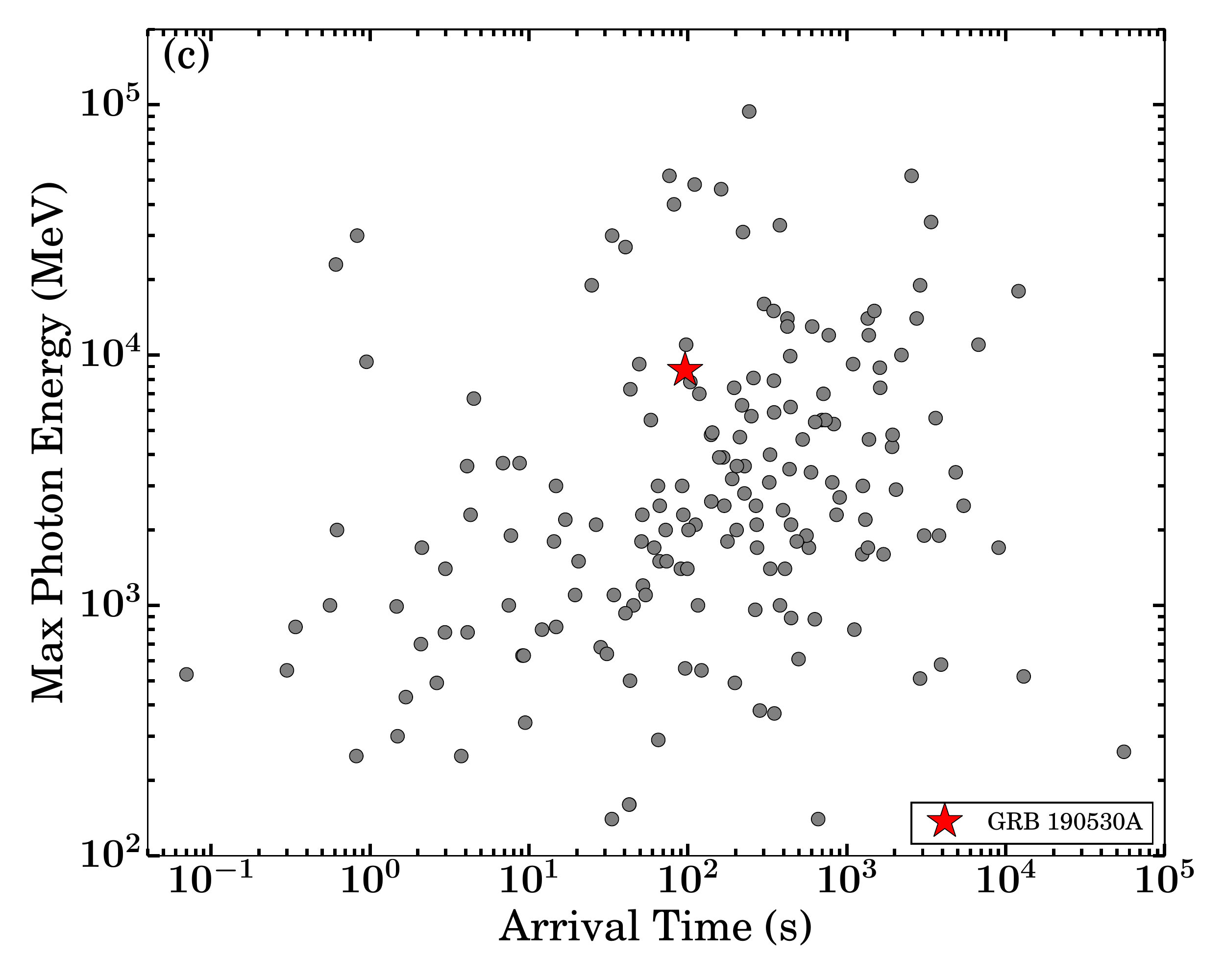}
\includegraphics[scale=0.35]{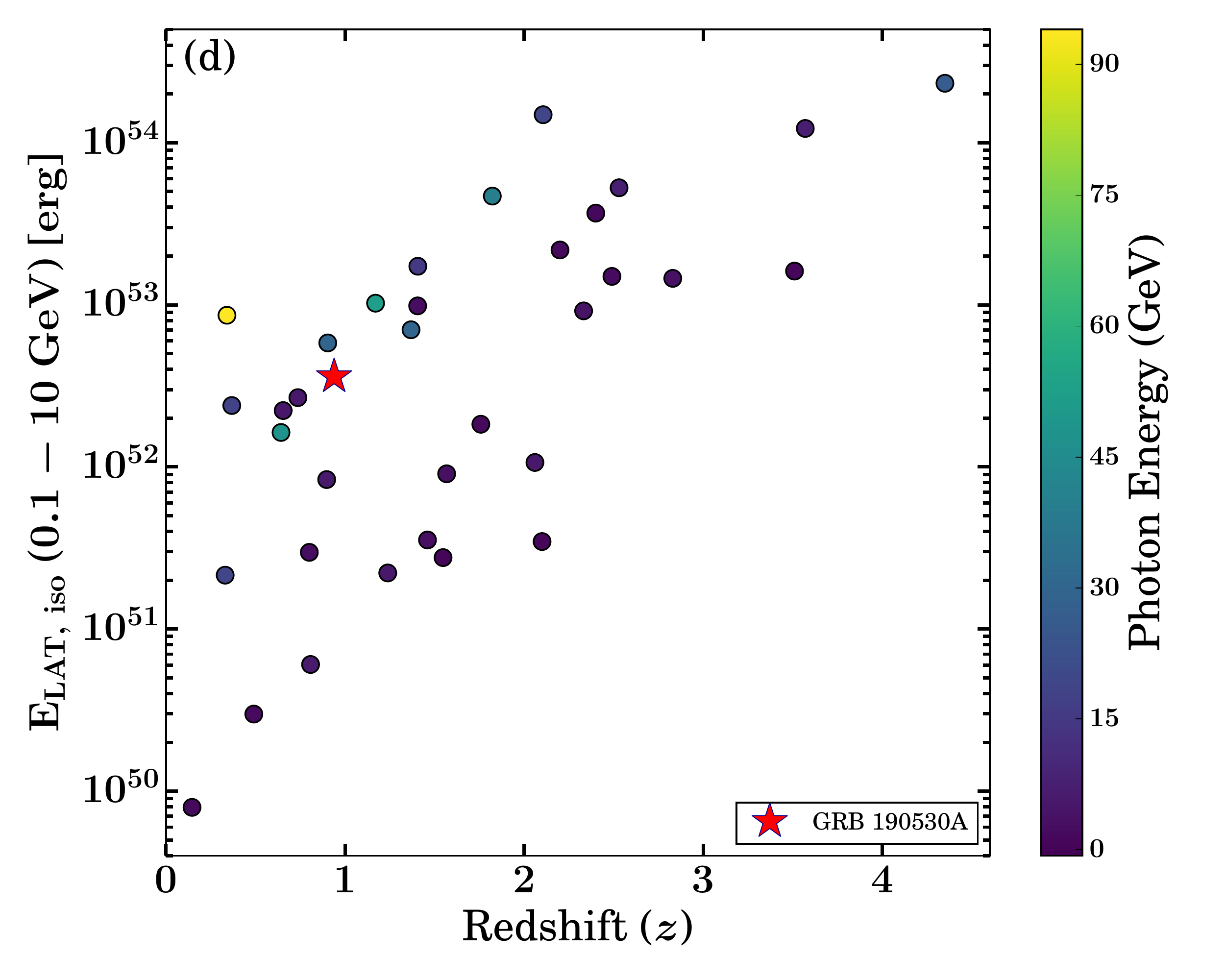}
\caption{{\bf Comparison of \thisgrb with second \fermi-LAT GRB catalogue(2FLGC):} (a) The distribution of energy fluence in the 10-1000 \keV and 0.1 - 100 GeV energy range for a temporal window of \tninty duration since \fermiT for \thisgrb (shown with a red star) along with LAT detected GRBs taken from the 2FLGC. The grey solid, dashed-dotted and dashed lines show the equal GBM-LAT fluence, the observed fluence changed by factors of 10 and the fluence changed by factors of 100, respectively. (b) Test Statistics (TS) histogram for \thisgrb along with other 138 GRBs detected by \fermi-LAT instrument (\protect\url{https://fermi.gsfc.nasa.gov/ssc/observations/types/grbs/lat_grbs/table.php}). There are eight bursts with significant TS $>$ 500. Here, for computing the TS value of \thisgrb, we consider the photon energy from 100 MeV to 100 GeV, and a duration of 25 s since \fermiT, a TS value of 240 (shown in the Figure) is obtained. For 0-18.4 s (from \fermiT to \tninty), the TS value is 189. (c) Maximum photon energy as a function of arrival time for the highest energy photon observed using \fermi LAT for \thisgrb (shown with a red star) and other data points taken from 2FLGC. (d) The distribution of $E_{\rm LAT, iso}$ (100 MeV - 10 GeV) as a function of redshift for \thisgrb along with various GRBs taken from the 2FLGC. Colours show the photon energy of the highest energy for each burst with an association probability greater than 90\%.}
\label{fig:LAT Comparison}
\end{figure*}

\begin{table*}
\begin{small}
\caption{The high energy emission ($>100$ MeV) observed by the \fermi LAT instrument in different temporal bins fit a power-law model for \thisgrb.}
\label{tab:lat_sed}
\begin{center}
\begin{tabular}{|c|c|c|c|c|c|c|c}
\hline
Sr. no.& Time   & LAT spectral index           & Energy flux                        & Photon flux              & Test Statistic  \\
&(s)    &                &( $\rm 10^{-8} ~ergs$  $c\rm m^{-2}$ $\rm s^{-1}$)   & ($\rm \times 10^{-6} ~photons$   $\rm cm^{-2}$ $\rm s^{-1}$) &  (TS)\\ \hline
(0)& 0 - 8   & $-5.33\pm2.19$ & $ 8.01 \pm 3.86$& $  385 \pm 174$& 23 \\
(1)& 8 - 11 & $-2.86\pm0.69$ & $37.8 \pm 18.9$ &    $1110 \pm 423$  &48\\ 
(2)& 11 - 13 & $-3.31\pm 0.58$ &$81.9 \pm 25.8 $&$2900  \pm 845$ &97 \\
(3) & 13 - 15  &$-5.02 \pm1.52$& $32.0 \pm 14.7$& $1500 \pm 640$ &44\\
(4)& 15 - 22 &$-3.1\pm 1.49$& $14.6 \pm 12.0$& $482 \pm  207$& 22\\
(5) & 22 - 30 &$-1.9 \pm0.26$&$ 52.8 \pm 21.8 $& $701 \pm 199$& 70 \\
(6)& 30 - 60 &$-2.07\pm0.18$ & $ 29.8 \pm$ 8.13& $ 437 \pm 84.0$& 123  \\
(7) & 60 - 100 &$-2.07\pm0.13$& $ 51.7 \pm 9.88$& $  764 \pm 104$& 267  \\
(8)& 100 - 268 &$-2.33\pm0.16$& $ 41.5 \pm 7.37$& $  821 \pm 133$& 241  \\
(9) & 3912 - 5981 &$-2$ (fixed)& $ <0.20$& $ < ~2.66$& 14  \\
(10) & 9624 - 10000 &$-2$ (fixed)& $ <0.57$& $ < ~7.68$& 7  \\
\hline 
\end{tabular}
\end{center}
\end{small}
\end{table*}

\begin{table*}
\scriptsize
\caption{The Joint GBM - LAT best fit (shown with boldface) spectral model parameters for the time-integrated spectrum (0 - 25 s) of \thisgrb.}
\begin{center}
\label{tab:TAS}
\begin{tabular}{|c|c|c|c|c|c|c|c|}
\hline
\textbf{Model} & \multicolumn{4}{c|}{\textbf{Parameters}} & \textbf{-Log(Likelihood)} & \textbf{AIC} & \textbf{BIC} \\ \hline 
Band & $\it \alpha_{\rm pt}$=  -0.99$^{+0.01}_{-0.01}$ & \multicolumn{2}{c|}{$\it \beta_{\rm pt}$= -3.15$^{+0.04}_{-0.03}$} & \Ep = 822.00$^{+7.67}_{-10.36}$ & 4368.60 & 8757.68 & 8798.82\\ \hline 
\sw{SBPL} & $\alpha_{1}$=  -1.05$^{+0.01}_{-0.01}$ & \multicolumn{2}{c|}{$\alpha_{2}$= -3.23$^{+0.03}_{-0.03}$} & $E_{0}$= 854.47$^{+16.73}_{-18.20}$ & 4462.02 & 8944.52 & 8985.66\\ \hline 
CPL & $\it \alpha_{\rm pt}$=  -1.00$^{+0.01}_{-0.01}$ & \multicolumn{3}{c|}{$E_{0}$ =849.58$^{+11.40}_{-10.01}$} & 4569.47 & 9157.33 & 9194.39\\ \hline 
\sw{bknpow} & $\it \alpha_{1,2}$= 1.12$^{+0.01}_{-0.01}$, 2.70$^{+0.02}_{-0.02}$ & \multicolumn{3}{c|}{$E_{\rm b1}$= 596.18$^{+5.24}_{-5.34}$} & 5115.49 & 10251.45 & 10292.59\\ \hline 
\textbf{\sw{bkn2pow}} & \bf $\it \bf \alpha_{1,2,3}$= 1.03$\bf^{+0.01}_{-0.01}$, 1.42$\bf^{+0.01}_{-0.01}$, 3.04$\bf^{+0.02}_{-0.02}$ & \multicolumn{3}{c|}{$\bf E_{\rm \bf b1,b2}$= \bf 136.65$^{\bf+2.90}_{-2.88}$, 888.36$\bf^{+12.71}_{-11.94}$} &\bf 4331.84 &\bf 8688.35 &\bf 8737.61\\ \hline 
{Band+BB} & $\alpha_{\rm pt}$= -1.01$^{+0.01}_{-0.01}$ & $\beta_{\rm pt}$ = -3.19$^{+0.04}_{-0.04}$ & \Ep= 871.78$^{+11.10}_{-11.94}$ & $\rm k{\it T}_{\rm BB}$= 35.38$^{+2.28}_{-2.10}$ & 4334.66 & 8694.00 & 8743.26\\ \hline 
\sw{SBPL}+BB & $\alpha_{1}$= -1.07$^{+0.01}_{-0.01}$ & $\alpha_{2}$= -3.38$^{+0.04}_{-0.04}$ & $E_{0}$= 1017.30$^{+25.61}_{-26.37}$ & $\rm k{\it T}_{\rm BB}$= 32.66$^{+1.42}_{-1.50}$ & 4378.32 & 8781.31 & 8830.57\\ \hline 
\sw{bknpow}+BB & $\it \alpha_{1,2}$= 1.17$^{+0.01}_{-0.01}$, 2.90$^{+0.03}_{-0.05}$ & $E_{\it b1}$= 757.92$^{+8.84}_{-11.17}$ & \multicolumn{2}{c|}{$\rm k{\it T}_{\rm BB}$ = 42.38$^{+0.82}_{-0.76}$} & 4445.97 & 8916.61 & 8965.87\\ 
\hline
\end{tabular}
\end{center}
\end{table*}

\begin{table*}
\caption{Results of time-resolved spectral fitting of \thisgrb for \sw{Band} and \sw{Band}+ \sw{BB} functions using \fermi GBM data. Temporal binning are performed based on constant binning method of 1 s. The best fit model is shown in bold for each bin. Flux values (in erg $\rm cm^{-2}$ $\rm s^{-1}$) are calculated in 8 \keV-30 MeV energy range.}
\label{TRS_Table_coarser}
\begin{scriptsize}
\begin{center}
\begin{tabular}{|c|c|c|c|c|c|c|c|c|c|c|}
\hline
\bf Sr. no. & $\rm \bf t_1$,$\rm \bf t_2$ \bf (s) & \boldmath $\it \alpha_{\rm pt}$ & \boldmath $\it \beta_{\rm pt}$ & \boldmath \Ep (\keV) &  \bf (Flux $\times 10^{-06}$)  &\bf \sw{-Log(likelihood)/BIC} &\textbf{GoF}& \boldmath ${\rm k}{\it T} ~(keV)$ & \bf \sw{-Log(likelihood)/BIC}&\textbf{GoF} \\
\hline
1& 0,1&$-1.20_{-0.03}^{+0.03}$&$-2.42_{-0.18}^{+0.19}$&$308.32_{-25.30}^{+24.23}$& 7.43 &$1041.25/2107.14$& 0.76 & 30.76$_{-1.34}^{+1.45}$& $1041.25/2119.46$& 0.68\\
2& 1,2&$-1.03_{-0.03}^{+0.03}$&$-2.86_{-0.24}^{+0.24}$&$211.55_{-9.47}^{+9.43}$&  8.14&$\bf 1114.92/2254.47$ & 0.02 & $36.52_{-5.60}^{+4.12}$& $1112.82/2262.60$& 0.02 \\
3& 2,3&$-1.12_{-0.04}^{+0.04}$&$-2.77_{-0.28}^{+0.29}$&$147.51_{-9.69}^{+9.22}$&4.29  &$\bf 1000.59/2025.82$ & 0.64 &$6.60_{-2.35}^{+0.76}$ & $999.21/2035.37$& 0.61\\
4& 3,4&$-1.11_{-0.10}^{+0.10}$&$-2.44_{-0.18}^{+0.18}$&$90.78_{-10.22}^{+10.23}$& 2.27 &$\bf 910.28/1845.20$ & 0.97 & $4.91_{-1.64}^{+0.48}$& $909.70/1856.35$& 0.97\\
5& 4,5&$-1.22_{-0.09}^{+0.08}$&$-4.38_{-0.16}^{+2.09}$&$75.42_{-4.83}^{+5.32}$&  1.04&$\bf 895.46/1815.56$ &0.50 & $25.94_{-1.40}^{+1.42}$ & $895.43/1827.82$& 0.49 \\
6& 5,6&$-1.28_{-0.08}^{+0.08}$&unconstrained&$79.70_{-6.01}^{+6.06}$& 0.93 &$\bf 890.51/1805.65$& 0.41 &$4.23_{-0.18}^{+0.21}$ & $890.51/1817.97$& 0.38\\
7& 6,7&$-1.05_{-0.09}^{+0.09}$&unconstrained&$67.87_{-3.61}^{+3.63}$&0.91  &$\bf 888.94/1802.51$ &0.54 & $10.15_{-5.24}^{+19.83}$& $888.94/1814.83$& 0.49\\
8& 7,8&$-1.33_{-0.07}^{+0.08}$&unconstrained&$73.01_{-1.41}^{+1.27}$& 0.94 &$\bf 913.64/1851.92$ & 0.02 & $9.29_{-3.09}^{+0.96}$& $912.21/1861.37$& 0.04 \\
9& 8,9&$-1.02_{-0.01}^{+0.01}$&$-2.74_{-0.13}^{+0.13}$&$1127.57_{-51.46}^{+51.26}$&52.85 &$\bf 1323.85/2672.33$& 0.03 &$0.61_{-1.28}^{+0.31}$ & $1323.85/2684.65$& 0.03 \\
10& 9,10&$-0.94_{-0.01}^{+0.01}$&$-2.88_{-0.12}^{+0.12}$&$923.78_{-33.56}^{+32.53}$& 62.25 &$1368.76/2762.15$ & 0.06 &$0.82_{-0.41}^{+1.40}$& $1368.76/2774.47$& 0.04\\
11& 10,11&$-0.88_{-0.02}^{+0.02}$&$-2.46_{-0.09}^{+0.09}$&$419.46_{-17.23}^{+17.69}$&28.97  &$\bf 1249.27/2523.18$& 0.35 & $19.69_{-4.22}^{+3.18}$& $1240.81/2518.58$& 0.47 \\
12& 11,12&$-1.23_{-0.02}^{+0.02}$&$-2.43_{-0.21}^{+0.20}$&$349.15_{-30.28}^{+29.87}$& 11.14 &$\bf 1114.14/2252.92$& 0.20 & $5.86_{-1.19}^{+0.71}$& $1112.30/2261.55$& 0.20 \\
13& 12,13&$-1.01_{-0.01}^{+0.01}$&unconstrained&$637.91_{-25.47}^{+27.10}$& 17.90 &$\bf 1163.90/2352.43$& 0.65 & $57.64_{-1.34}^{+1.37}$& $1163.90/2364.75$& 0.57\\
14& 13,14&$-0.84_{-0.01}^{+0.01}$&unconstrained&$677.40_{-19.82}^{+20.03}$&  28.52&$\bf 1229.01/2482.66$& 0.53 &$34.06_{-23.64}^{+19.49}$ & $1228.96/2494.88$& 0.48\\
15& 14,15&$-0.92_{-0.01}^{+0.01}$&unconstrained&$709.16_{-21.16}^{+20.59}$& 33.00 &$1349.78/2724.20$& 0.01 &$1.25_{-0.91}^{+0.48}$ & $1349.79/2736.54$& 0.01\\
16& 15,16&$-0.75_{-0.01}^{+0.01}$&$-4.20_{-0.35}^{+0.38}$&$1066.21_{-19.44}^{+19.53}$&85.46  &$\bf 1393.07/2810.78$& 0.07 & $17.67_{-3.99}^{+2.60}$& $1389.33/2815.61$& 0.11\\
17& 16,17&$-0.70_{-0.01}^{+0.01}$&unconstrained&$1029.19_{-17.56}^{+16.12}$& 85.83 &$\bf 1427.67/2879.98$ & 0.01 &$4.30_{-2.75}^{+0.51}$ & $1423.82/2884.59$& 0.01 \\
18& 17,18&$-0.69_{-0.01}^{+0.01}$&unconstrained&$894.11_{-15.05}^{+16.03}$&  73.49&$\bf 1393.67/2811.97$& 0.01 & $6.20_{-4.42}^{+1.05}$& $1392.84/2822.64$& 0.01\\
19& 18,19&$-0.71_{-0.01}^{+0.01}$&unconstrained&$714.18_{-12.50}^{+12.58}$&  60.07&$\bf 1399.09/2822.81$& 0.01 & unconstrained& $1399.09/2835.13$& 0.01\\
20& 19,20&$-1.10_{-0.02}^{+0.02}$&$-3.45_{-0.69}^{+0.69}$&$549.74_{-27.86}^{+28.03}$&15.46  &$\bf 1158.58/2341.79$& 0.29 & $16.11_{-5.23}^{+2.29}$& $1151.61/2340.18$& 0.45 \\

\hline
\end{tabular}
\end{center}
\end{scriptsize}
\end{table*}

\begin{table*}
\caption{Results of time-resolved spectral fitting of \thisgrb for \sw{bkn2power} function using \fermi GBM data. Temporal binning is performed based on the constant binning method of 1 s. The best fit model is shown in bold for each bin.}
\label{TRS_Table_coarser_bkn2pow}
\begin{scriptsize}
\begin{center}
\begin{tabular}{|c|c|c|c|c|c|c|c|c|}
\hline
\bf Sr. no. & $\rm \bf t_1$,$\rm \bf t_2$ \bf (s) & \boldmath $\it \alpha_{\rm 1}$ & \boldmath $\it \alpha_{\rm 2}$ & \boldmath $\it \alpha_{\rm 3}$ & \boldmath $E_{\rm break, 1}$ (\keV) & \boldmath \Ep or $E_{\rm break, 2}$(\keV)  &\bf \sw{-Log(likelihood)/BIC}&\textbf{GoF} \\
\hline
1& 0, 1&unconstrained&$1.39_{-0.02}^{+0.02}$&$2.34_{-0.12}^{+0.11}$& $14.99_{-0.79}^{+1.23}$ &$223.22_{-20.67}^{+19.58}$ &$\bf 1028.51/2093.97$& 0.89\\
2& 1, 2&$0.58_{-0.21}^{+0.28}$&$1.27_{-0.02}^{+0.02}$&$2.38_{-0.06}^{+0.06}$& $15.75_{-1.62}^{+2.05}$ &$143.91_{-6.83}^{+6.73}$ &$ 1112.09/2261.14$& 0.04\\
3& 2,3&$1.17_{-0.06}^{+0.06}$&$1.61_{-0.05}^{+0.05}$&$2.68_{-0.16}^{+0.16}$& $38.47_{-5.16}^{+5.39}$ &$164.67_{-15.90}^{+16.42}$ &$ 1000.42/2037.80$& 0.61 \\
4& 3,4&$1.10_{-0.17}^{+0.19}$&$1.56_{-0.07}^{+0.07}$&$2.38_{-0.11}^{+0.11}$& $21.47_{-4.33}^{+4.86}$ &$85.06_{-9.91}^{+10.27}$ &$ 909.36/1855.68$& 0.96\\
5& 4,5&$0.84_{-0.13}^{+1.11}$&$1.79_{-0.04}^{+0.09}$&$3.00_{-0.36}^{+0.37}$& $19.27_{-1.17}^{+8.86}$ &$110.85_{-15.33}^{+18.06}$ &$ 893.04/1823.03$& 0.53 \\
6& 5,6&unconstrained&$1.67_{-0.04}^{+0.04}$&$3.33_{-0.30}^{+0.30}$& $11.38_{-0.21}^{+0.19}$ &$113.85_{-0.98}^{+1.06}$ &$ 890.09/1817.14$& 0.33\\
7& 6,7&$0.17_{-0.18}^{+0.84}$&$1.62_{-0.05}^{+0.05}$&$3.61_{-0.38}^{+0.39}$& $13.53_{-0.92}^{+2.03}$ &$98.90_{-7.70}^{+8.33}$ &$ 891.95/1820.86$& 0.37\\
8& 7,8&unconstrained&$1.72_{-0.04}^{+0.04}$&$3.20_{-0.54}^{+0.52}$& $10.37_{-0.79}^{+0.81}$ &$115.27_{-16.75}^{+16.55}$ &$ 916.47/1869.89$& 0.02 \\
9& 8,9&$1.01_{-0.01}^{+0.01}$&$1.34_{-0.02}^{+0.02}$&$2.62_{-0.08}^{+0.07}$& $111.75_{-10.65}^{+10.17}$ &$1038.47_{-62.60}^{+62.84}$ &$1319.88/2676.71$& 0.04\\
10& 9,10&$0.92_{-0.01}^{+0.01}$&$1.34_{-0.02}^{+0.02}$&$2.75_{-0.07}^{+0.07}$& $107.54_{-6.94}^{+7.05}$ &$933.16_{-42.24}^{+43.84}$ &$\bf 1354.17/2745.29$& 0.31\\
11& 10,11&$0.90_{-0.02}^{+0.03}$&$1.33_{-0.03}^{+0.02}$&$2.39_{-0.05}^{+0.05}$& $66.84_{-5.32}^{+5.17}$ &$357.54_{-19.22}^{+18.32}$ &$ 1247.16/2531.28$& 0.44\\
12& 11,12&$0.31_{-0.20}^{+0.23}$&$1.38_{-0.02}^{+0.02}$&$2.16_{-0.06}^{+0.06}$& $14.81_{-0.45}^{+0.50}$ &$200.63_{-14.20}^{+14.73}$ &$ 1113.07/2263.10$& 0.24\\
13& 12,13&$0.89_{-0.13}^{+0.13}$&$1.18_{-0.01}^{+0.01}$&$2.83_{-0.11}^{+0.11}$& $20.61_{-3.89}^{+4.10}$ &$478.51_{-20.66}^{+20.83}$ &$ 1185.90/2408.75$& 0.20\\
14& 13,14&$0.93_{-0.01}^{+0.01}$&$1.41_{-0.04}^{+0.04}$&$3.34_{-0.17}^{+0.17}$& $162.73_{-13.59}^{+13.44}$ &$752.28_{-37.88}^{+37.04}$ &$ 1235.15/2507.26$& 0.46\\
15& 14,15&$0.87_{-0.02}^{+0.02}$&$1.46_{-0.02}^{+0.02}$&$3.77_{-0.33}^{+0.32}$& $98.70_{-4.95}^{+4.95}$ &$1138.55_{-104.06}^{+98.93}$ &$\bf 1302.00/2640.95$& 0.01 \\
16& 15,16&$0.78_{-0.01}^{+0.01}$&$1.13_{-0.02}^{+0.02}$&$3.02_{-0.06}^{+0.06}$& $123.96_{-9.19}^{+9.10}$ &$920.13_{-24.13}^{+24.74}$ &$ 1415.49/2867.93$& 0.02\\
17& 16,17&$0.81_{-0.01}^{+0.01}$&$1.16_{-0.03}^{+0.03}$&$3.41_{-0.08}^{+0.08}$& $201.73_{-15.79}^{+15.57}$ &$970.73_{-25.05}^{+24.58}$ &$ 1443.64/2924.23$& 0.01 \\
18& 17,18&$0.76_{-0.01}^{+0.01}$&$1.17_{-0.02}^{+0.02}$&$3.42_{-0.09}^{+0.09}$& $144.39_{-9.66}^{+9.79}$ &$890.79_{-22.25}^{+22.15}$ &$ 1404.30/2845.56$& 0.01 \\
19& 18,19&$0.74_{-0.01}^{+0.01}$&$1.39_{-0.02}^{+0.02}$&$3.61_{-0.12}^{+0.12}$& $137.24_{-5.39}^{+5.23}$ &$941.15_{-28.94}^{+27.47}$ &$ 1400.04/2837.04$& 0.01\\
20& 19,20&$1.10_{-0.02}^{+0.02}$&$1.50_{-0.03}^{+0.03}$&$3.40_{-0.36}^{+0.37}$& $81.36_{-8.00}^{+8.15}$ &$735.67_{-59.17}^{+61.97}$ &$1150.73/2338.42$ & 0.43\\
\hline
\end{tabular}
\end{center}
\end{scriptsize}
\end{table*}

\begin{table*}
\caption{Results of time-resolved spectral fitting of \thisgrb for \sw{Band} and \sw{Band}+ \sw{BB} functions using \fermi GBM data. Temporal binning are performed based on Bayesian Block algorithm. The best fit model is shown in bold for each bin.  Flux values (in erg $\rm cm^{-2}$ $\rm s^{-1}$) are calculated in 8 \keV-30 MeV energy range.}
\label{TRS_Table_band}
\begin{scriptsize}
\begin{center}
\begin{tabular}{|c|c|c|c|c|c|c|c|c|c|c|}
\hline
\bf Sr. no. & $\rm \bf t_1$,$\rm \bf t_2$ \bf (s) & \boldmath $\it \alpha_{\rm pt}$ & \boldmath $\it \beta_{\rm pt}$ & \boldmath \Ep (\keV) &  \bf (Flux $\times 10^{-06}$)  &\bf \sw{-Log(likelihood)/BIC} & \textbf{GoF} &\boldmath ${\rm k}{\it T} ~(keV)$ & \bf \sw{-Log(likelihood)/BIC}& \textbf{GoF} \\
\hline
1& 0.000106, 0.282062&$-1.30_{-0.07}^{+0.08}$&$-2.83_{-0.99}^{+0.68}$&$263.00_{-61.64}^{+61.92}$& 2.62 &$\bf 81.66/187.95$& 0.35 & $8.33_{-1.69}^{+1.10}$ & 78.77/194.49 &  0.36 \\
2&0.282062, 0.509457  &$-1.25_{-0.05}^{+0.06}$&unconstrained&$413.58_{-78.62}^{+82.23}$& 5.70 &$\bf -12.61/-0.58$& 0.89 & $6.69_{-0.80}^{+0.72}$ & -17.11/2.74& 0.88\\
3&0.509457, 1.148330 &$-1.13_{-0.03}^{+0.03}$&$-2.85_{-0.41}^{+0.40}$&$291.50_{-19.82}^{+19.42}$& 8.39 &$804.74/1634.12$& 0.23  & $69.08_{-7.85}^{+6.26}$ & 799.99/1636.93& 0.27\\
4& 1.148330, 1.947446 &$-1.02_{-0.03}^{+0.03}$&$-2.66_{-0.19}^{+0.18}$&$203.43_{-10.91}^{+10.90}$& 8.70 &$\bf  993.30/2011.23$& 0.03  & $32.70_{-5.85}^{+3.70}$ & 992.12/2021.19& 0.02 \\
5& 1.947446, 2.336272 &$-1.08_{-0.05}^{+0.05}$&$-3.21_{-0.52}^{+0.51}$&$174.37_{-11.61}^{+12.26}$& 5.54 &$\bf 374.64/773.92$& 0.70 & $6.87_{-2.86}^{+0.89}$ & 374.08/785.12& 0.74 \\
6& 2.336272, 3.005074 &$-1.09_{-0.06}^{+0.06}$&$-2.61_{-0.22}^{+0.23}$&$122.03_{-9.57}^{+9.76}$& 3.64 &$\bf 724.79/1474.21$& 0.57 & $5.21_{-1.97}^{+0.47}$ & 723.80/1484.55& 0.55 \\
7& 3.005074, 3.657240 &$-1.15_{-0.09}^{+0.09}$&$-2.37_{-0.17}^{+0.16}$&$96.51_{-11.12}^{+10.73}$& 2.65 &$\bf 669.09/1362.82$& 0.70 & $29.66_{-18.38}^{+7.65}$ & 668.95/1374.86& 0.69 \\
8& 3.657240, 8.002803 &$-1.19_{-0.06}^{+0.06}$&$-2.67_{-0.19}^{+0.18}$&$71.82_{-4.42}^{+4.33}$& 1.19 &$\bf 1982.52/3989.68$& 0.85 & $4.47_{-0.98}^{+0.41}$ & 1980.80/3998.55& 0.84\\
9&8.002803, 8.122972 &$-1.33_{-0.08}^{+0.08}$&unconstrained&$393.70_{-115.97}^{+116.96}$& 4.10 &$\bf -494.69/-964.75$& 0.47 & $9.17_{-2.64}^{+1.13}$ & -496.56/-956.16& 0.46\\
10& 8.122972, 8.225416  &$-1.18_{-0.05}^{+0.05}$&$-2.50_{-0.52}^{+0.44}$&$790.88_{-201.06}^{+206.51}$& 16.24  &$\bf -482.81/-940.98$& 0.47 & $21.75_{-14.98}^{+6.32}$ & -482.99/-929.02& 0.45\\
11& 8.225416, 8.436160&$-1.06_{-0.03}^{+0.03}$&$-2.58_{-0.37}^{+0.36}$&$653.78_{-85.17}^{+87.65}$& 22.99 &$\bf 105.11/234.87$& 0.36  &$21.57_{-7.45}^{+3.53}$ & 103.22/243.39& 0.40 \\
12&8.436160, 8.609198  &$-0.99_{-0.02}^{+0.02}$&unconstrained&$1052.68_{-74.37}^{+79.83}$& 41.42 &$\bf 49.78/124.20$&  0.53 & $0.50_{-0.16}^{+1.31}$ & 49.78/136.52& 0.49 \\
13&8.609198, 8.704119  &$-0.99_{-0.02}^{+0.02}$&$-4.01_{-0.54}^{+1.27}$&$1493.73_{-124.56}^{+113.07}$& 85.37 &$\bf -273.71/-522.78$& 0.05  &$1.10_{-0.58}^{+1.46}$ & -273.70/-510.45& 0.03\\
14&8.704119, 8.947630  &$-0.96_{-0.01}^{+0.01}$&$-3.24_{-0.37}^{+0.36}$&$1220.97_{-73.90}^{+73.97}$& 95.91 &$\bf 436.00/896.63$& 0.16  &$41.08_{-11.41}^{+7.67}$ & 427.60/892.15& 0.31\\
15&8.947630, 9.661208  &$-0.92_{-0.01}^{+0.01}$&$-3.00_{-0.16}^{+0.14}$&$1029.21_{-36.11}^{+38.79}$& 71.96 & $1151.30/2327.23$& 0.62  & $1.01_{-0.39}^{+6.23}$ & 1151.30/2339.55& 0.59  \\
16&9.661208, 9.912693  &$-0.93_{-0.03}^{+0.02}$&$-2.54_{-0.16}^{+0.16}$&$624.17_{-45.10}^{+45.71}$& 47.13 &$\bf 362.97/750.59$& 0.02  & $23.45_{-6.42}^{+4.00}$ & 359.19/755.34& 0.02\\
17&9.912693, 11.082932  &$-0.90_{-0.02}^{+0.02}$&$-2.43_{-0.08}^{+0.08}$&$415.68_{-17.25}^{+16.54}$& 28.87 &$\bf 1393.56/2811.75$& 0.25  & $102.00_{-1.37}^{+1.38}$ & 1393.56/2824.07& 0.18 \\
18&11.082932, 12.220607 &$-1.23_{-0.02}^{+0.02}$&$-2.40_{-0.20}^{+0.19}$&$381.01_{-34.18}^{+34.91}$& 11.27 &$\bf 1227.44/2479.51$& 0.48  & $24.67_{-9.10}^{+4.71}$ & 1226.42/2489.79& 0.50\\
19&12.220607, 12.748669 &$-1.06_{-0.02}^{+0.02}$&$-4.46_{-0.08}^{+2.12}$&$627.33_{-35.72}^{+38.90}$& 17.42 &$\bf 758.81/1542.25$& 0.02  & $10.02_{-6.52}^{+1.85}$ & 758.51/1553.98& 0.02 \\
20&12.748669, 13.396796 &$-0.87_{-0.02}^{+0.02}$&unconstrained&$657.71_{-23.09}^{+24.19}$& 25.08 &$\bf 960.57/1945.77$& 0.15  & $53.05_{-1.46}^{+1.32}$ & 960.57/1958.09& 0.15  \\
21&13.396796, 13.542326 &$-0.72_{-0.03}^{+0.03}$&$-3.83_{-0.65}^{+1.28}$&$520.89_{-30.95}^{+29.20}$& 30.26 &$\bf -101.70/-178.76$& 0.09 & $84.16_{-10.83}^{+9.60}$ & -106.21/-175.46& 0.07 \\
22&13.542326, 13.836101  &$-0.97_{-0.02}^{+0.02}$&unconstrained&$827.80_{-57.81}^{+55.25}$& 27.72 &$\bf 381.11/786.85$& 0.46 & $21.90_{-6.74}^{+3.59}$ & 379.73/796.42& 0.52\\
23&13.836101, 14.219772  &$-0.71_{-0.02}^{+0.02}$&unconstrained&$593.45_{-20.75.64}^{+20.34}$& 35.13 &$\bf 623.29/1271.22$& 0.30 & $39.07_{-10.68}^{+7.37}$  & 619.54/1276.03& 0.38 \\
24&14.219772, 14.658330 &$-0.88_{-0.03}^{+0.03}$&$-2.75_{-0.29}^{+0.28}$&$491.14_{-35.24}^{+35.56}$& 29.07 &$713.56/1451.76$& 0.03 & $27.52_{-2.55}^{+2.53}$ & {\bf 688.39/1413.74}& 0.24\\
25&14.658330, 15.192231  &$-0.89_{-0.01}^{+0.01}$&$-4.60_{-0.07}^{+1.53}$&$1233.27_{-43.23}^{+34.78}$& 72.45 &$954.15/1932.93$& 0.01 & $141.81_{-1.42}^{+1.36}$ & 954.15/1945.25& 0.01 \\
26& 15.192231, 15.338963 &$-0.51_{-0.02}^{+0.02}$&unconstrained&$1430.37_{-39.35}^{+40.02}$& 203.24 &$186.79/398.22$& 0.01 & $8.02_{-1.04}^{+1.13}$ & {\bf 162.21/361.37}& 0.01 \\
27& 15.338963, 15.687352 &$-0.59_{-0.01}^{+0.02}$&$-4.47_{-0.21}^{+1.27}$&$785.94_{-25.83}^{+17.10}$& 79.11 &$\bf 660.06/1344.76$ & 0.40 & $1.19_{-0.65}^{+1.22}$ & 660.06/1357.07& 0.38 \\
28& 15.687352, 16.251414 &$-0.93_{-0.03}^{+0.03}$&$-2.15_{-0.07}^{+0.07}$&$286.27_{-24.07}^{+24.34}$& 26.66 &$847.89/1720.42$& 0.11 & $20.19_{-1.75}^{+1.75}$ & 825.21/1687.38& 0.27 \\
29&16.251414, 16.407026 &$-0.69_{-0.02}^{+0.02}$&$-3.81_{-0.50}^{+0.54}$&$992.80_{-42.60}^{+43.88}$& 88.26 &$\bf 67.38/159.39$& 0.67 & $9.87_{-1.17}^{+1.12}$ & 60.67/158.29& 0.79 \\
30& 16.407026, 16.524165 &$-0.59_{-0.02}^{+0.02}$&unconstrained&$984.19_{-34.64}^{+36.28}$& 125.52 &$\bf -70.94/-117.25$& 0.83 & $32.06_{-13.17}^{+5.73}$ & -72.26/-107.56& 0.83 \\
31& 16.524165, 16.705623 &$-0.59_{-0.02}^{+0.02}$&unconstrained&$1069.46_{-32.93}^{+32.73}$& 115.88 &$\bf 229.87/484.38$& 0.38 & $10.15_{-2.78}^{+2.08}$ & 228.01/492.97& 0.36\\
32&16.705623, 16.787528   &$-0.50_{-0.02}^{+0.02}$&unconstrained&$1185.83_{-47.53}^{+43.98}$& 168.56 &$\bf -250.30/-475.96$& 0.01 & $4.81_{-3.38}^{+0.53}$ & -250.74/-464.53& 0.01 \\
33&16.787528, 17.114033  &$-0.60_{-0.01}^{+0.02}$&$-4.47_{-0.25}^{+1.13}$&$805.07_{-26.23}^{+17.72}$& 80.52 &$\bf 632.33/1289.29$& 0.07  & $0.87_{-0.48}^{+1.25}$ & 632.33/1301.61& 0.09\\
34& 17.114033, 17.332842  &$-0.71_{-0.02}^{+0.02}$&unconstrained&$867.86_{-33.23}^{+33.24}$& 62.57 &$\bf 300.86/626.35$&  0.02 & $9.84_{-1.60}^{+1.41}$&  296.81/630.57& 0.02\\
35& 17.332842, 17.924038 &$-0.69_{-0.01}^{+0.01}$&$-4.86_{-0.03}^{+1.09}$&$970.42_{-23.81}^{+17.70}$& 82.93 &$\bf 1050.02/2124.69$& 0.10  & $3.90_{-2.84}^{+0.62}$ & 1050.18/2137.31& 0.08\\
36&17.924038, 18.441228  &$-0.69_{-0.01}^{+0.01}$&$-3.85_{-0.48}^{+0.51}$&$806.72_{-23.57}^{+22.02}$& 71.72 &$\bf 931.43/1887.50$& 0.15  & unconstrained & 931.43/1899.82& 0.12\\
37&18.441228, 18.542279  &$-0.78_{-0.02}^{+0.02}$&unconstrained&$538.64_{-1.40}^{+1.32}$& 35.51 &$-299.09/-573.54$& 0.04  & $25.50_{-3.02}^{+2.62}$ & {\bf -319.09/-601.22}& 0.37\\
38& 18.542279, 18.880490&$-0.61_{-0.02}^{+0.02}$&$-3.97_{-0.51}^{+0.55}$&$648.72_{-18.48}^{+18.02}$& 66.69 &$625.26/1275.16$& 0.53  & $60.01_{-5.89}^{+5.98}$ & {\bf 615.12/1267.20}& 0.74\\
39&18.880490, 19.209439  &$-0.82_{-0.02}^{+0.02}$&$-2.89_{-0.24}^{+0.26}$&$459.88_{-25.13}^{+24.29}$& 36.18 &$523.32/1071.28$& 0.29  & $25.45_{-1.92}^{+1.88}$ & {\bf 490.26/1017.47}& 0.95\\
40& 19.209439, 19.316852 &$-0.70_{-0.09}^{+0.09}$&$-2.26_{-0.14}^{+0.14}$&$194.21_{-22.45}^{+24.49}$& 18.01 &$\bf -419.57/-814.50$& 0.56   & $0.88_{-0.70}^{+0.28}$ & -419.58/-802.20& 0.47 \\
41& 19.316852, 19.790851&$-1.22_{-0.06}^{+0.06}$&$-1.99_{-0.11}^{+0.10}$&$271.00_{-64.36}^{+64.39}$& 13.81 &$586.79/1198.21$& 0.35  & $11.74_{-1.98}^{+1.52}$ & 578.98/1194.92& 0.38 \\
\hline
\end{tabular}
\end{center}
\end{scriptsize}
\end{table*}

\begin{table*}
\caption{Results of time-resolved spectral fitting of \thisgrb for \sw{bkn2power} function using \fermi GBM data. Temporal binning are performed based on Bayesian Block algorithm. The best fit model is shown in bold for each bin. Flux values (in erg $\rm cm^{-2}$ $\rm s^{-1}$) are calculated in 8 \keV-30 MeV energy range.}
\label{TRS_Table_bkn2pow}
\begin{scriptsize}
\begin{center}
\begin{tabular}{|c|c|c|c|c|c|c|c|c|c|}
\hline
\bf Sr. no. & $\rm \bf t_1$,$\rm \bf t_2$ \bf (s) & \boldmath $\it \alpha_{\rm 1}$ & \boldmath $\it \alpha_{\rm 2}$ & \boldmath $\it \alpha_{\rm 3}$ & \boldmath $E_{\rm break, 1}$ (\keV) & \boldmath \Ep or $E_{\rm break, 2}$(\keV) &  \bf (Flux $\times 10^{-06}$)  &\bf \sw{-Log(likelihood)/BIC}& \textbf{GoF} \\
\hline 
1& 0.000106, 0.282062&$0.57_{-0.20}^{+0.47}$&$1.58_{-0.06}^{+0.07}$&$2.57_{-0.44}^{+0.43}$& $20.78_{-2.58}^{+4.44}$ &$267.78_{-2.32}^{+2.29}$ & 2.26&$77.87/192.70$& 0.40\\
2&0.282062, 0.509457  &$0.47_{-0.19}^{+0.48}$&$1.44_{-0.04}^{+0.05}$&$2.71_{-0.36}^{+0.37}$& $19.43_{-2.18}^{+3.75}$ &$323.15_{-63.26}^{+64.76}$ & 4.93&$-17.23/2.50 $& 0.88\\
3&0.509457, 1.148330 &unconstrained&$1.33_{-0.02}^{+0.02}$&$2.52_{-0.12}^{+0.13}$& $13.87_{-0.13}^{+0.14}$ &$215.70_{-14.95}^{+15.66}$ &7.88 &$\bf 794.98 /1626.91 $& 0.42\\
4& 1.148330, 1.947446 &$0.46_{-0.18}^{+0.21}$&$1.27_{-0.02}^{+0.02}$&$2.30_{-0.05}^{+0.06}$& $15.30_{-0.31}^{+0.31}$ &$131.78_{-7.07}^{+6.49}$ & 9.53 &$ 989.30/2015.56 $& 0.09\\
5& 1.947446, 2.336272 &$0.35_{-0.21}^{+0.56}$&$1.35_{-0.03}^{+0.03}$&$2.67_{-0.14}^{+0.15}$& $13.58_{-1.39}^{+2.24}$ &$148.30_{-11.90}^{+11.91}$ & 4.85&$ 376.09/789.13 $& 0.73\\
6& 2.336272, 3.005074 &$1.22_{-0.08}^{+0.07}$&$1.63_{-0.09}^{+0.10}$&$2.51_{-0.14}^{+0.14}$& $40.00_{-8.46}^{+8.53}$ &$125.82_{-17.05}^{+17.46}$ & 3.70&$ 725.85/1488.66 $& 0.60\\
7& 3.005074, 3.657240 &$1.24_{-0.16}^{+0.17}$&$1.59_{-0.07}^{+0.07}$&$2.37_{-0.13}^{+0.12}$& $25.92_{-5.48}^{+6.05}$ &$93.44_{-12.14}^{+11.48}$ & 2.46&$668.18 /1373.31 $& 0.67\\
8& 3.657240, 8.002803 &$0.91_{-0.20}^{+0.26}$&$1.57_{-0.04}^{+0.04}$&$2.43_{-0.06}^{+0.06}$& $15.20_{-1.60}^{+2.01}$ &$64.82_{-4.58}^{+4.38}$ & 1.22&$1984.51 /4005.98 $& 0.75\\
9&8.002803, 8.122972 &$0.46_{-0.19}^{+0.65}$&$1.51_{-0.06}^{+0.06}$&$2.93_{-0.61}^{+0.59}$& $18.39_{-2.65}^{+4.69}$ &$330.09_{-63.33}^{+61.71}$ &3.10 &$ -496.51/-956.07 $& 0.28\\
10&8.122972, 8.225416  &$1.08_{-0.15}^{+0.21}$&$1.34_{-0.06}^{+0.06}$&$2.38_{-0.29}^{+0.29}$& $34.64_{-16.86}^{+21.45}$ &$538.39_{-105.60}^{+108.55}$ & 13.99&$ -482.93/-928.91 $& 0.40\\
11& 8.225416, 8.436160&$1.05_{-0.05}^{+0.05}$&$1.43_{-0.05}^{+0.05}$&$2.64_{-0.23}^{+0.23}$& $84.96_{-19.03}^{+19.23}$ &$687.39_{-46.84}^{+45.03}$ &22.30 &$ 102.61/242.19 $& 0.40 \\
12&8.436160, 8.609198  &unconstrained&$1.12_{-0.01}^{+0.02}$&$3.18_{-0.23}^{+0.23}$& $10.75_{-1.53}^{+4.53}$ &$797.24_{-54.77}^{+57.77}$ & 13.75&$ 58.22/153.39 $& 0.41\\
13&8.609198, 8.704119  &$0.83_{-0.13}^{+0.14}$&$1.17_{-0.03}^{+0.03}$&$2.85_{-0.20}^{+0.20}$& $41.93_{-12.84}^{+13.76}$ &$1132.80_{-123.51}^{+120.41}$ & 79.60&$ -274.05/-511.15 $& 0.03\\
14&8.704119, 8.947630  &$0.96_{-0.02}^{+0.02}$&$1.42_{-0.04}^{+0.04}$&$3.05_{-0.16}^{+0.17}$& $167.56_{-17.24}^{+18.23}$ &$1479.67_{-124.99}^{+131.35}$ &98.18 &$ 431.44/899.83 $& 0.36 \\
15&8.947630, 9.661208  &$0.90_{-0.02}^{+0.02}$&$1.30_{-0.02}^{+0.02}$&$2.76_{-0.07}^{+0.08}$& $109.80_{-8.23}^{+8.46}$ &$988.01_{-46.76}^{+49.80}$ & 74.45&$\bf 1142.16/2321.27 $& 0.81\\
16&9.661208, 9.912693  &$0.92_{-0.03}^{+0.03}$&$1.36_{-0.04}^{+0.04}$&$2.55_{-0.12}^{+0.11}$& $83.94_{-10.83}^{+10.50}$ &$620.87_{-58.94}^{+57.32}$ & 47.42&$ 361.55/760.06 $& 0.06 \\
17&9.912693, 11.082932  &$0.92_{-0.02}^{+0.02}$&$1.36_{-0.03}^{+0.02}$&$2.38_{-0.05}^{+0.05}$& $68.22_{-4.90}^{+4.75}$ &$365.79_{-18.82}^{+19.24}$ & 29.69&$ 1389.78/2816.52 $& 0.31\\
18&11.082932, 12.220607 &$0.48_{-0.22}^{+0.33}$&$1.37_{-0.02}^{+0.02}$&$2.14_{-0.06}^{+0.05}$& $15.13_{-1.13}^{+1.57}$ &$209.76_{-15.81}^{+15.44}$ &11.61 &$1229.89 /2496.73 $& 0.49\\
19&12.220607, 12.748669 &$0.74_{-0.19}^{+0.22}$&$1.23_{-0.02}^{+0.02}$&$2.67_{-0.16}^{+0.15}$& $19.54_{-3.14}^{+3.46}$ &$465.04_{-33.77}^{+31.40}$ & 17.03&$ 766.07/1569.10 $& 0.02 \\
20&12.748669, 13.396796 &$0.83_{-0.17}^{+0.18}$&$1.05_{-0.01}^{+0.01}$&$2.70_{-0.09}^{+0.09}$& $19.90_{-6.33}^{+6.79}$ &$442.12_{-17.33}^{+18.50}$ &25.81 &$977.60/1992.16 $& 0.03\\
21&13.396796, 13.542326 &unconstrained&$0.92_{-0.02}^{+0.02}$&$2.56_{-0.12}^{+0.12}$& $6.55_{-1.28}^{+1.19}$ &$317.46_{-18.10}^{+18.37}$ &32.84 &$ -101.63/-166.30$& 0.06\\
22&13.542326, 13.836101  &$0.97_{-0.04}^{+0.04}$&$1.30_{-0.04}^{+0.04}$&$3.19_{-0.29}^{+0.30}$& $88.41_{-16.93}^{+17.60}$ &$821.11_{-75.92}^{+78.51}$ &28.17 &$ 381.38/799.71 $& 0.50\\
23&13.836101, 14.219772  &$0.77_{-0.02}^{+0.02}$&$1.38_{-0.04}^{+0.04}$&$3.16_{-0.17}^{+0.18}$& $129.86_{-7.05}^{+6.81}$ &$662.85_{-39.75}^{+39.29}$ &37.68 &$ 626.86/1290.67 $& 0.34 \\
24&14.219772, 14.658330 &$0.88_{-0.02}^{+0.02}$&$1.56_{-0.03}^{+0.03}$&$3.34_{-0.35}^{+0.35}$& $99.75_{-6.54}^{+6.87}$ &$878.35_{-94.82}^{+89.30}$ &28.31 &$688.86/1414.67 $& 0.29\\
25&14.658330, 15.192231  &$0.83_{-0.02}^{+0.02}$&$1.20_{-0.02}^{+0.02}$&$3.29_{-0.14}^{+0.14}$& $85.71_{-8.28}^{+8.65}$ &$1242.21_{-55.87}^{+52.35}$ & 73.55 &$\bf 944.50/1925.95 $& 0.03\\
26& 15.192231, 15.338963 &$0.19_{-0.06}^{+0.06}$&$0.71_{-0.01}^{+0.01}$&$3.17_{-0.10}^{+0.10}$& $7.17_{-1.74}^{+1.75}$ &$1021.15_{-29.69}^{+30.46}$ & 203.54&$183.75 /404.45 $& 0.01 \\
27& 15.338963, 15.687352 &$0.70_{-0.02}^{+0.02}$&$1.34_{-0.05}^{+0.05}$&$3.26_{-0.13}^{+0.13}$& $204.82_{-14.66}^{+14.01}$ &$838.77_{-39.94}^{+40.87}$ &80.39 &$671.73/1380.42 $& 0.28\\
28& 15.687352, 16.251414 &$0.99_{-0.03}^{+0.03}$&$1.65_{-0.03}^{+0.03}$&$4.32_{-0.73}^{+0.76}$& $74.92_{-4.73}^{+4.59}$ &$952.13_{-5.48}^{+5.40}$ &19.15 &$\bf 822.72/1682.39 $& 0.34\\
29&16.251414, 16.407026 &$0.22_{-0.20}^{+0.32}$&$0.89_{-0.02}^{+0.02}$&$2.95_{-0.12}^{+0.12}$& $18.36_{-2.64}^{+3.46}$ &$702.41_{-30.15}^{+31.25}$ & 78.26&$ 65.37/167.70 $& 0.84 \\
30& 16.407026, 16.524165 &$0.66_{-0.03}^{+0.03}$&$1.27_{-0.05}^{+0.05}$&$3.96_{-0.27}^{+0.28}$& $189.76_{-20.48}^{+19.34}$ &$1235.85_{-73.37}^{+71.17}$ & 126.94&$ -63.64/-90.33 $&  0.63 \\
31& 16.524165, 16.705623 &$0.69_{-0.03}^{+0.03}$&$0.97_{-0.04}^{+0.04}$&$3.26_{-0.13}^{+0.13}$& $148.00_{-31.18}^{+31.97}$ &$892.21_{-38.97}^{+37.92}$ & 116.25&$ 228.55/494.06 $& 0.66 \\
32&16.705623, 16.787528   &$0.60_{-0.03}^{+0.03}$&$1.04_{-0.07}^{+0.07}$&$3.42_{-0.19}^{+0.18}$& $201.81_{-37.64}^{+37.06}$ &$1105.90_{-65.19}^{+61.52}$ & 168.73&$-243.39 /-449.83 $& 0.01 \\
33&16.787528, 17.114033  &$0.64_{-0.02}^{+0.02}$&$1.20_{-0.04}^{+0.04}$&$3.16_{-0.11}^{+0.12}$& $137.79_{-10.87}^{+11.02}$ &$791.75_{-33.91}^{+34.40}$ & 82.75 &$633.13 /1303.21 $& 0.19\\
34& 17.114033, 17.332842  &$0.69_{-0.11}^{+0.11}$&$0.98_{-0.02}^{+0.02}$&$3.74_{-0.24}^{+0.23}$& $42.63_{-12.47}^{+12.16}$ &$794.57_{-32.43}^{+31.65}$ & 58.16&$296.58 /630.11 $& 0.01\\
35& 17.332842, 17.924038 &$0.77_{-0.01}^{+0.01}$&$1.14_{-0.03}^{+0.03}$&$3.43_{-0.11}^{+0.10}$& $157.94_{-15.52}^{+15.05}$ &$943.25_{-29.66}^{+28.16}$ & 82.62&$ 1056.42/2149.80 $& 0.09\\
36&17.924038, 18.441228  &$0.75_{-0.01}^{+0.02}$&$1.30_{-0.03}^{+0.03}$&$3.27_{-0.13}^{+0.12}$& $155.01_{-10.38}^{+10.62}$ &$885.85_{-37.61}^{+36.76}$ &73.37 &$ 934.56/1906.07 $& 0.23 \\
37&18.441228, 18.542279  &$0.63_{-0.06}^{+0.06}$&$1.53_{-0.04}^{+0.04}$&$4.50_{-1.32}^{+1.29}$& $85.41_{-7.61}^{+7.68}$ &$1160.60_{-173.62}^{+167.14}$ & 39.28&$-318.51/-600.06 $& 0.33\\
38&  18.542279, 18.880490&$0.71_{-0.02}^{+0.01}$&$1.62_{-0.05}^{+0.05}$&$3.85_{-0.28}^{+0.28}$& $199.46_{-9.29}^{+9.43}$ &$1039.63_{-65.84}^{+65.72}$ &68.04 &$622.30 / 1281.55$& 0.67 \\
39&18.880490, 19.209439  &$0.79_{-0.03}^{+0.03}$&$1.50_{-0.03}^{+0.03}$&$3.14_{-0.17}^{+0.17}$& $85.44_{-5.37}^{+5.04}$ &$687.54_{-18.57}^{+18.94}$ & 36.01&$492.69/1022.33 $& 0.91\\
40& 19.209439, 19.316852 &$0.70_{-0.15}^{+0.16}$&$1.26_{-0.10}^{+0.10}$&$2.25_{-0.11}^{+0.11}$& $36.06_{-7.83}^{+8.49}$ &$158.09_{-21.90}^{+21.12}$ &16.70 &$ -419.90/-802.85 $& 0.53 \\
41& 19.316852, 19.790851&$1.04_{-0.08}^{+0.07}$&$1.61_{-0.03}^{+0.03}$&$3.01_{-1.14}^{+1.10}$& $34.35_{-3.75}^{+3.95}$ &$803.15_{-253.48}^{+253.73}$ & 9.72&$\bf  577.01/1190.97 $& 0.44\\
\hline
\end{tabular}
\end{center}
\end{scriptsize}
\end{table*}

\begin{table*}
\begin{scriptsize}
\begin{center}
\caption{UV and optical observations \& photometry of \thisgrb afterglow obtained using \swift UVOT as part of the present analysis. Magnitudes are not corrected for Galactic extinction in the direction of the burst.}
\label{table:UVOT_190530A}
\begin{tabular}{|c|c|c|c||c|c|c|c|}
\hline
\bf $\rm \bf  T_{\rm \bf  mid}$  &\bf Exp.& {\bf  Magnitude}  &\bf Filter  & \bf $\rm \bf T_{\rm \bf mid}$ &\bf Exp. & {\bf  Magnitude} & \bf Filter   \\
\bf  (s) &   \bf  (s)&  &  &\bf  (s) &  \bf  (s)&  & \\
\hline
34251.38 & 44.76 & $17.28^{+0.29}_{-0.23}$& $V$ &209070.58 &6401.85& $21.05^{+0.20}_{-0.17}$ & $U$\\ 56633.25 & 201.77 & $18.51^{+0.40}_{-0.29}$& $V$ & 219088.65 &6943.20& $21.42^{+0.22}_{-0.18}$ & $U$\\
128976.37&5751.40  & $> 18.11$& $V$ &33895.94 &89.76 & $18.32^{+0.12}_{-0.10}$ & $UVW1$\\
137010.08&194.477  & $> 18.64$& $V$ & 37930.50 &393.76 & $18.47^{+0.06}_{-0.06}$ & $UVW1$\\
34017.00 & 44.77 & $17.46^{+0.15}_{-0.13}$& $B$ & 43669.72&393.76 & $18.67^{+0.07}_{-0.07}$ & $UVW1$\\
38412.04&155.81  & $17.84^{+0.12}_{-0.11}$& $B$ & 49739.81&1046.14 & $18.95^{+0.10}_{-0.09}$ & $UVW1$\\
44171.59& 196.78 & $18.09^{+0.13}_{-0.12}$& $B$ &55095.24 &404.76 & $18.96^{+0.08}_{-0.08}$ & $UVW1$\\
55610.68& 201.74 & $18.41^{+0.18}_{-0.15}$& $B$ & 122877.43&6039.23 & $20.53^{+0.18}_{-0.15}$ & $UVW1$\\
122994.91&5945.01 & $19.16^{+0.32}_{-0.25}$& $B$ &132635.14 &6168.04 & $20.53^{+0.15}_{-0.13}$ & $UVW1$\\
133137.85&5962.255 & $19.56^{+0.29}_{-0.23}$& $B$ &34134.76 &179.78& $18.67^{+0.09}_{-0.08}$ & $UVW2$\\
33967.37& 44.77 & $17.67^{+0.11}_{-0.10}$& $U$ & 44442.89&334.09 & $19.08^{+0.08}_{-0.07}$ & $UVW2$\\
38230.71&196.78& $18.03^{+0.07}_{-0.07}$& $U$ &56121.94 &808.77 & $19.49^{+0.06}_{-0.06}$ & $UVW2$\\
43969.93& 196.77 & $18.28^{+0.09}_{-0.08}$& $U$ &123087.80 &6015.96 & $21.00^{+0.25}_{-0.21}$ & $UVW2$\\
50319.19&101.49 & $18.26^{+0.11}_{-0.10}$& $U$ & 133634.00& 6545.15& $21.33^{+0.16}_{-0.14}$ & $UVW2$\\
55403.49&201.78 & $18.61^{+0.12}_{-0.10}$& $U$ & 34312.60&69.33 & $18.44^{+0.15}_{-0.14}$ & $UVM2$\\
122949.03&5965.98& $19.72^{+0.24}_{-0.19}$& $U$ &129027.69 &5784.32 & $> 20.39$ & $UVM2$\\
132936.03&5966.08 & $19.99^{+0.25}_{-0.20}$& $U$ &137147.16 &70.35 & $20.38^{+0.48}_{-0.33}$ & $UVM2$\\
\hline
\end{tabular}
\end{center}
\end{scriptsize}
\end{table*}

\begin{table*}
\begin{scriptsize}
\begin{center}
\caption{Log of optical follow-up observations \& photometry of \thisgrb afterglow using various ground-based telescopes.}
\label{optical_table}
\begin{tabular}{|c|c|c|c||c|c|c|c|c|c|}
\hline
\bf $\rm \bf  T_{\rm \bf  mid}$  &\bf Exp.& {\bf  Magnitude}  &\bf Filter& \bf Telescope  & \bf $\rm \bf T_{\rm \bf mid}$ &\bf Exp. & {\bf  Magnitude} & \bf Filter & \bf Telescope    \\
\bf  (s) &   \bf  (s)&  &  & &\bf  (s) &  \bf  (s)&  &  &    \\
\hline
38384&120 & 17.80$\pm$0.04& B& OSN &38112 & 120& 16.55$\pm$0.13& I& OSN\\
39052&120 & 17.81$\pm$0.13 &B & OSN & 38916 &120 & 16.62$\pm$0.19&I & OSN\\
39614&120 & 17.93$\pm$0.05&B & OSN & 39455 &120 & 16.67$\pm$0.20 & I&OSN \\
40181&120 & 17.92$\pm$0.13 &B & OSN& 40045 & 120& 16.69$\pm$0.21& I& OSN\\
40723&120 & 17.99$\pm$0.08 &B & OSN & 40588 &120 & 16.70$\pm$0.17 &I & OSN\\
123069& 360&$>$19.70 &B & OSN & 122995.5 & 360 & 19.03$\pm$0.26& I& OSN\\
124709.5&360 & 1988$\pm$0.14 &B & OSN  & 124634.5 & 360& 19.45$\pm$0.18 & I& OSN\\ 
126429.5&360 & 19.98$\pm$0.15 &B & OSN & 126235 & 360 & 19.59$\pm$0.18 & I& OSN\\
37970&120 & 17.59$\pm$0.09 & V& OSN & 16262 &180 & 16.68$\pm$0.36& unfiltered & MASTER-Amur\\
38783&120 & 17.68$\pm$0.20& V& OSN & 26443 & 180& 16.88$\pm$0.18 & unfiltered & MASTER-Tunka\\
39322&120 & 17.75$\pm$0.10& V& OSN &29085 & 1260& 17.15$\pm$0.04&unfiltered &MASTER-Kislovodsk \\
39911&120 & 17.66$\pm$0.14 & V& OSN & 29086 & 1260 & 17.05$\pm$0.05& unfiltered& MASTER-Kislovodsk\\
40454&120 & 17.71$\pm$0.09 & V& OSN &31268 &1800 & 17.15$\pm$0.02& unfiltered&MASTER-Kislovodsk \\
122861&360 &$>$19.37 &V & OSN & 31268&  1800 &17.19$\pm$0.03&unfiltered & MASTER-Kislovodsk\\
124501.5&360 & 20.08$\pm$0.55 & V& OSN & 31283& 1800 & 17.03$\pm$0.05&unfiltered &MASTER-Tavrida \\
126161.5&360 & 20.11$\pm$0.15 & V& OSN & 31490&  1440& 17.07$\pm$0.06&unfiltered& MASTER-Tavrida\\
37784& 120& 17.25$\pm$0.21 & R&OSN & 33875&2700  & 17.29$\pm$0.04&unfiltered & MASTER-Tavrida\\
38643 &120 & 17.32$\pm$0.14 & R& OSN &35073 & 1800& 17.36$\pm$0.03& unfiltered&MASTER-Kislovodsk\\
39183 &120 & 17.33$\pm$0.20 &R & OSN &35257 & 1620& 17.26$\pm$0.06& unfiltered& MASTER-Tavrida\\
39772 & 120& 17.26$\pm$0.22 &R & OSN& 35278&  1800&17.43$\pm$0.04 &unfiltered & MASTER-Kislovodsk\\
40315 & 120 & 17.39$\pm$0.22 & R& OSN & 37330& 2160& 17.53$\pm$0.07& unfiltered& MASTER-Kislovodsk\\
124087 &360 & 19.34$\pm$0.25 &R & OSN & 37740& 2160& 17.37$\pm$0.09& unfiltered& MASTER-Kislovodsk\\
126024.5 &360 & 19.68$\pm$0.24 &R &OSN& 38878& 1800& 17.40$\pm$0.04&unfiltered & MASTER-IAC\\
298088 & 3720 &21.71$\pm$0.30 &R & OSN & 117140 &4320 & 19.45$\pm$0.08& unfiltered & MASTER-Kislovodsk\\
384315.5 & 3000 & $>$21.85& R & OSN & 117242&3960 & 19.53$\pm$0.11& unfiltered &MASTER-Kislovodsk \\
116899.2 & 300x11 & 19.39$\pm$0.05 & R & SAO & 190548.23 & 2600 &20.29$\pm$0.06 &r &GIT \\
275525.6& 60x15 & 21.30$\pm$0.30 & R & HCT & 279387.04& 4800 &21.03$\pm$0.15 &r & GIT\\
1160748 & 1950 &$>$ 22.70 & R & CAHA &122667 & 3600&20.36$\pm$0.37 &r &RC80 \\
209088& 3600 &20.15$\pm$0.11 & r& RC80 & & & & & \\
\hline
\end{tabular}
\label{tab:opticaldata}		
\end{center}
\end{scriptsize}
\end{table*}

\begin{table*}
  \caption{Log of the observations for the deep search of the host galaxy of \thisgrb using 3.6m DOT and 10.4m GTC telescopes. The limiting magnitudes are in the AB system and have not been corrected for foreground extinction.}
  \begin{tabular}{|c|c|c|c|c|}
  \hline
  \bf Date of Observations & \bf Exposure (s) & \bf Limiting magnitude & \bf Filter & \bf Telescope\\ \hline
  14.10.2020   &   2x600   & $> 24.0$  &  U  & 3.6m DOT \\
  14.10.2020    &   1x600   & $> 23.4$  &  B  & 3.6m DOT \\
  14.10.2020    &   1x600    & $> 23.1$  &  V  & 3.6m DOT \\
  14.10.2020    &   1x600    & $> 22.7$ &  R  & 3.6m DOT \\
  14.10.2020    &   1x600    & $> 22.3$  &  I  & 3.6m DOT \\ \hline \hline
  23.05.2021    &   10x90    & $> 23.2$  &  g & 10.4m GTC \\
  23.05.2021    &   9x60    & $> 23.4$  &  r & 10.4m GTC \\
  23.05.2021    &   5x60    & $> 22.6$  &  i & 10.4m GTC \\
  23.05.2021    &   8x50    & $> 22.1$  &  z & 10.4m GTC \\
  \hline
  \end{tabular}
  \label{host}
\end{table*}

\bsp	
\label{lastpage}
\end{document}